\documentclass[aip,jcp,reprint,amsmath,longbibliography]{revtex4-1}
\usepackage{mycommands}
\renewcommand{\S}[1]{\ensuremath{\mathrm{S}_{#1}}}
\newcommand{\T}[1]{\ensuremath{\mathrm{T}_{#1}}}
\newcommand{\C}[1]{\ensuremath{C_\mathrm{#1}}}

\usepackage{booktabs}
\usepackage{chemformula}
\newcommand{\activestate}{n}
\usepackage{amssymb}
\usepackage{physics}
\usepackage{bm}%
\usepackage{enumerate}
\usepackage{makecell}
\usepackage{booktabs}
\usepackage{textgreek}

\usepackage{chemmacros}
\usepackage{hyperref}

\chemsetup{modules=reactions}
\chemsetup{
  formula = mhchem,
  reactions/tag-open=(R,
  reactions/tag-close=),
}

\makeatletter
\@ifundefined{ignorespacesafterend}{\def\ignorespacesafterend{\global\@ignoretrue}}{}
\newenvironment{subreactions}{%
  \refstepcounter{reaction}%
  \protected@edef\theparentequation{\thereaction}%
  \setcounter{parentequation}{\value{reaction}}%
  \setcounter{reaction}{0}%
  \def\thereaction{\theparentequation\alph{reaction}}%
  \ignorespaces
}{%
  \setcounter{reaction}{\value{parentequation}}%
  \ignorespacesafterend
}

\begin{document}
\title{A MASH simulation of the photoexcited dynamics of cyclobutanone}
\author{Joseph E. Lawrence}
\email{joseph.lawrence@nyu.edu}
\affiliation{Department of Chemistry and Applied Biosciences, ETH Zurich, 8093 Zurich, Switzerland}
\affiliation{Simons Center for Computational Physical Chemistry, New York University, New York, NY 10003, USA}
\affiliation{Department of Chemistry, New York University, New York, NY 10003, USA}
\author{Imaad M. Ansari}
\affiliation{Department of Chemistry and Applied Biosciences, ETH Zurich, 8093 Zurich, Switzerland}
\author{Jonathan R. Mannouch}
\affiliation{Hamburg Center for Ultrafast Imaging, Universit\"at Hamburg and the Max Planck Institute for the Structure and Dynamics of Matter, Luruper Chaussee 149, 22761 Hamburg, Germany}
\author{Meghna A. Manae}
\author{Kasra Asnaashari}
\affiliation{Department of Chemistry and Applied Biosciences, ETH Zurich, 8093 Zurich, Switzerland}
\author{Aaron Kelly}
\affiliation{Hamburg Center for Ultrafast Imaging, Universit\"at Hamburg and the Max Planck Institute for the Structure and Dynamics of Matter, Luruper Chaussee 149, 22761 Hamburg, Germany}
\author{Jeremy O. Richardson}
\email{jeremy.richardson@phys.chem.ethz.ch}
\affiliation{Department of Chemistry and Applied Biosciences, ETH Zurich, 8093 Zurich, Switzerland}
\date{2nd April 2024}
\begin{abstract}
    In response to a community prediction challenge, we simulate the nonadiabatic dynamics of cyclobutanone using the mapping approach to surface hopping (MASH). %
We consider the first 500\,fs of relaxation following photo-excitation to the \S2 state and predict the corresponding time-resolved electron-diffraction signal that will be measured by the planned experiment.
397 ab-initio trajectories were obtained on the fly with state-averaged complete active space self-consistent field (SA-CASSCF) using a (12,11) active space. To obtain an estimate of the potential systematic error 198 of the trajectories were calculated using an aug-cc-pVDZ basis set and 199 with a 6-31+G* basis set.
MASH is a recently proposed independent trajectory method for simulating nonadiabatic dynamics, originally derived for two-state problems.
As there are three relevant electronic states in this system, we used a newly developed multi-state generalisation of MASH for the simulation: the uncoupled spheres multi-state MASH method (unSMASH).
This study therefore serves both as an investigation of the photo-dissociation dynamics of cyclobutanone, and also as a demonstration of the applicability of unSMASH to ab-initio simulations. 
In line with previous experimental studies%
, we observe that the simulated dynamics is dominated by three sets of dissociation products, \ch{C3H6 + CO}, \ch{C2H4 + C2H2O} and \ch{C2H4 + CH2 + CO}, and we interpret our predicted electron-diffraction signal in terms of the key features of the associated dissociation pathways.

\end{abstract}
\maketitle

\section{Introduction}

In recent years, there has been a significant increase in experimental capabilities making it possible to follow ultrafast photochemical processes in real time. %
Nonetheless, obtaining a clear mechanistic interpretation of molecular quantum dynamics often requires theoretical calculations to be performed in tandem.
Computer simulations of photochemistry are complicated by the breakdown of the Born--Oppenheimer approximation.
We thus require methods that can describe nonadiabatic dynamics involving transitions between electronic states.

In any theory, it is desirable to minimize the complexity of the description as much as possible, in order to obtain a simple intuitive picture of the key processes at play. Nonadiabatic approaches based on independent semiclassical trajectories achieve just that, of which Tully's fewest-switches surface hopping (FSSH)\cite{Tully1990hopping} is the most commonly used. In addition, the favourable computational scaling of independent trajectories with system size means that a high-level of electronic-structure theory can be employed, which is crucial for making quantitative predictions of experiments. %
Unfortunately, FSSH has a number of well-known problems due to inconsistency and overcoherence.\cite{Subotnik2016review}

The mapping approach to surface hopping (MASH)\cite{MASH} is a recently proposed alternative to the FSSH algorithm.
It has a rigorous derivation based on mapping approaches\cite{Meyer1979nonadiabatic,Stock1997mapping,spinmap} but instead of using a mean-field force, it hops between adiabatic states, similarly to FSSH\@.
The key difference between MASH and FSSH is that MASH's dynamics are deterministic rather than stochastic.
This has an important benefit, ensuring consistency at all times between the electronic variables and the active surface.
Also, the MASH derivation uniquely determines how the momentum should be rescaled at attempted hops.  One should rescale along the direction of the nonadiabatic coupling vector, and reflect in the case of all forbidden hops.
Although this prescription is identical to Tully's original concept,\cite{HammesSchiffer1994FSSH}
many alternative suggestions have been made,\cite{Mueller1997FSSH,Jasper2003FSSH,Subotnik2016review}
and in practice, approximations are often taken.\cite{Mai2018SHARC}
Previous work has shown that MASH is often more accurate than FSSH for a range of model systems\cite{MASH} and there is reason to believe it can even be more accurate than ab initio multiple spawning (AIMS)\cite{BenNun2002AIMS,Curchod2018review} for photochemical problems.\cite{Molecular_Tully_JM}
MASH was shown to correctly recover Marcus theory rates,\cite{MASHrates} where FSSH is known to require complicated decoherence corrections.\cite{Landry2011hopping,Landry2012hopping}
Additionally, unlike other mapping approaches,\cite{Meyer1979nonadiabatic,identity,spinmap,multispin,Miller2016Faraday} 
MASH rigorously captures the detailed balance necessary to thermalize to the correct equilibrium distribution.\cite{thermalization}

In this work, we simulate the photochemistry of cyclobutanone, in response to a community prediction challenge initiated by the Journal of Chemical Physics (JCP). In the experiment that we seek to predict, cyclobutanone is initially photoexcited using a 200\,nm pump pulse, which is assumed to excite the molecule from the \S0 to the \S2 adiabatic electronic state. The resulting nonequilibrium dynamics are then measured using time-resolved electron diffraction. Prior to the community challenge, all previous theoretical studies of cyclobutanone have instead considered the dynamics starting in \S1,\cite{Kao_ringstrain,diau_femtochemistry_2001,Xia_cyclobuitanone_mscaspt2,Liu_cyclobutanone_AIMS} meaning that the present study covers new ground. Simulating the photoexcited dynamics initialized in \S2 also poses additional theoretical challenges. \S2 is known to be a Rydberg state\cite{cyclobutanone_rydberg_spectrum, Kuhlman_cyclobutanone_mctdh_modelPES} and therefore requires a set of diffuse electronic basis functions to correctly describe the dynamics.

The original MASH formulation was limited to two-state problems and is therefore not directly applicable to the present study.
Although a multi-state version of MASH has recently been proposed by Runeson and Manolopoulos,\cite{Runeson2023MASH,Runeson2024MASH}
we note that this does not reduce to the original two-state version and thus loses some of the key benefits of the MASH approach. This is particularly significant for photochemical applications, where the dynamics is expected to largely be a succession of effective two-state nonadiabatic transitions. 
We therefore developed a new multi-state generalization of MASH which does recover the original version in the case that two states are uncoupled from all others.
This is the method employed for the present study.
This new approach will be described in detail alongside application to a series of benchmarks in a forthcoming publication.\cite{MultiMASH} 

The present work describes the first implementation of our new multi-state MASH method using ab initio electronic-structure methods.
Due to the time constraints imposed by JCP's challenge, we were not able to implement the most powerful version of the algorithm and thus we limit ourselves to studying the internal conversion between singlet states only and employing an initial distribution obtained from a vertical transition according to the Franck--Condon principle.
We note, however, that the MASH formalism can be rigorously extended to treat intersystem crossing to triplet states and to describe the excitation pulse explicitly.\cite{JonathanUnpublished}
Future work will test the impact of this more complete description of the nonadiabatic process.
This study therefore serves to provide a proof of principle that MASH can be used for realistic simulations of photochemistry and can compete with more established methods such as FSSH and AIMS.

\section{Methods}

Before describing the algorithm used for generating MASH trajectories, we first turn to the question of sampling initial conditions.
Most nonadiabatic trajectory simulations are initialized using a Wigner function based on a harmonic approximation around the ground-state equilibrium geometry.
Ordinarily, we would have used this standard approach within a MASH simulation.
However, in the \S0 state, cyclobutanone has a low-frequency puckering mode around a \C{2v} geometry, which is very anharmonic and depending on the electronic-structure method used may even be predicted to be a double well with a low barrier.
It is therefore clear that the harmonic approximation is not valid for this mode.
At the level of density-functional theory (B3LYP/def2-TZVP as implemented in ORCA),\cite{Neese2012orca} we located the minimum-energy pathway between the two minima using the nudged-elastic band method \cite{Jonsson1998NEB} and performed a ``stream-bed walk''\cite{Nichols1990mep} (in Cartesian coordinates) up the other side.
We call this the puckering path and from now on we employ mass-weighted coordinates using atomic masses of the most common isotopes, unless otherwise stated.
A one-dimensional discrete variable representation (DVR) calculation \cite{Light1985DVR} was carried out to obtain the nuclear wavefunctions along the puckering path.
The predicted fundamental vibrational transition is 33~cm$^{-1}$ in good agreement with the experimental value of 35~cm$^{-1}$ from far-infrared spectroscopy.\cite{Durig1966cyclobutanone}
There is also reasonable agreement with the higher excited vibrational states (supplementary material).%
\footnote{Such good agreement was not achieved when using a simple scan of the potential along the normal mode corresponding to the puckering motion.}
We learned that the experiment will slightly heat the sample to avoid condensation.\cite{WolfPrivate}
Therefore, one-dimensional positions and momenta were sampled according to the thermal Wigner function at 325\,K, which can itself be evaluated in terms of the DVR wavefunctions (supplementary material).%
\footnote{We note that a classical Boltzmann sampling along the puckering path would probably have been sufficient due to the fact that the thermal energy is sufficiently larger than the energy-level spacing of the puckering vibrations.  However, our approach is more generally applicable at any temperature and is consistent with the quantum-mechanical treatment of the perpendicular modes.}
Hessians were computed at a set of points along the puckering path and interpolated (in Cartesian coordinates).
Translational and rotational modes along with the vector tangential to the path were projected out.
It was found that the perpendicular frequencies are much larger than the energy-level spacing of the puckering vibrations and vary relatively slowly along the puckering path (supplementary material), which justifies our approach.
Finally, the modes perpendicular to the puckering path were sampled using the thermal Wigner function within the standard harmonic approximation, and angular momentum was sampled from a classical Boltzmann distribution.
More elaborate schemes for sampling Wigner distributions have been proposed, but these are not yet applicable to such large systems.\cite{Ple2019wigner}

Now that the initial conditions for the nuclei are specified,
we discuss the electronic-structure method used for the dynamical simulations.
In order to capture the excited-state manifold and the bond-breaking dynamics after photo-excitation, a multireference method is required; we employ the state-averaged complete active space self-consistent field (SA-CASSCF) method in order to simultaneously describe the $\S0$, $\S1$ and $\S2$ states.
In SA-CASSCF, the ground and electronic excited states are optimized simultaneously with a common set of orbitals but different configuration-interaction (CI) coefficients, which are constrained to form an orthonormal set.
This ensures that the electronic states are orthogonal, which is particularly important when using the overlaps of the electronic wavefunction in the dynamics. 
The common set of orbitals also allow for efficient implementations of analytic
gradients and nonadiabatic coupling vectors (NACV).
All CASSCF calculations were performed using Molpro 2023,\cite{MOLPRO2023} {and used a Slater basis with projection onto the singlet space. This is recommended as the default option within Molpro for CASSCF calculations due to increased computational efficiency over using configurational state functions. }

The excitation from the ground electronic state, \S0, to the first excited state, \S1, is locally characterized by a transition from a non-bonding $n$ orbital to an antibonding $\pi^{*}$ orbital of the carbonyl\cite{cyclobutanone_uv-vis, diau_femtochemistry_2001, Kao_ringstrain, Liu_cyclobutanone_AIMS} while the excitation to $\S2$ is characterized by a transition to a Rydberg $3s$ orbital.\cite{cyclobutanone_rydberg_spectrum, Kuhlman_cyclobutanone_mctdh_modelPES}
{Additionally, previous experimental\cite{Kao_ringstrain, diau_femtochemistry_2001} and theoretical studies\cite{Liu_cyclobutanone_AIMS} indicate that C--C bonds are cleaved during the relaxation dynamics.
It is therefore critical to choose a basis set that includes diffuse orbitals to accurately describe the Rydberg orbital, and an active space that is able to characterize the excited state manifold while simultaneously allowing for a good description of the possible C--C bond breaking processes.}

{
Due to the time constraints imposed by the prediction challenge, we were restricted by the size of the basis set and active space  that could be used. 
We considered two different basis sets with diffuse functions, the Dunning aug-cc-pVDZ basis and the Pople 6-31+G* basis and three different active spaces
(detailed descriptions along with illustrations of the active spaces are provided in the supplementary information).
We benchmarked vertical excitation energies (Table I in the supplementary information) and found that the aug-cc-pVDZ with a (12,11) active space resulted in the best balance between accurate vertical excitation energies, computational cost, and being large enough to accurately describe the bond-breaking dynamics. 
Over the course of the dynamics, this (12,11) active space has sufficient flexibility to be able to describe three simultaneous bond breaking events.
The aug-cc-pVDZ calculations gave a vertical excitation energy (\SI{6.231}{\electronvolt}) that was closer to the experimental \S2 peak maximum  (\SI{6.4}{\electronvolt})\cite{cyclobutanone_rydberg_spectrum} than with 6-31+G* (\SI{6.846}{\electronvolt})\footnote{This is true even after accounting for the width of the rather wide absorption peak observed in \cite{cyclobutanone_rydberg_spectrum}, with a full width half maximum of around \SI{0.3}{\electronvolt}.} and also much closer to the pump laser frequency of the planned experiment (\SI{6.2}{\electronvolt}).}

{On the basis of this data, unless otherwise stated, we choose to present the calculations performed using the aug-cc-pVDZ basis with a (12,11) active space in the main text. 
Analogous calculations using the 6-31+G* basis are however provided in the supplementary material, and differences between the two sets of calculations are used to help assess the sensitivity of the predicted results to details of the electronic structure.}

At the \C{2v} geometry, the $n \to \pi^{*}$ transition is electric-dipole forbidden while the $n \to 3s$ is electric-dipole allowed.
In addition, the pump-pulse energy is on resonance with the \S2 excitation.
We therefore utilized the Franck--Condon approximation to initialize the electronic state in \S2 (according to the MASH procedure described below). %
In this way, we allowed vertical transitions for the entire initial distribution and do not take account of the bandwidth of the laser pulse, as it is not obvious how to do this in a rigorous way without explicitly simulating the light field.
We optimized the structures of relevant minimum-energy conical intersections (MECI) and crossing points (MECP) and present their relative energies in the supplementary material.
Both the \S2/\S1 MECI and the \S2/\T2 MECP are below the Franck--Condon energy, implying that they are both energetically accessible.  However, the spin-orbit coupling (SOC) at the S$_2$/T$_2$ MECP is only 5 cm$^{-1}$ which suggests that intersystem crossing may be quite unlikely. Similar conclusions are indicated for relaxation from the S$_1$ state, in which the SOC is zero at the \S1/\T1 MECP\@. %
Taking cues from the study by Liu and Fang,\cite{liu2016new} we identified three S$_1$/S$_0$ MECI structures, which are lower or comparable in energy to the S$_1$/T$_1$ MECP\@.
Together, these results suggest that intersystem crossing is not important for the photodynamics of cyclobutanone.
This is in agreement with previous theoretical studies of photochemistry of cyclic ketones on the \S1 state, which show that intersystem crossing only plays a role in molecules with rings of 5 or 6 carbon atoms.
\cite{xia2015excited}

Our approximation of neglecting the triplet states in the dynamics of cyclobutanone is further tested by selecting 10 trajectories from the final 
aug-cc-pVDZ 
 set, along which we simulated the electronic dynamics including 3 triplet states (i.e., a 6-state SA-CASSCF) with a (12,11) active space, starting on the \S2 state. 
The resulting electronic dynamics showed %
less than 0.6\% population transfer to the triplet manifold.
Although this simple test is not completely reliable, especially considering the small number of trajectories considered, it nonetheless lends further weight to justify our neglect of the triplet states.
Further details of these calculations and plots of the triplet populations over time for a couple of representative trajectories are given in the supplementary material.

We now turn to our choice of dynamics method, MASH\@.
For clarity, we give here a brief discussion of the important features of the method, highlighting key differences to FSSH\@.
For a more detailed discussion see the supplementary material as well as  Refs.~\citenum{MASH}, \citenum{MASHrates} and \citenum{MultiMASH}. Before discussing the multi-state generalisation it is instructive to introduce the key ideas behind the original two-state implementation of MASH.\cite{MASH} In this case there are two key differences to FSSH: first, how the active surface is determined, and second, how the initial electronic variables (wavefunction coefficients) are chosen. Unlike FSSH, in MASH the active state is obtained deterministically. In the two-state case, the active state is determined by the sign of the $z$ component of the Bloch sphere, $S^{(2,1)}_z=|c_2|^2-|c_1|^2$, such that when $S^{(2,1)}_z>0$ the active state is $\activestate=2$, and when $S^{(2,1)}_z<0$ the active state is $\activestate=1$. That the dynamics can be fully deterministic might at first seem surprising, particularly given that it is the stochastic nature of the hopping in FSSH that allows it to capture wavepacket bifurcation. However, the stochastic nature of the FSSH hops is effectively replaced in MASH by an initial sampling of the wavefunction coefficients. For example, to initialize a system in a pure state on adiabat 2, in FSSH one chooses $S^{(2,1)}_z=1$ and hence $S^{(2,1)}_x=S^{(2,1)}_y=0$.  However, in MASH $S^{(2,1)}_z$ is instead sampled from the probability density $\rho_2(S^{(2,1)}_z)=2 h(S^{(2,1)}_z) |S^{(2,1)}_z|$ (where $h(x)$ is the Heaviside step function), with $S^{(2,1)}_x$ and $S^{(2,1)}_y$ chosen uniformly from the corresponding circle on the Bloch sphere.
It is due to this ensemble that MASH is able to describe wavepacket bifurcation.

MASH has been shown to offer a number of formal improvements over FSSH\@. Firstly, unlike FSSH it can be rigorously derived as a short-time approximation to the quantum--classical Liouville equation (QCLE), meaning that it can in principle be systematically improved towards this limit. Secondly, and perhaps most importantly, MASH does not suffer from the inconsistency error of FSSH\@.
{This is because the deterministic nature of the MASH algorithm means that the electronic variables are always directly related to the current active state. In contrast, in FSSH the wavefunction coefficients can become completely inconsistent with the active state.}
 These methodological improvements are expected to lead to more reliable predictions for no extra computational cost.  In fact, for a series of different model systems, MASH has been shown to be as accurate or more accurate than FSSH at reproducing quantum-mechanical benchmark results.\cite{MASH,MASHrates}
Decoherence corrections can be derived rigorously for MASH,\cite{MASH} although in the vast majority of cases, the dynamics are already accurate enough without them.\cite{Molecular_Tully_JM}

While the original MASH method was derived for two-state systems only, we have recently proposed an $N$-state generalisation of MASH ideally suited for simulating photochemical systems, which we call the uncoupled-spheres multi-state MASH method (unSMASH).\cite{MultiMASH} 
The unSMASH method generalises the original two-state MASH by treating possible transitions between pairs of adiabatic states independently. This is done by introducing $N-1$ independent effective two-state Bloch spheres between the current active state and each of the other states: $\bm{S}^{(n,j)}$ for $j\neq n$, $j=1,\dots,N$. Each sphere then evolves as it would in the original two-state MASH for the truncated electronic space consisting of the two corresponding adiabatic states.
 Attempted hops occur when one of the $S_z^{(n,j)}$ changes sign. As in the two-state theory, the hops are accepted or rejected according to whether there is enough kinetic energy in the direction of the NACV between the active state and the possible new state. The component of the momentum along the NACV is then either rescaled to conserve energy in the case of allowed hops, or reflected in the case of rejected (frustrated) hops. The unSMASH method is a rigorous short-time approximation to the QCLE when there is only coherence between one pair of adiabatic states at a given time. It is therefore well suited to photochemical problems involving a series of successive separate transitions between adiabatic surfaces, as one would expect in the photochemical relaxation of a typical organic molecule  such as cyclobutanone.

The integrator used to evolve the unSMASH equations of motion is closely related to those suggested previously for FSSH.\cite{Meek2014FSSHIntegrator,Jain2016AFSSH,Mai2020FSSHChapter,Jain2022FSSH}  A full mathematical description of the integrator is given in the supplementary material; here we simply give some of its key features. The basic structure of the integrator involves first propagating the nuclear positions and momenta from $t$ to $t+\delta t$ using velocity Verlet, before the electronic variables are evolved from $t$ to $t+\delta t$ using {a unitary operator based on} information calculated at geometries $\bm{q}(t)$ and $\bm{q}(t+\delta t)$. Finally any attempted hops are treated, along with their associated momentum rescalings. As with FSSH propagation schemes, one must contend with the fact that close to conical intersections the NACV is very sharply peaked. This means that algorithms that rely solely on the NACV can require arbitrarily small time steps in order to capture the corresponding electronic transition. For this reason it is common to use an effective time-averaged nonadiabatic coupling that can be calculated from the overlap between the adiabatic wavefunctions at successive time steps rather than to compute the NACVs explicitly.\cite{Meek2014FSSHIntegrator,Jain2016AFSSH,Mai2020FSSHChapter,Jain2022FSSH} This can be implemented for MASH in the same way as it is for FSSH\@. However, in the present work we found that the calculation of NACVs was significantly less computationally expensive than computing the overlaps. For this reason, the integrator we employ uses a mixed scheme, only calculating overlaps when the adiabatic surfaces come close together ($|\Delta V|<2000\,{\rm cm}^{-1}$), and otherwise using the NACVs. We note that using the NACVs rather than the overlaps has an additional advantage when combined with CASSCF, as it means that discontinuities in the active space do not lead to spurious nonadiabatic transitions. This is similar to the advantage of calculating the forces analytically rather than by finite difference, which can lead to unphysical sudden large changes in the momenta when encountering a discontinuity in the energy.

\begin{figure}[t]
    \centering
    \includegraphics[width=\linewidth]{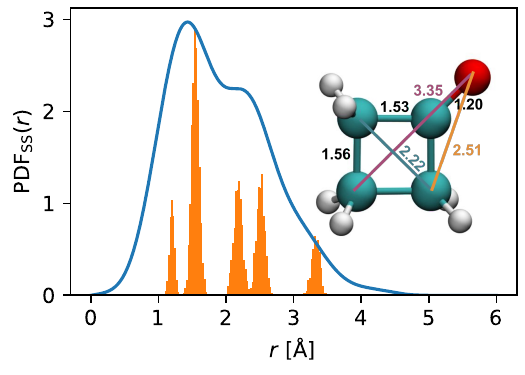}
    \caption{Calculated steady-state PDF (in $\AA^{-2}$) for the initial distribution.  Also shown in orange is a histogram of the atom pair distribution (carbons and oxygens only).
    The inset shows the atom pair distances (in $\AA$) for the $\C{2v}$ geometry.
    }
    \label{fig:steadystate}
\end{figure}
\section{Results}

\begin{figure*}[t]
    \centering
    \includegraphics[width=\textwidth]{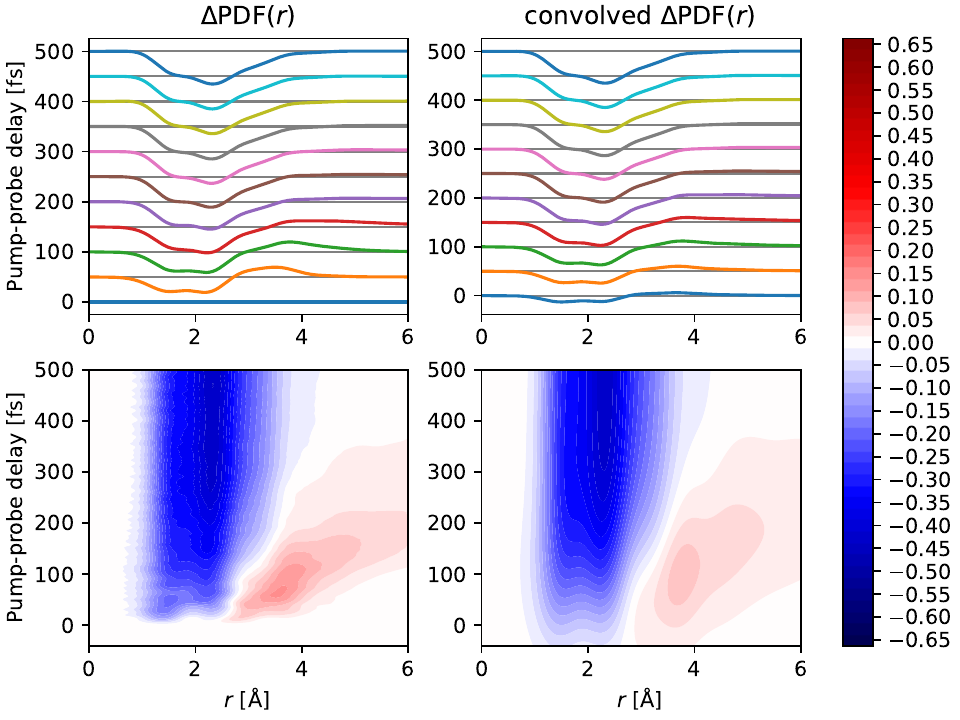}
    \caption{Simulated ultrafast electron-diffraction results.
    The panels on the left show the change in the probability density function relative to the initial configuration.
    The panels on the right show the same data convolved with a 160\,fs (FWHM) Gaussian to simulate the instrument response function.
    Blue is loss, red is gain, with equally spaced contour levels showing the height of the $\Delta \mathrm{PDF}(r)$ signal relative to the maximum peak height in the steady-state PDF.    
    }
    \label{fig:gued}
\end{figure*}

In total we sampled 200 sets of initial positions and momenta from the Wigner function at 325\,K\@. In each case the initial active state was \S2, and the initial spheres were randomly sampled as described above. These initial samples were used to launch 200 separate unSMASH trajectories 
(using the aug-cc-pVDZ basis)
with a time step of 0.5\,fs. Of these initial 200, a total of 198 trajectories were run for the desired total time of 500\,fs.
For {123} of these trajectories, spin contamination in the excited-state manifold meant that the SCF step in the SA-CASSCF cycle did not converge within the maximum number of allowed iterations, which was chosen as 160. This typically only occurred at later times after the molecule had already reached \S0, and had already undergone the primary dissociation step, with the majority {(77)} having already reached at least 200\,fs before the SA-CASSCF failed to converge. Rather than discarding these trajectories (which would bias the results), we elected to finish those that had already reached \S0 by running them for the remaining time on the ground state. This was done using state-specific CASSCF (SS-CASSCF) with the same basis and active space size as for the three-state SA-CASSCF calculations. {Note, the initial momentum and position for the SS-CASSCF part of the trajectory were simply taken from the end of the previous convergent SA-CASSCF step, and the trajectory continued for the remaining time using velocity Verlet integration.}  
The 2 trajectories that could not be completed were those that crashed due to spin contamination while on an excited electronic state; these trajectories are excluded from all analysis that follows. 

To compare our simulated results with the results of the proposed experiment, the final set of {198} trajectories (of length 500\,fs) were used to generate electron-diffraction signals. This was done based on elastic-scattering %
calculations within the 
independent-atom model \cite{Centurion2022review} using the ELSEPA program\cite{Salvat2021elsepa} as described in the supplementary material.
Note that it would in principle be possible to go beyond this approximation, in the spirit of Ref.~\citenum{Yang2020UED}, using the ab initio two-electron density of the MASH active state, provided by the CASSCF calculations.
However, in this work, we assume the independent-atom model to be sufficient.
The electron-diffraction signal was transformed {(including a Gaussian damping function)\cite{Centurion2022review}} to obtain the atomic pair distribution functions (PDF).
Then $\Delta$PDF was defined as the difference between the PDF at time $t$ and the steady-state PDF as shown in Fig.~\ref{fig:steadystate}. %
As experimental electron-diffraction results are typically presented with arbitrary units,\cite{Wolf2019CHD}
all results are given relative to the maximum peak height in the steady-state PDF, i.e.
\begin{equation} \label{PDF}
    \Delta \mathrm{PDF}(r,t) =\frac{ \mathrm{PDF}(r,t)-\mathrm{PDF}_{\rm SS}(r)}{\max(\mathrm{PDF}_{\rm SS}(r))}
\end{equation}
Finally, the results were convoluted with a Gaussian (160\,fs FWHM) to simulate the instrument response function (accounting for both the width of the pump pulse as well as the detector).

In Fig.~\ref{fig:gued}, we present the the electron-diffraction signal predicted by our MASH simulation, both before and after convolution. From these results one immediately obtains a qualitative picture of the dynamics after photo-excitation.
In the unconvolved signal, after only 50\,fs the predicted electron-diffraction signal shows a significant positive peak at around $3.25\,\AA$, along with a corresponding negative peak in the region 1.25--2.5\,$\AA$. %
At 100\,fs the positive peak has broadened and shifted towards $4\,\AA$ while the negative peak has deepened. This trend carries on until around 350\,fs after which point the negative peak approaches a steady state and the positive peak has broadened and shifted to such large distances as to become almost invisible. 
Although in the convolved signal the locations and heights of the peaks are somewhat modified, the same qualitative behaviour can be observed. This behaviour is clearly indicative of a rapid dissociation following the photoexcitation of cyclobutanone. The dissociation leads to a depletion of short bond distances due to bond breaking, with a corresponding increase at continually larger and larger distances as the resulting fragments move apart. In contrast a simple ring-opening reaction, without dissociation, would result in a persistent positive signal between $4\,\AA<r<6\,\AA$, as seen for example in the photo-induced ring-opening of cyclohexadiene.\cite{Wolf2019CHD} From these results we can ascertain that the majority of the dissociation occurs within the first 250--300\,fs, with the onset of dissociation occurring very rapidly at around 50\,fs. 

Although the electron diffraction signal contains a large amount of information, and gives an immediate qualitative picture of the nuclear dynamics, it is nevertheless hard to immediately extract detailed mechanistic information from the signal alone. This is why, despite the ever increasing resolution of experiments, molecular simulations are an important tool in understanding complex photochemical processes. In the following we consider what the additional information available from our molecular simulations tells us about the dissociation process before returning to discuss how signatures of these features could be observed in the experimental electron-diffraction signal.

\subsection{Electronic dynamics}
\begin{figure}[t]
    \centering
    \includegraphics{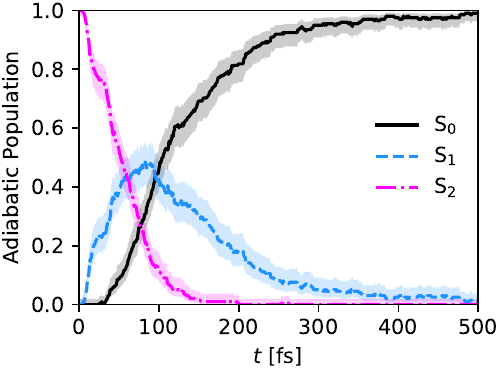}
    \caption{Average (unconvolved) adiabatic populations as a function of time for the 198 trajectories. Shaded region shows an approximate 95\% confidence interval (the Wilson score interval).\cite{Wilson1927BinomialErrors}  
    }
    \label{fig:pops}
\end{figure}

We first consider what our simulation predicts about the electronic dynamics after excitation. Figure \ref{fig:pops} shows the average population on each adiabatic state as a function of time. From this we can clearly see that the system undergoes rapid electronic relaxation, with a half life on \S2 of about {50\,fs}. It appears that the system primarily undergoes a sequential transition, first from \S2 to \S1 and then from \S1 to \S0. The resulting half life for the combined excited-state manifold (\S2 + \S1) is predicted to be about {100\,fs}, with about {90\%} of the molecules having relaxed to \S0 by {250\,fs}. This timescale matches closely the dynamics seen in the electron-diffraction signal, indicating that the energy released into the nuclear degrees of freedom by the electronic relaxation leads rapidly to dissociation.  

\subsection{Reaction products}
To obtain a clearer quantitative picture of the photodissociation process it is helpful to analyse the trajectories according to the fragments formed in the dissociation process. We define a molecular fragment as a series of atoms that form a connected graph, where nodes of the graph correspond to atoms and edges of the graph %
indicate that the
 distance between the corresponding pair of atoms %
is within a cutoff distance of $r<2.0\,\AA$.{\footnote{{Note the same cutoff was used for all pairs of atoms}}}  Table \ref{tab:product-yields} shows the total yield of all observed molecular fragments at 500\,fs. We note that, due to the presence of secondary dissociation processes that occur after 500\,fs, this is not likely to be equivalent to the final product distribution. We can see that the most abundant product is carbon monoxide (\ch{CO}), closely followed by cyclopropane/propene (\ch{C3H6}). There are also significant numbers of ethene (\ch{C2H4}) as well as ketene (\ch{C2H2O}) and the highly reactive methene (\ch{CH2}), along with a handful of other fragments that are only observed in a small number of trajectories. Most notable of these are the {7} remaining  \ch{C4H6O} molecules that have not yet dissociated, of which 3 have already undergone ring-opening, 3 remain as cyclobutanone and 1 has undergone a rearrangement to form cyclopropanal. We note that, based on the analysis that follows, we expect the majority of them to eventually dissociate to give \ch{C3H6 + CO}.

Further insight can be gained by grouping the trajectories according to the molecular fragments they produce. Table \ref{tab:reaction-products} shows the three most common sets of products present at 500\,fs, along with the frequency at which they are observed. That these three reaction products should dominate is not a surprise based on previous theoretical and experimental work\cite{Kao_ringstrain,diau_femtochemistry_2001,Xia_cyclobuitanone_mscaspt2,Liu_cyclobutanone_AIMS,Denschlag1968Cyclobutanone,Trentelman1990Cyclobutanone} such as that of Trentelman \emph{et al.}\cite{Trentelman1990Cyclobutanone}~who rationalised their %
measurements on the \ch{CO} produced 
after photo-excitation of cyclobutanone with 193\,nm light in terms of these 3 possible reaction products. Following this earlier work, it is helpful to distinguish reactions according to whether the  primary dissociation event produces ethene (\ch{C2H4}), labelled the C2 pathway, or cyclopropane/propene (\ch{C3H6}), labelled the C3 pathway.
The C3 pathway is the simplest, 
\begin{reaction}
    \ch{C4H6O} \rightarrow \ch{C3H6} + \ch{CO}, \label{rxn:C3}
\end{reaction}
involving the cleavage of two carbon--carbon bonds to form carbon monoxide and a \ch{C3H6} diradical, which typically rapidly forms highly vibrationally excited cyclopropane.
In a small number of trajectories the \ch{C3H6} radical was observed to undergo a rearrangement to form the more stable propene, and we note that on longer timescales one would expect the excited cyclopropane to also undergo this rearrangement.
 The C2 pathway is more complicated. In their work Trentelman \emph{et al.}~considered this to consist of a primary dissociation step,
\begin{reaction}
    \ch{C4H6O} \rightarrow \ch{C2H4} + \ch{C2H2O}, \label{rxn:C2a}
\end{reaction}
forming ethene and ketene (ethenone), with a possible secondary dissociation step,
\begin{subreactions}
\begin{reaction}
    \ch{C2H2O} \rightarrow \ch{CH2} +\ch{CO}, \label{rxn:ketene_dissociation}
\end{reaction}  
\end{subreactions}
in which ketene dissociates to form carbon monoxide and methene. Here, however, we also consider the possibility of a third process in which both (R\ref{rxn:C2a}) and (R\ref{rxn:ketene_dissociation}) occur in a single primary dissociation step,
\setcounter{reaction}{2}
\begin{subreactions}
\setcounter{reaction}{1}
\begin{reaction}
    \ch{C4H6O} \rightarrow \ch{C2H4} + \ch{CH2} +\ch{CO}. \label{rxn:primary-methene-formation}
\end{reaction}
\end{subreactions}

\begin{table}[t]
\centering
\caption{Total product yields  500\,fs after initial excitation. Note the total number of initial \ch{C4H6O} molecules is {198}. Fragments are identified by using a cutoff radius of $2\AA$.}
\label{tab:product-yields}
\begin{ruledtabular}
\begin{tabular}{lrr}
\textbf{Fragment} & \textbf{Count} & \textbf{Yield} \\
 \hline \vspace{-0.3cm} \\
 \ch{CO} & 158 & 80\% \\
 \ch{C3H6} & 141 & 71\% \\
 \ch{C2H4} & 45 & 23\% \\
 \ch{C2H2O} & 30 & 15\% \\
 \ch{CH2} & 15 & 7.6\% \\
 \ch{C4H6O} & 7 & 3.5\% \\
 \ch{H} & 5 & 2.5\% \\
 \ch{C4H5O} & 3 & 1.5\% \\
\ch{C3H5} & 2 & 1.0\% 
\end{tabular}
\end{ruledtabular}
\end{table}

\begin{table}[t]
    \centering
    \caption{Main reaction products at 500\,fs. Reaction products are identified by using a cutoff radius of $2\AA$. %
}
    \label{tab:reaction-products}
    \begin{ruledtabular}
    \begin{tabular}{lc|c|c}
         & I & II & III  \\
         \textbf{Products} & \makecell{\ch{C3H6 + CO}} & \makecell{\ch{C2H4 + C2H2O}} & \makecell{ \ch{C2H4 + CH2 + CO}} \\
         \textbf{Count} & 141 (71\%)  & 30 (15\%) & 15 (7.6\%) \\
    \end{tabular}
    \end{ruledtabular}
\end{table}

\subsection{Time-dependent fragment formation}
\begin{figure}[t]
    \centering
    \includegraphics{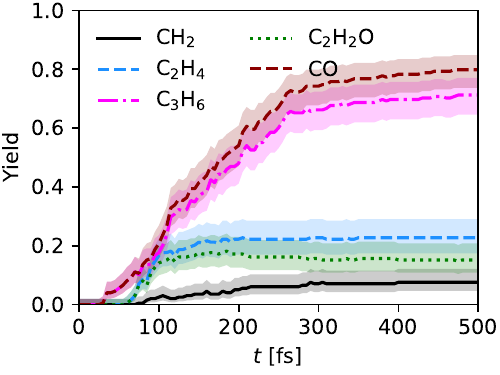}
    \caption{Average (unconvolved) fragment yields, for the 5 most common fragments, as a function of time for the {198} trajectories. Shaded region shows an approximate 95\% confidence interval (the Wilson score interval).\cite{Wilson1927BinomialErrors}  
    }
    \label{fig:fragment_yields_t_dep}
\end{figure}

To obtain a more detailed understanding of these dissociation pathways, 
 in Fig.~\ref{fig:fragment_yields_t_dep} we plot the time-dependent yield of each of the 5 major reaction products. It is instructive to first consider the yields of \ch{C2H4} and \ch{C3H6}. As was observed in the electron-diffraction signal, the onset of dissociation occurs at around 50\,fs, where the number of observed fragments begins to increase sharply, and the vast majority of the primary dissociation is over by around 300\,fs where the yields of both \ch{C2H4} and \ch{C3H6} are greater than 90\% of their final value. It is notable that although the onset of formation of \ch{C3H6} begins slightly earlier than \ch{C2H4}, the timescale associated with the formation of \ch{C2H4} is significantly shorter. In fact, while the yield of \ch{C2H4} is essentially constant between 200 and 500\,fs there is a notable $50\%$ increase in the yield of \ch{C3H6} over this timescale. %
This can be understood by noting that (as shown in the supplementary material Fig.~S24), trajectories which form \ch{C2H4} stay on \S2 for slightly longer but reach \S0 earlier than those which form \ch{C3H6}.
 The rapid formation of \ch{C2H4} is thus associated with a sudden successive relaxation from \S2 to \S1 and then \S1 to \S0, releasing a large amount of energy and resulting in a rapid (almost concerted) bond breaking. While \ch{C3H6} can also form in this way, we see that there is another slower formation mechanism. This slower mechanism involves trajectories becoming temporarily trapped on \S1 without dissociating; when they eventually relax to \S0 they then predominantly form \ch{C3H6} rather than \ch{C2H4}. It seems likely that this is because the formation of \ch{C2H4} requires a greater amount of energy in the corresponding carbon--carbon bond stretch, and that this slower pathway results in a more even distribution of the energy released from the relaxation from \S2 to \S1.

Returning to Fig.~\ref{fig:fragment_yields_t_dep} we note that, in contrast to \ch{C2H4}, the yield of \ch{CH2} continues to increase beyond 200\,fs. %
Specifically, the yield of \ch{CH2} increases from around {4.5}\% 
 at $t=200$\,fs to around {7.6}\% at $t=500$\,fs corresponding to an approximately {$70$}\% increase in population. This can be understood as arising from a secondary dissociation of ketene, as defined by (R\ref{rxn:ketene_dissociation}). 
This is confirmed by the concomitant decrease in the yield of ketene over the same period. However, we note that  
the appearance of methene before 100\,fs 
 indicates that secondary dissociation is not the only pathway to its formation. If it were, one would expect the rate of formation of \ch{CH2} to be proportional to the population of ketene. We therefore conclude that a significant fraction 
of methene is formed via 
 the primary dissociation reaction shown in (R\ref{rxn:primary-methene-formation}).   

 Although there are differences in the relative fractions of each product formed, the 6-31+G* calculations show similar qualitative behaviour to these aug-cc-pVDZ calculations. In the next subsection we will discuss these quantitative differences further.

\begin{figure}[t]
    \centering
    \includegraphics{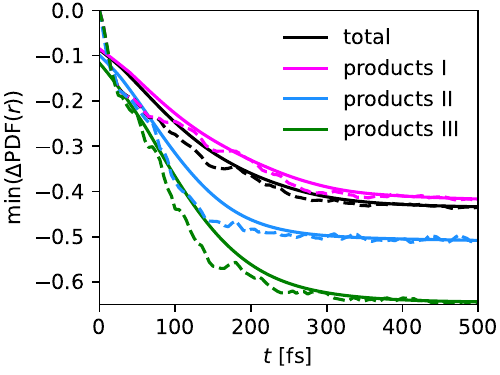}
    \caption{Minimum value of the simulated electron-diffraction signal given as a change in probability density function relative to the initial configuration. 
    Dashed lines show the unconvolved results from the {198} trajectories, and solid lines show the convolved results using a 160\,fs (FWHM) Gaussian to simulate the instrument response function.
    Note the height of the $\Delta \mathrm{PDF}(r)$ signal in all cases is given relative to the maximum peak height in the steady-state PDF\@. 
    }
    \label{fig:min_vals_elec_diff}
\end{figure}

\begin{figure*}
    \centering
    \includegraphics[width=\textwidth]{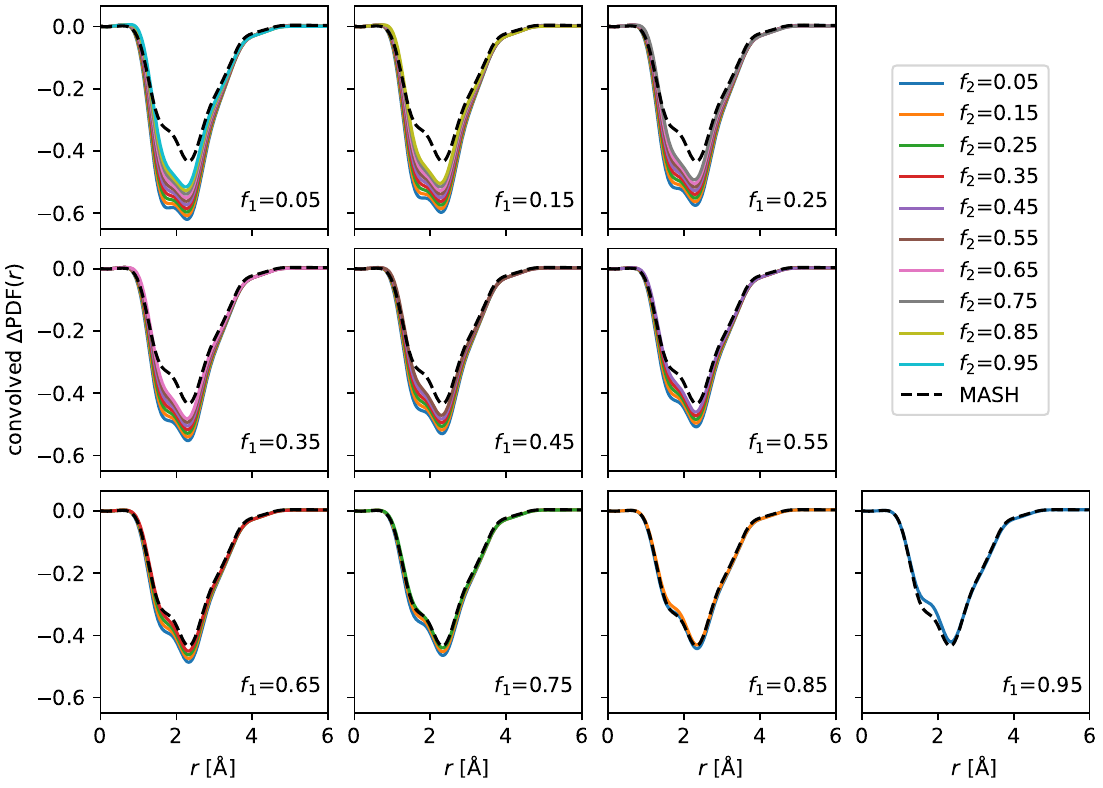}
    \caption{Weighted average over the three dominant reaction products of the simulated ultrafast electron-diffraction results at 500\,fs. $f_1$ corresponds to the fraction of products I and $f_2$ to the fraction of products II used in the weighted average, with the remaining fraction corresponding to products III.  
    Results are shown here  for the signal convoluted with with a 160\,fs (FWHM) Gaussian to simulate the instrument response function. Dashed line shows the prediction from the aug-cc-pVDZ trajectories. %
    }
    \label{fig:gued-weighted-averages}
\end{figure*}

\subsection{C3 vs C2 ratio}
These observations, along with directly visualising the trajectories, confirm %
that the dominant reaction pathways are those given in (R\ref{rxn:C3}) to (R\ref{rxn:primary-methene-formation}), and allow us to make a prediction of the relative yields of the C3 and C2 pathways (the C3/C2 ratio). 
 To do so we choose in all cases to identify reaction products I as belonging to the C3 pathway and reaction products II and III to the C2 pathway. \footnote{We note that in 2 cases we observe that \ch{C2H4} is formed via dissociation of a very short lived \ch{C3H6} radical. However, given that in both cases \ch{CO} is still in close proximity, and given that it would be difficult to distinguish these trajectories experimentally we feel it is reasonable to include these trajectories with the C2 pathway, effectively considering them as being described by (R\ref{rxn:primary-methene-formation}).} 
The trajectories not in groups I, II or III can be categorised as those that have undergone hydrogen dissociation and those that have not yet dissociated. We choose to exclude those reactions that have undergone hydrogen dissociation from the discussion of the C3/C2 ratio. For the 7 remaining undissociated \ch{C4H6O},  the fact that the yield of \ch{C3H6} in Fig.~\ref{fig:fragment_yields_t_dep} still shows a significant positive slope at 500\,fs makes it seem likely that the majority will eventually dissociate according to (R\ref{rxn:C3}).  
For this reason we additionally categorise these trajectories as following the C3 pathway.
On this basis, of the trajectories that follow either the C3 or C2 pathways, we observe that $77\%$ dissociate via the C3 pathway, and can estimate the influence of statistical error using a 95\% Wilson score interval\cite{Wilson1927BinomialErrors} to give an lower and upper bound to our prediction of $(70\%,82\%)$.
The corresponding C3/C2 ratio is found to be 3.3 and propagating the 95\% Wilson score interval gives upper and lower bounds on the statistical error as $(2.35,4.6)$. 
 Following the same analysis of the 6-31+G* product yields (given in the supplementary material) one arrives at a somewhat different C3/C2 ratio. There we find the fraction of all C3 and C2 trajectories that dissociate via C3 to be 0.5 with a 95\% Wilson score confidence interval of $(43\%,57\%)$ corresponding to a C3/C2 ratio of 1 with the statistical confidence interval at 95\% of $(0.75,1.35)$. The two sets of simulations therefore show a statistically significant difference in this quantity, and we will return to discuss how this influences our estimation of the systematic error in our predicted electron diffraction signal in Sec.~\ref{sec:systematic_errors}. 
As far as we are aware there does not exist a definitive experimental measurement of the C3/C2 ratio at 200\,nm.
The only attempt at a direct measurement that we could find in the literature was that of Shortridge \emph{et al.}\cite{Shortridge1969Cyclobutanone}~who obtained a value of 1.2. This experiment used a full arc zinc lamp rather than laser excitation, observing only the 202.6\,nm line showed appreciable absorption, and had a buffer gas of cyclopropane that somewhat complicated the interpretation of their results.
 In later work, Trentelman \emph{et al.}~chose to ignore this result and instead extrapolated the measurements of Denschlag and Lee\cite{Denschlag1968Cyclobutanone} to obtain a ratio of 1.3 at 193\,nm. It is, however, questionable whether this extrapolation is valid, given that it was based on excitation at wavelengths between 318\,nm and 248\,nm, which correspond to excitation to \S1 (centered at 280\,nm) rather than \S2. We additionally note that, if one takes the C3/C2 ratios for the lowest two wavelengths reported by Denschlag and Lee (253.7\,nm and 248\,nm) then a linear extrapolation in energy results in a significantly higher C3/C2 ratio of 2.5, in much closer agreement with the aug-cc-pVDZ ratio prediction. In any case what is clear from the existing experimental results is that within the \S1 band the C3/C2 ratio has a strong energetic dependence. Given this, systematic errors in the electronic structure and approximations used in the initial conditions might be expected to have a profound effect on the ratio seen in the simulations, which would certainly be consistent with the observed difference between the 6-31+G* and aug-cc-pVDZ results.

 \subsection{How will the different reaction pathways influence the experimental signal?}
Analysing our calculated trajectories has allowed us to give a detailed prediction of what happens during the dissociation. However, in order for our predictions to be testable, we need to connect them to observable features of the planned experimental signal. Hence, in the following we return to consider how key features of the different reaction pathways could be observed in the experimental electron-diffraction signal.
To do this we begin by calculating three hypothetical electron-diffraction signals corresponding to the trajectories that produce reaction products I, II and III\@. The full signals are given in the supplementary material. In the following we connect key features of these signals to the total signal and analyse how sensitive the planned experiment is to the relative fractions of reactions that follow each pathway.  

In Fig.~\ref{fig:min_vals_elec_diff} we show the minimum value of the $\Delta \mathrm{PDF}$ curve as a function of time for each of the reactions alongside the total signal. The minimum value occurs in all reactions and at all times close to $2.5\,\AA$. This corresponds to the distance between the oxygen and the two \textbeta-carbons in cyclobutanone (Fig.~\ref{fig:steadystate}), and also to the distance between the oxygen and the \textbeta-carbon in ketene. Hence, this minimum is closely associated with the formation of \ch{CO}\@. We note, however, that drawing a direct connection between the depth and the amount of \ch{CO} is complicated somewhat, both by the presence of two rather than just one \textbeta-carbons in cyclobutanone, and also by the fact that the distance between the carbons and the four hydrogens on the neighbouring carbons in cyclopropane is about 2.5\,$\AA$ (although the lighter H atoms have a weaker signal). Nevertheless, comparing the curves for products~I and products~II we can see a clear parallel with the \ch{C3H6} and \ch{C2H4} curves in Fig.~\ref{fig:fragment_yields_t_dep}. The curve for products~II has clearly plateaued by around 350\,fs while products~I continues to decrease gradually. Using this connection we can attribute the notably shorter timescale of the curve for products~II vs.~products~I to the ``concerted'' vs.~``sequential'' nature of the dissociation reactions. In a similar manner, the slope at long time for products~III can be understood as arising from the secondary dissociation of ketene to form methene and carbon monoxide.

Although these qualitative differences in the slopes are interesting, the most notable difference between the curves is their magnitude at 500\,fs.
This would therefore appear a good way of using the experimental result to judge the accuracy of our predicted product yields (and C3/C2 ratio). 
We remind the reader, that in order to compare the results shown here to the future experimental results, one should bear in mind that the results here are given relative to the maximum peak height of the steady-state PDF [\eqn{PDF}].\footnote{To obtain this quantity from the experimental measurements one would of course need to take into account the fraction of molecules excited by the pump pulse.} %
If there were only two dominant sets of reaction products, the depth of the minimum alone would allow one to get a good estimate of the ratio of products, which could be done by fitting a weighted average of the corresponding peak depths. Since there are three dominant reaction products, determining the relative ratios of all three from the electron-diffraction signal is not so straightforward. However, by considering the full $\Delta\mathrm{PDF}(r)$ signal at 500\,fs one can additionally make use of the lineshape to help distinguish the reaction products present. Figure \ref{fig:gued-weighted-averages} does exactly that. Alongside the signal obtained from our MASH simulation (dashed line), it shows the hypothetical signal that would be obtained for different fractions $f_1$ and $f_2$ of products~I and products~II respectively, with the remaining contribution from products~III ($f_1+f_2+f_3=1$). [Note the individual signals that are being averaged can be seen in Figs.~S25-S27 in the supplementary material.] This serves as a relatively simple testable quantitative prediction that one can use to assess the relative fractions of these key reaction products. While it is not highly sensitive, there is only a relatively small part of the parameter space which is consistent with the aug-cc-pVDZ results, i.e., $0.65\lesssim f_1 \lesssim 0.85$. Note the equivalent plot for the 6-31+G* calculations is given in the supplementary material and shows that in this case only the region $0.35\lesssim f_1 \lesssim 0.65$ would be consistent with the predicted result. This demonstrates that, although the predicted electron-diffraction signal is not radically different in the two cases, with a low enough experimental error one would be able to distinguish between them.

In case it is difficult to accurately determine the depth from the experimental signal, in the supplementary material we also consider how the lineshape at 500\,fs alone could be used to distinguish different reaction products. Figure S28 shows the possible signals at 500\,fs normalised so that the depth at their minimum is $-1$. From this one can see that the most notable change to the line shape with varying fractions, $f_1$, of reaction products I, is the depth of the shoulder at about 1.8\,$\AA$. Although the changes are relatively subtle compared to the absolute differences of Fig.~\ref{fig:gued-weighted-averages}, we see that, with a small enough relative error, it should nevertheless be possible to distinguish between different fractions, $f_1$, of reaction products I, and hence different C3/C2 ratios.
This could therefore also be used as a direct test of our prediction of the product ratio.

\subsection{Potential systematic errors} \label{sec:systematic_errors}
Finally, having focused on how sensitive the experiment would need to be to distinguish between different possible reaction pathways, it is natural to ask how confident we are in our predicted signal and how large our systematic errors are likely to be. There are four potential sources of error:
\begin{enumerate}
    \item  electronic-structure theory
    \item  initial conditions/treatment of the pulse
    \item  dynamics (including zero-point energy leakage)
    \item  calculation of the electron diffraction signal.
\end{enumerate}
The established sensitivity of the C3/C2 ratio on the excitation wavelength,  means that we expect that the results will be particularly sensitive to errors in the electronic-structure theory and initial conditions. Given the accuracy of MASH in previous benchmark tests and the established nature of the independent atom approximation, we therefore focus here on assessing the errors due to the electronic-structure theory and initial conditions. 
We have therefore performed a number of additional calculations for which additional figures are given in the supplementary material. 

Ideally to confirm the accuracy of our prediction we would perform the same simulations at a higher level of electronic-structure theory, e.g., CASPT2 with the same (12,11) active space. Given the time constraints of the challenge, this is not possible. %
However, we have calculated CASPT2 energies with the aug-cc-pVDZ basis along linear interpolated (in internal coordinates) paths (Fig.~S12), to examine the effect of dynamical correlation. Along the path that leads from the Franck-Condon geometry to the \S2/\S1 MECI and the \S1 minimum, CASPT2 energies are very similar to CASSCF. This is also observed along the path connecting the \S1 minimum to the first \S1/\S0 MECI\@. However, the barriers to access the second and third \S1/\S0 MECIs increase by 0.7\,eV and 0.8\,eV, respectively, at the CASPT2 level of theory. Since, the third \S1/\S0 MECI seems to lead to the C2 products (Fig.~S20), it is likely that repeating our aug-cc-pVDZ simulations at the CASPT2 level would further increase the C3/C2 ratio.

Another thing that might influence the C3/C2 ratio is the initial conditions. Hence, to assess the extent to which the initial conditions may influence the predicted electron diffraction signal, we considered how the vertical excitation energy of each initially sampled geometry correlated with the final reaction products. Under the assumption that the strong wavelength dependence observed in the product ratio at longer wavelengths (within the \S1 window) implies a strong wavelength dependence when exciting to \S2, then one would expect the initial excitation energy in our simulation to correlate with the final product. To test this hypothesis we performed a two-sample z test on the mean of the vertical excitation energy for products I and products II + III\@. This resulted in a difference in the means that was not statistically significant, with a p value of 0.55. This goes against the hypothesis stated above. This implies that the C3/C2 ratio is not so sensitive at these shorter wavelengths,
probably because the large excess of energy makes all conical intersections accessible, in contrast to excitation to \S1 where shorter wavelengths may open new channels significantly altering the product distribution. 
 Hence, it appears from these tests at least that the initial conditions are perhaps not such a large source of systematic error.
 Of course, there are other features of the initial conditions, such as the dependence of transition-dipole moments which we have not included but could affect the product ratios.

Our most direct information about possible systematic errors comes from the difference between the aug-cc-pVDZ calculations and the 6-31+G* calculations. As discussed above, when comparing the dynamics within a particular channel both sets of calculations give very similar results. However, where they differ is in their predicted C3/C2 ratio (statistical intervals giving 2.35 to 4.6 for aug-cc-pVDZ vs.~0.75 to 1.35 for 6-31+G*). We have shown that this difference could in principle be observed in the electron-diffraction signal, but the overall difference is small. If one were forced to choose between the two without considering prior photochemical experiments, one would say that aug-cc-pVDZ should be preferred since it is a slightly larger basis than 6-31+G*. It was for this reason, in addition to the more accurate \S0 to \S2 vertical excitation energy that we chose to focus on the aug-cc-pVDZ results in the main paper. We note however, that the aug-cc-pVDZ are not expected to be so much more accurate than the 6-31+G* calculations that we should discount them entirely. Given that prior experimental results indicate that the C3/C2 ratio should be around 1.2--1.3,\cite{Shortridge1969Cyclobutanone,Denschlag1968Cyclobutanone,Trentelman1990Cyclobutanone} one might well suggest that the most accurate prediction could be obtained by taking an average of the two sets of calculations. We note, however, that the electron-diffraction signal that results from averaging (shown in supplementary material) is difficult to distinguish by eye from the signal in Fig.~\ref{fig:gued} or the {6-31+G*} result (also in supplementary material).
 Hence we can conclude that, while there is some uncertainty in our predicted C3/C2 ratio, it seems reasonable to assume that the fraction of trajectories following the C3 pathway is between about 0.5 and 0.85. Overall, given the sensitivity of the electron-diffraction signal observed in Fig.~\ref{fig:gued-weighted-averages} and additional figures in the supplementary material we can therefore be confident that our results are likely to accurately reproduce the timescales and other key features of the planned experimental signal.
 
 {The only caveat to this is that, since writing our original manuscript, we have become aware that there do exist previous measurements of time resolved photoelectron and mass spectra of cyclobutanone after excitation with a 200\,nm pulse.\cite{Kuhlman2012cyclobutanone} These measurements were used to estimate the timescale of \S2 decay, and found a timescale of about 740\,fs. If correct, this would indicate that the timescale for \S2 decay obtained in the current study is too fast. We note that our scans between the Franck--Condon point and the \S2/\S1 MECI (shown in supporting information Figs.~S12 and S13) do show a significant barrier at the CASPT2 level that is not present in CASSCF, which could contribute to such an error in the timescale of \S2 decay.}

\section{Conclusion}

In conclusion, we have used molecular simulations to investigate the dynamics of cyclobutanone after photo-excitation to the \S2 state. We have focused here on details of the dynamics most relevant to JCP's community challenge. In particular we have made quantitative predictions of the electron-diffraction signal that will be observed in the planned experiment and explained this signal in terms of the main fragmentation reactions observed in our simulation.

In addition to understanding the dissociation of cyclobutanone, this study also serves as the first application of the newly proposed multi-state MASH method, called unSMASH, to an ab initio simulation. By choosing to use MASH for our simulations instead of the more commonly used FSSH, we expect to have gained the following advantages. Firstly, our simulations require no ad hoc decoherence corrections. This is because the deterministic dynamics of MASH means that the electronic variables always remain consistent with the active surface, therefore fixing the inconsistency problem of FSSH.\cite{MASH,MASHrates} This is likely to be important for accurately describing the photodissociation process, where it is known that the inconsistency error can lead to suppressed product yields.\cite{Molecular_Tully_JM} Secondly, MASH is able to correctly describe the effects of the nonadiabatic force through its uniquely determined momentum rescaling algorithm. This is crucial for accurately describing the electronic population dynamics, which is known to be particularly sensitive to how the momentum rescaling is performed.\cite{Toldo2024,Molecular_Tully_JM} While FSSH can in principle describe this effect by rescaling the momenta along the NACV at a hop and reflecting in the case of a frustrated hop, in practice {it is common to use} an isotropic momentum rescaling in ab initio FSSH simulations.\cite{Mai2018SHARC}

While we expect the present study to have captured the most important details of the system, there are a number of areas in which the simulation could be improved. Firstly, we could include intersystem crossing between the singlet manifold and the triplet manifold. This is something that can be rigorously achieved within the MASH framework, as will be described in an forthcoming publication.\cite{JonathanUnpublished}  We note, however, that given the rapid nature of the dissociation observed in the present study and the weak spin--orbit coupling, it is unlikely there would be time for sufficient population transfer to the triplet manifold to significantly influence the dynamics. %
Secondly, we may seek to improve the accuracy of our initial conditions.
The branching ratio between C3 and C2 channels may 
depend sensitively on the amount of energy imparted to the system by the photo-excitation.\cite{Denschlag1968Cyclobutanone,Trentelman1990Cyclobutanone} The Franck--Condon approach we have used, taking the initial distribution as the ground state nuclear Wigner distribution placed on the \S2 state and mimicking the finite width of the exciting laser pulse by convolution with a Gaussian, is a standard way of initialising a simulation in photo-excited systems,\cite{Wolf2019CHD} however, it is not without approximation. It will, therefore, be interesting in future studies to explore the sensitivity of the results to the initial conditions used, or even explicitly simulate the pulse within the MASH dynamics.\cite{JonathanUnpublished} %

We note, however, that despite the possible improvements that could be made to the initial conditions, at present they would all rely on the use of a Wigner transformed initial density. This has the advantage that it introduces zero-point energy into the initial distribution. However, this is not without its limitations. Specifically, since the underlying nuclear dynamics is classical, the zero-point energy will eventually (due to anharmonicity) become evenly distributed among the molecular degrees of freedom. This so called ``zero-point energy leakage'' is a well-known problem.\cite{Habershon2009water} In condensed-phase problems in thermal equilibrium this issue has been largely solved by imaginary-time path-integral methods such as ring-polymer molecular dynamics (RPMD).\cite{RPMDcorrelation,Habershon2013RPMDreview,Lawrence2020rates} There has therefore been significant interest in the development of a nonadiabatic version of RPMD.\cite{Shushkov2012RPSH,mapping,Kretchmer2016KCRPMD,Duke2015MVRPMD,Chowdhury2017CSRPMD,Tao2019RPSH,Lawrence2019isoRPMD} However, for photochemical problems that are inherently far from the linear-response regime, where imaginary-time path-integral methods have been shown to give very accurate results,\cite{Lawrence2018Wolynes,inverted,thiophosgene,Lawrence2020FeIIFeIII,Lawrence2020Improved} it is unclear whether a nonadiabatic RPMD would be the final solution to this problem. This is because RPMD effectively assumes a rapid decoherence of vibrational modes that may introduce additional errors in such low-pressure gas-phase systems.\cite{JoeFaraday,RPMDcavity} The search for an optimal method for such problems therefore continues. In future studies it would be interesting to assess the importance of zero-point energy leakage in this system by initialising all modes from a classical Boltzmann distribution, and comparing the resulting simulations. If the classical nuclear limit is valid, then one has the additional advantage that MASH has the correct detailed balance to thermalise to the correct distribution.\cite{MASH}
The aspect that will probably have the most significant impact on our dynamics
is the accuracy of the electronic-structure theory.
In the present study we have chosen a level of theory that minimises the cost while still being sufficient to allow the simulation to describe the key qualitative features of the electronic subspace.
We used two different basis sets and found that although the qualitative description of the electron-diffraction pattern is similar, the C3/C2 product ratio is significantly different.
In future work it will therefore be interesting to investigate the system using even more accurate electronic-structure theory methods, such as using larger basis sets and including dynamic correlation via CASPT2, MRCI or coupled cluster theory.\cite{kjonstad2023CCSD} 
Although this would significantly increase the cost of an on-the-fly simulation, by  exploiting modern machine-learning techniques\cite{Westermayr2019ML,MLNACs} one may hope to make such calculations tractable. %

There are more advantages of developing a machine-learned model than just making the use of high accuracy electronic-structure theory achievable. In particular, having trained a model, it would then be comparatively inexpensive to perform a systematic analysis of the sensitivity to different initial conditions. Furthermore, it would  give the opportunity to use more expensive dynamics methods that are impractical for on-the-fly calculations. Such studies would complement the aims of the JCP challenge, by providing an objective comparison of the accuracy of different dynamics methods and the approximations they make, and helping to push the boundaries of accuracy in excited-state simulations.  
Given the right potential-energy surfaces and couplings, we believe that MASH can be a very powerful simulation tool for obtaining reliable predictions of photochemistry.

\section*{Supplementary Material}
See the supplementary material for additional numerical results referred to in the main text.
\section*{Acknowledgements}
JEL was supported by an Independent Postdoctoral Fellowship at the Simons Center for Computational Physical Chemistry, under a grant from the Simons Foundation (839534, MT)\@.
JRM and AK are supported by the Cluster of Excellence ``CUI: Advanced Imaging of Matter'' of the Deutsche Forschungsgemeinschaft (DFG) – EXC 2056 – project ID 390715994.
\section*{Author Declarations}
\subsection*{Conflict of interest}
The authors have no conflicts to disclose.
\subsection*{Author Contributions}

\begin{comment}
\blue{Please add/modify the description of your contribution as appropriate

I would say let's include a detailed description in addition to using the somewhat limited Credit taxonomy:
Conceptualization 
 Data curation 
 Formal analysis 
 Funding acquisition 
 Investigation 
 Methodology 
 Project administration 
 Resources 
 Software 
 Supervision 
 Validation 
 Visualization 
 Writing - original draft 
 Writing - review \& editing 
(lead, equal or supporting)
}
\end{comment}

J.E.L. MASH dynamics code development (lead), data analysis (lead), writing manuscript first draft (lead), running trajectories (lead), Code testing (equal), development of interface between MASH code and molpro (supporting).
I.M.A. Selection of electronic-structure method (lead), development of interface between MASH code and molpro (equal), Initial Conditions (supporting), Triplet simulations (supporting), writing manuscript first draft (supporting), Code testing (equal), Analysis of accuracy of electronic-structure theory (supporting).
J.R.M. code testing (equal), writing manuscript first draft (supporting).
M.A.M. Analysis of accuracy of electronic-structure theory (lead), development of interface between MASH code and molpro (equal), writing manuscript first draft (supporting).
K.A. Triplet simulations (lead), writing manuscript first draft (supporting).
A.K. writing manuscript first draft (supporting).
J.O.R. Initial Conditions (lead), Calculation of electron-diffraction signal (lead), writing manuscript first draft (supporting), conceptualisation (lead), project management (lead).

\section*{Data Availability}
The data that supports the findings of this study are available within the article and its supplementary material. 

\bibliography{references,molpro,other,extra_refs} %

\end{document}

\title{Supplementary Material: A MASH simulation of the photoexcited dynamics of cyclobutanone}
\author{Joseph E. Lawrence}
\email{joseph.lawrence@nyu.edu}
\affiliation{Department of Chemistry and Applied Biosciences, ETH Zurich, 8093 Zurich, Switzerland}
\affiliation{Simons Center for Computational Physical Chemistry, New York University, New York, NY 10003, USA}
\affiliation{Department of Chemistry, New York University, New York, NY 10003, USA}
\author{Imaad M. Ansari}
\affiliation{Department of Chemistry and Applied Biosciences, ETH Zurich, 8093 Zurich, Switzerland}
\author{Jonathan R. Mannouch}
\affiliation{Hamburg Center for Ultrafast Imaging, Universit\"at Hamburg and the Max Planck Institute for the Structure and Dynamics of Matter, Luruper Chaussee 149, 22761 Hamburg, Germany}
\author{Meghna A. Manae}
\author{Kasra Asnaashari}
\affiliation{Department of Chemistry and Applied Biosciences, ETH Zurich, 8093 Zurich, Switzerland}
\author{Aaron Kelly}
\affiliation{Hamburg Center for Ultrafast Imaging, Universit\"at Hamburg and the Max Planck Institute for the Structure and Dynamics of Matter, Luruper Chaussee 149, 22761 Hamburg, Germany}
\author{Jeremy O. Richardson}
\email{jeremy.richardson@phys.chem.ethz.ch}
\affiliation{Department of Chemistry and Applied Biosciences, ETH Zurich, 8093 Zurich, Switzerland}
\maketitle

\section{Electronic Structure} %
\subsection{Basis set}
The \S2 state is a Rydberg state, which corresponds to an $n \to 3s$ transition from the \S0 state.
Diffuse orbitals are required to correctly describe the characteristic Rydberg molecular orbital of this state and for this, we chose the 6-31+G* from the Pople and the aug-cc-pVDZ from the Dunning's correlation-consistent family of basis sets.

\subsection{Active space} 
\begin{figure}
       \centering
       \includegraphics[width=\linewidth]{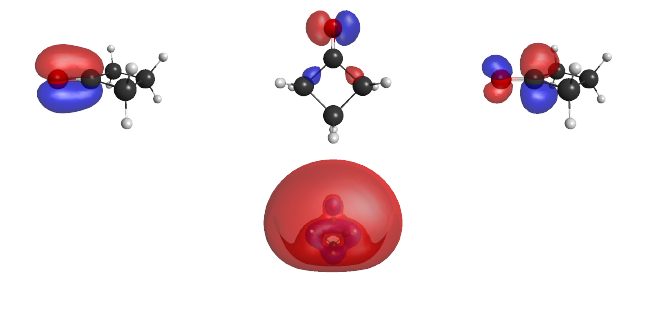}
       \caption{The natural orbitals of AS1, a minimal (4,4) active space with the aug-cc-pVDZ basis set, which contains the most important orbitals involved in the excitation manifold of the $\S0$, $\S1$ and $\S2$ states. Qualitatively similar orbitals were chosen for the 6-31+G* basis set.}
       \label{fig:AS1}
    \end{figure}

\begin{figure}
       \centering
       \includegraphics[width=\linewidth]{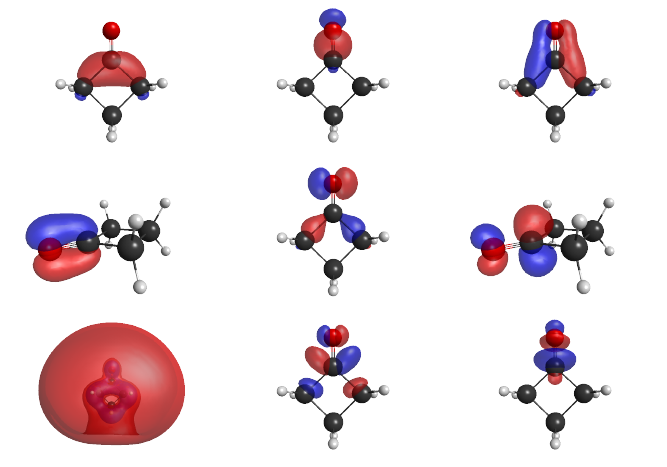}
       \caption{The natural orbitals of AS2, a (10,9) active space with the aug-cc-pVDZ basis set, which additionally accounts for the \textsigma-orbitals of the carbonyl and the \textalpha-carbon bonds. Qualitatively similar orbitals were chosen for the 6-31+G* basis set.}
       \label{fig:AS2}
    \end{figure}

\begin{figure}
       \centering
       \includegraphics[width=\linewidth]{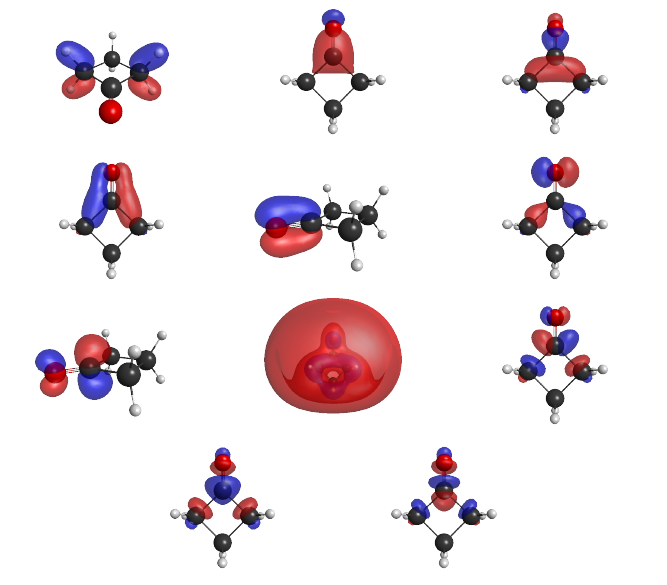}
       \caption{The natural orbitals of AS3, a (12,11) with the aug-cc-pVDZ basis set active space, which adds on C--H \textsigma-bonds and two C--C \textsigma$^{*}$-bonds.}
       \label{fig:AS3_aug-cc-pVDZ}
    \end{figure}

A three-state state-averaged complete active space self-consistent field (SA-CASSCF) was used to calculate the ground and excited states simultaneously. 
Three active spaces were tested, in increasing order of complexity, which we have called AS1, AS2 and AS3 and are shown in Figs.~\ref{fig:AS1}, \ref{fig:AS2} and \ref{fig:AS3_aug-cc-pVDZ} respectively.

\begin{figure}
       \centering
       \includegraphics[width=\linewidth]{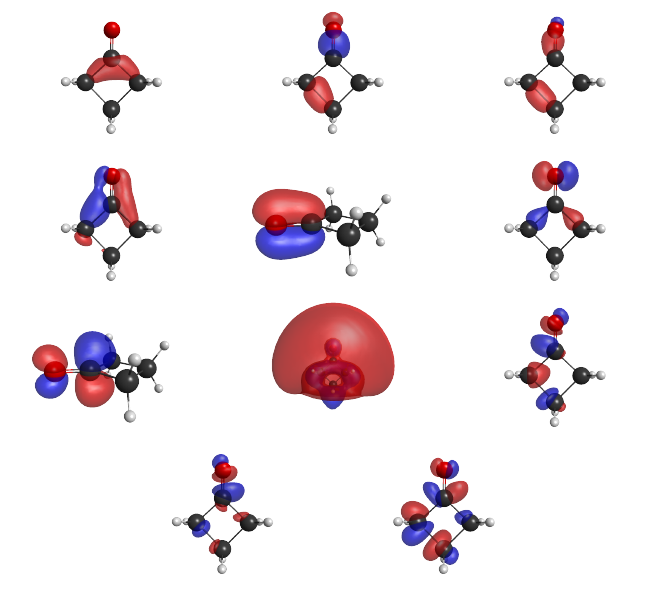}
       \caption{The natural orbitals of AS3, a (12,11) with the 6-31+G* basis set active space, which accounts for one additional C--C bond in addition to AS2.}
       \label{fig:AS3_6-31+Gd}
    \end{figure}

AS1 is a (4,4) space which contains orbitals that characterize the electronic states \S0, \S1 and \S2 ($n$, $\pi^{*}$ and $3s$) and have a very large contribution to the static correlation ($\pi$).
AS2 is a (10,9) space which adds $\sigma$ and $\sigma^{*}$ orbitals for the C-O and the $\alpha$-carbon bonds.
While the active spaces AS1 and AS2 are qualitatively similar between the aug-cc-pVDZ and 6-31+G* basis sets, the CASSCF optimization finds a slightly different set of active orbitals at the \C{2v} saddle-point geometry for AS3, a (12,11) space (Figs.~\ref{fig:AS3_aug-cc-pVDZ} and \ref{fig:AS3_6-31+Gd}).
For 6-31+G*, AS3 results in an active space which adds one set of $\sigma$ and $\sigma^{*}$ orbitals for the C-C bond between an $\alpha$ carbon and the carbon not attached to the oxygen atom, while for aug-cc-pVDZ, it completes the $\sigma, \sigma^{*}$ set of the C-O and the two $\alpha$-carbon bonds and adds one C-H $\sigma$ orbital.

\begin{figure}
       \centering
       \includegraphics[width=\linewidth]{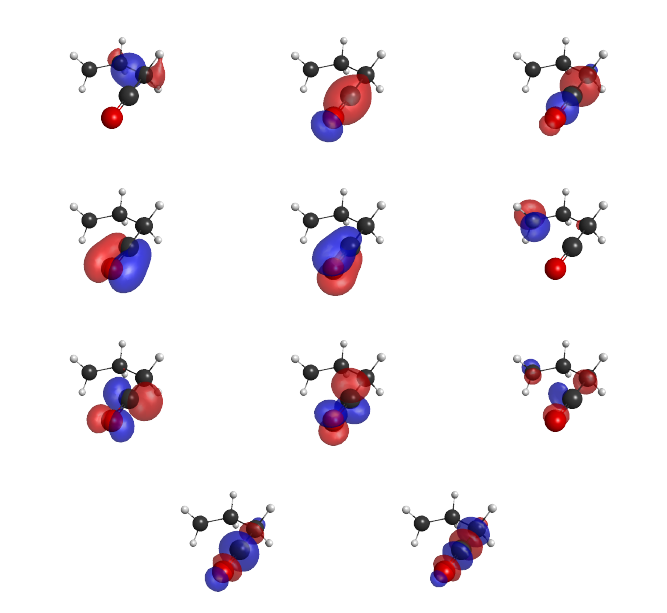}
       \caption{The natural orbitals of AS3 with the aug-cc-pVDZ basis set active space at an MECI between the \S0 and \S1 surfaces.}
       \label{fig:AS3_meci_s0s1_ci1_aug-cc-pVDZ}
    \end{figure}

\begin{figure}
       \centering
       \includegraphics[width=\linewidth]{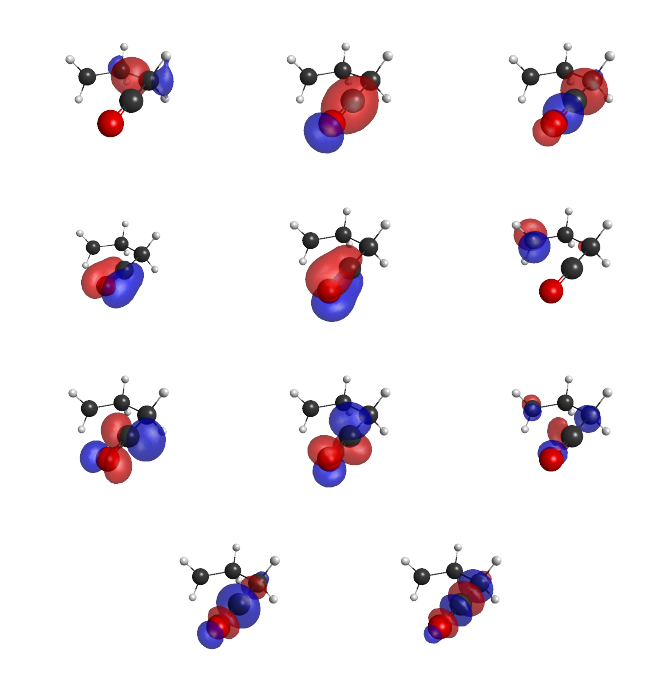}
       \caption{The natural orbitals of AS3 with the 6-31+G* basis set active space at an MECI between the \S0 and \S1 surfaces.}
       \label{fig:AS3_meci_s0s1_ci1_6-31+Gd}
    \end{figure}

We attribute these differences to the fact that the initial orbitals were chosen at the \S0 \C{2v} equilibrium geometry, where the wavefunction is not expected to have large multireference character and thus the active orbitals, especially ones with small occupation numbers, in a large (12,11) active space can be sensitive to parameters like the basis set. 
This is consistent with the observation that with AS1 and AS2, one gets similar active orbitals in the respective active spaces using either basis set.
This is further supported by analyzing the active orbitals at a geometry with a large multireference character in Figs.~\ref{fig:AS3_meci_s0s1_ci1_aug-cc-pVDZ} and~\ref{fig:AS3_meci_s0s1_ci1_6-31+Gd}, where the two basis sets give similar active orbitals.

\subsection{Results}

The calculated vertical excitation energies with the three active spaces are given in Table~\ref{tab:vert_energies_calc}, calculated at the $C_{2\mathrm{v}}$ saddle-point geometry obtained at the B3LYP/def2-TZVP level of theory.
These can be compared with theoretical and experimental results from the literature, summarized in Table~\ref{tab:vert_energies}. 
Note that the absorption peak of the $\S0 \to \S2$ observed in experiment\cite{cyclobutanone_rydberg_spectrum} is quite broad, with a full width at half maximum of approximately \SI{0.3}{\electronvolt}.

Using (3S+2T) state-averaged CASSCF, we also computed vertical excitation energies to the triplet states in addition to the singlets.  The results are presented in Table~\ref{tab:vee_and_aee}.
Minimum-energy crossing points and minimum-energy conical intersections were located and are presented in Table~\ref{tab:crossings}.

\begin{figure}
       \centering
       \includegraphics[width=\linewidth]{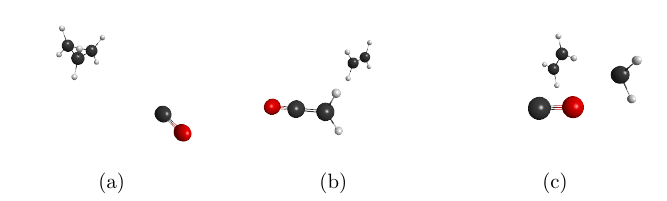}
       \caption{Geometries of (a) Products I, (b) Products II and (c) Products III, as defined in the main text. Note that this data is also available as an xyz file in the supplementary material.}
       \label{fig:products}
\end{figure}

\begin{figure}
       \centering
       \includegraphics[width=\linewidth]{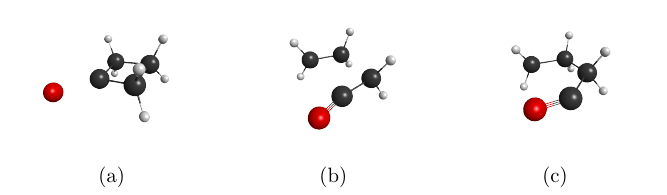}
       \caption{Geometries of the (a) \S1/\T1, (b) \S2/\T2, (c) \S0/\T1 MECPs.
       All geometries correspond to AS3/6-31+G*. Note that this data is also available as an xyz file in the supplementary material.}
       \label{fig:mecp}
\end{figure}

\begin{figure}
       \centering
       \includegraphics[width=\linewidth]{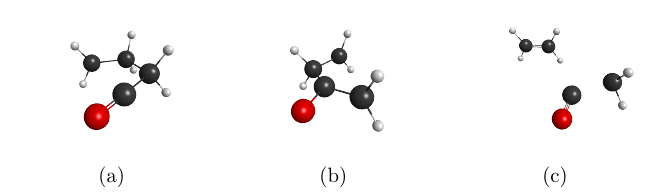}
       \caption{Geometries of the \S1/\S0 MECIs.
       All geometries correspond to AS3/6-31+G*. Note that this data is also available as an xyz file in the supplementary material.}
       \label{fig:meci_s1s0}
\end{figure}

\begin{figure}
       \centering
       \includegraphics[width=\linewidth]{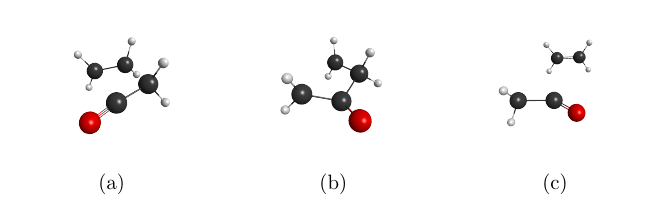}
       \caption{Geometries of the \S1/\S2 MECIs.
       All geometries correspond to AS3/6-31+G*. Note that this data is also available as an xyz file in the supplementary material.}
       \label{fig:meci_s1s2}
\end{figure}

\begin{figure}
       \centering
       \includegraphics[width=\linewidth]{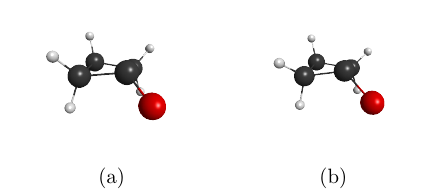}
       \caption{Geometries of the minima on (a) $\S1$ and (b) $\S2$. 
       All geometries correspond to AS3/aug-cc-pVDZ. Note that this data is also available as an xyz file in the supplementary material.}
       \label{fig:minima}
\end{figure}

Geometries of the products, MECPs, \S1/\S0 MECIs, \S1/\S2 MECIs and excited-state minima are shown in Figs.~\ref{fig:products}, \ref{fig:mecp}, \ref{fig:meci_s1s0}, \ref{fig:meci_s1s2} and \ref{fig:minima} respectively.

\begin{figure}
       \centering
       \includegraphics[scale=0.3]{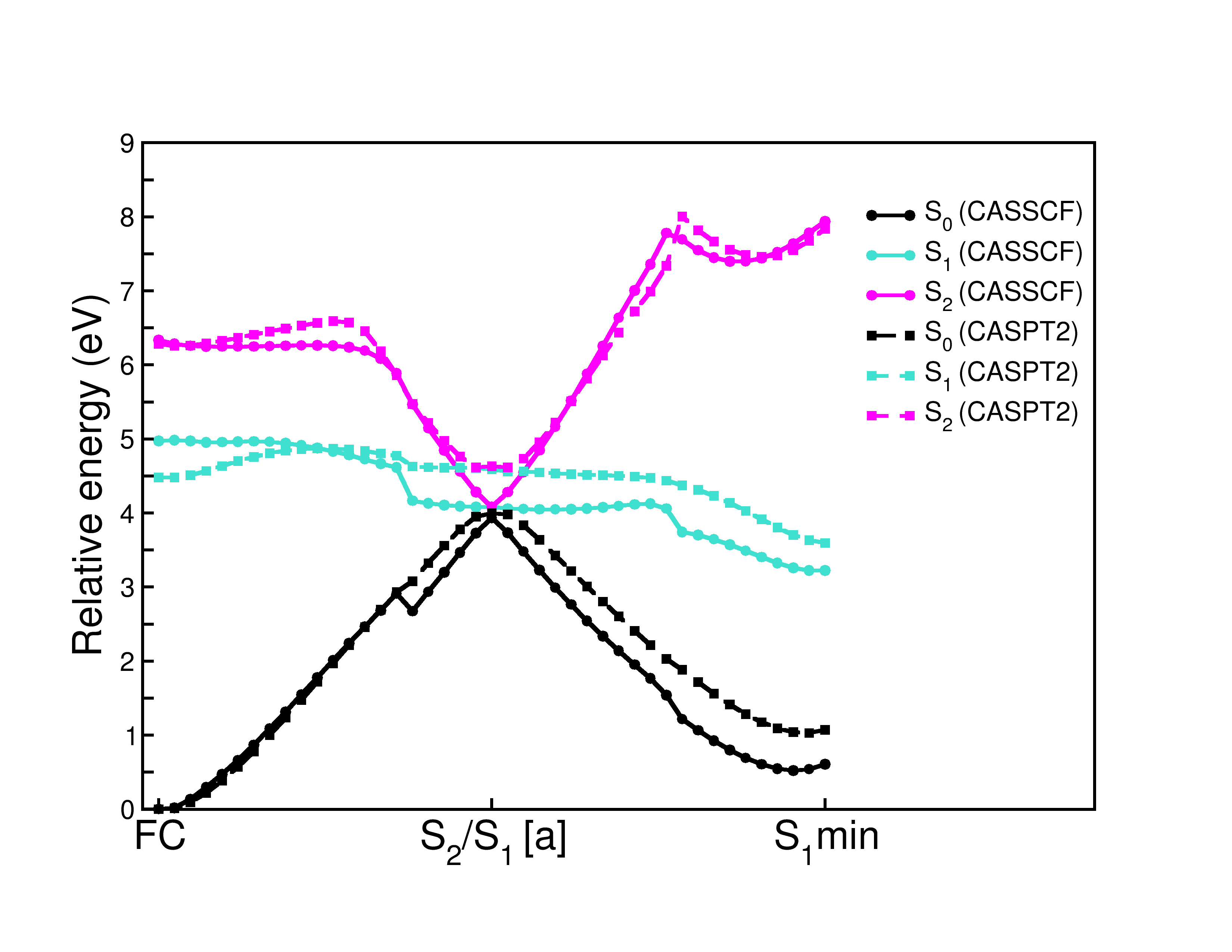}
       \includegraphics[scale=0.3]{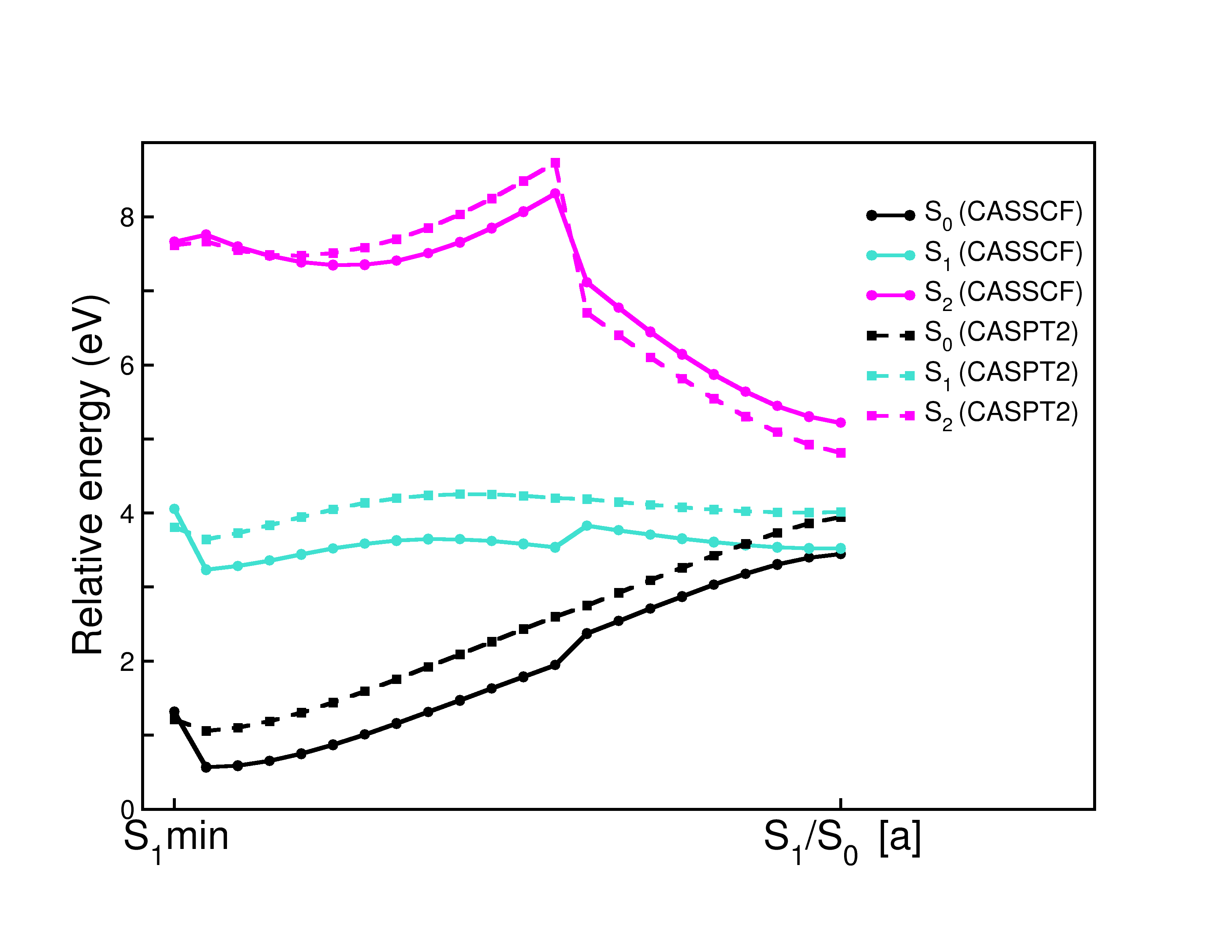}
       \includegraphics[scale=0.3]{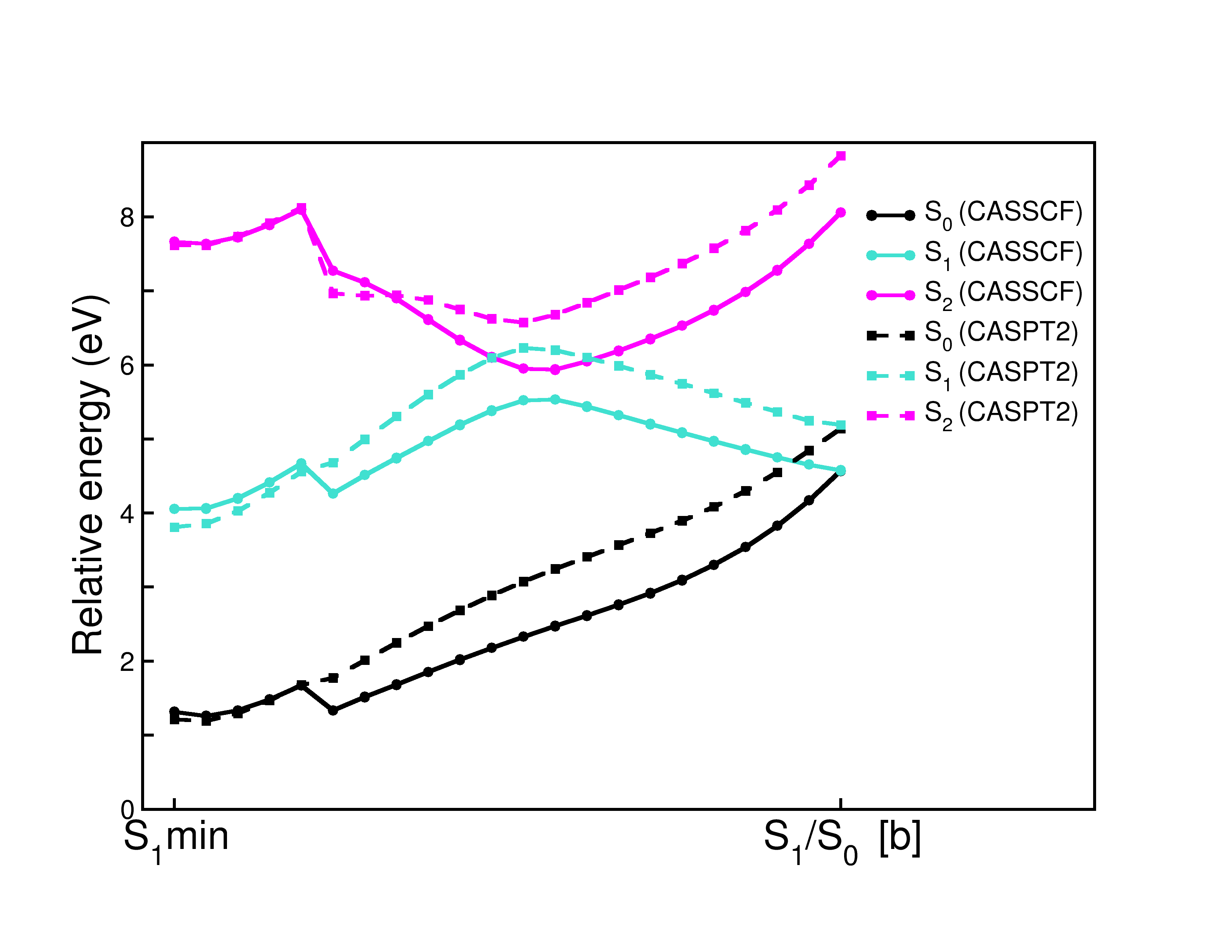}
       \includegraphics[scale=0.3]{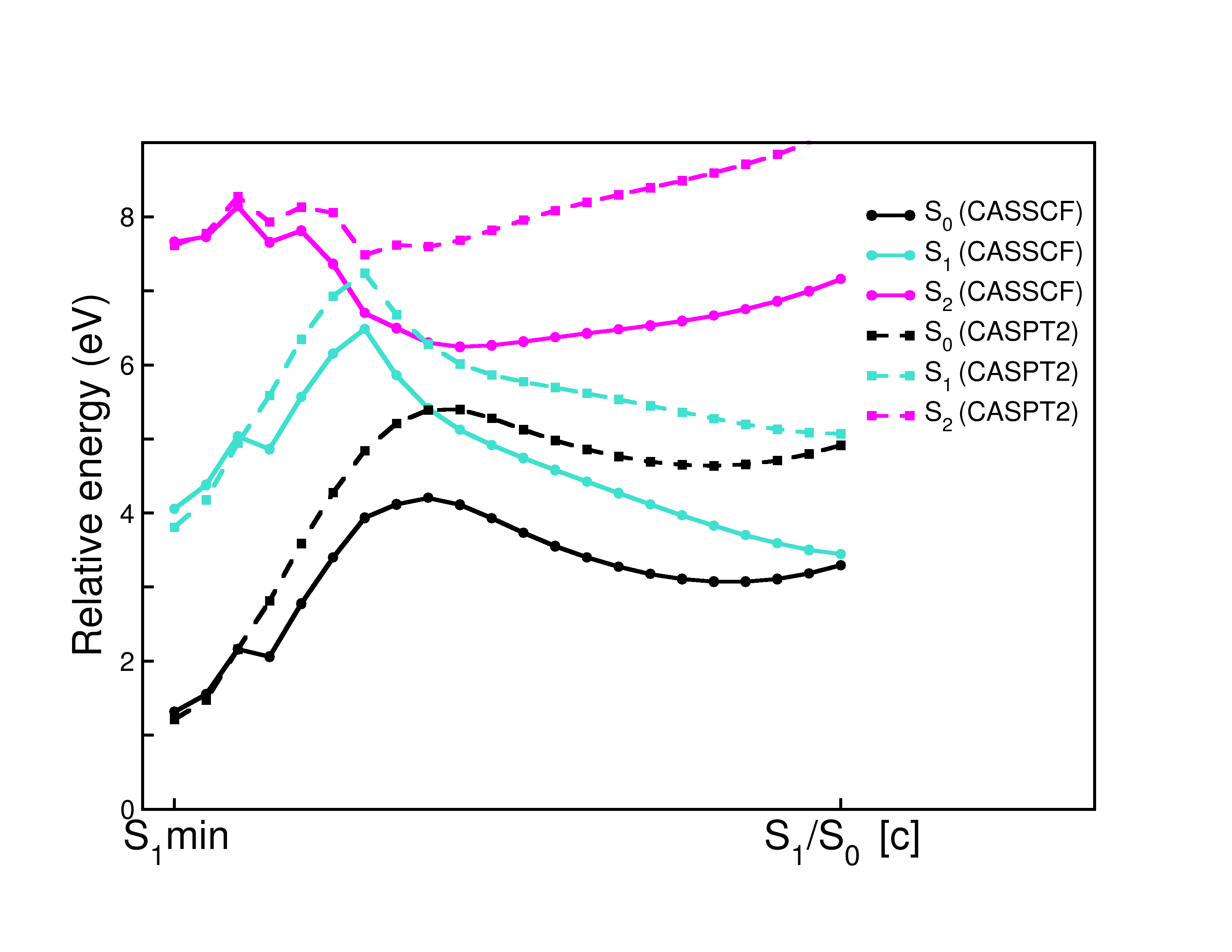}
       \caption{SA-CASSCF and XMS-CASPT2 single-point energies for intermediate structures generated between critical points using a linear interpolation in internal coordinates, calculated at the AS3/aug-cc-pVDZ level of theory. Energies have been shifted by the \S0 energy at the \C{2v} geometry at the respective level of theory. }
       \label{fig:caspt2}
\end{figure}

\begin{figure}
       \centering
       \includegraphics[scale=0.3]{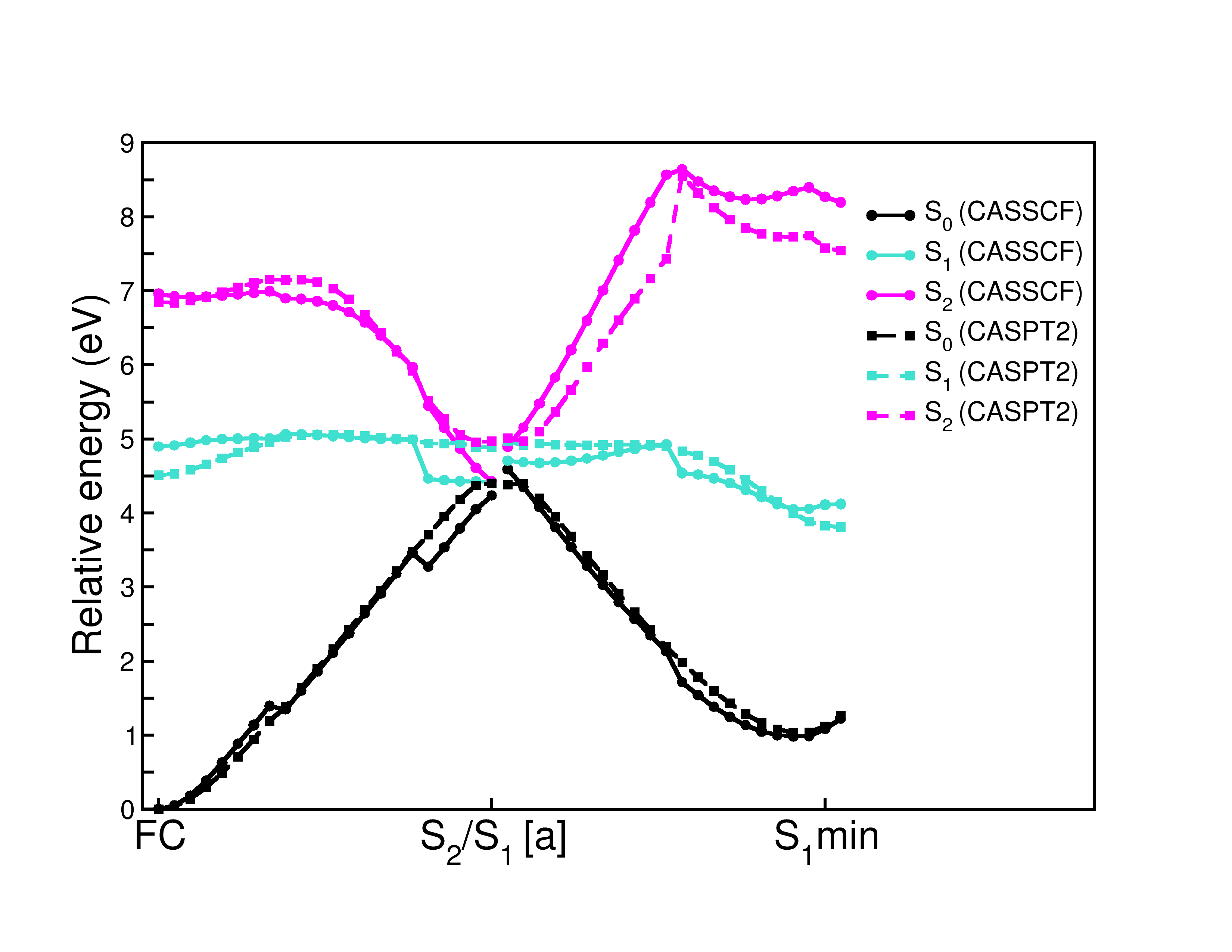}
       \includegraphics[scale=0.3]{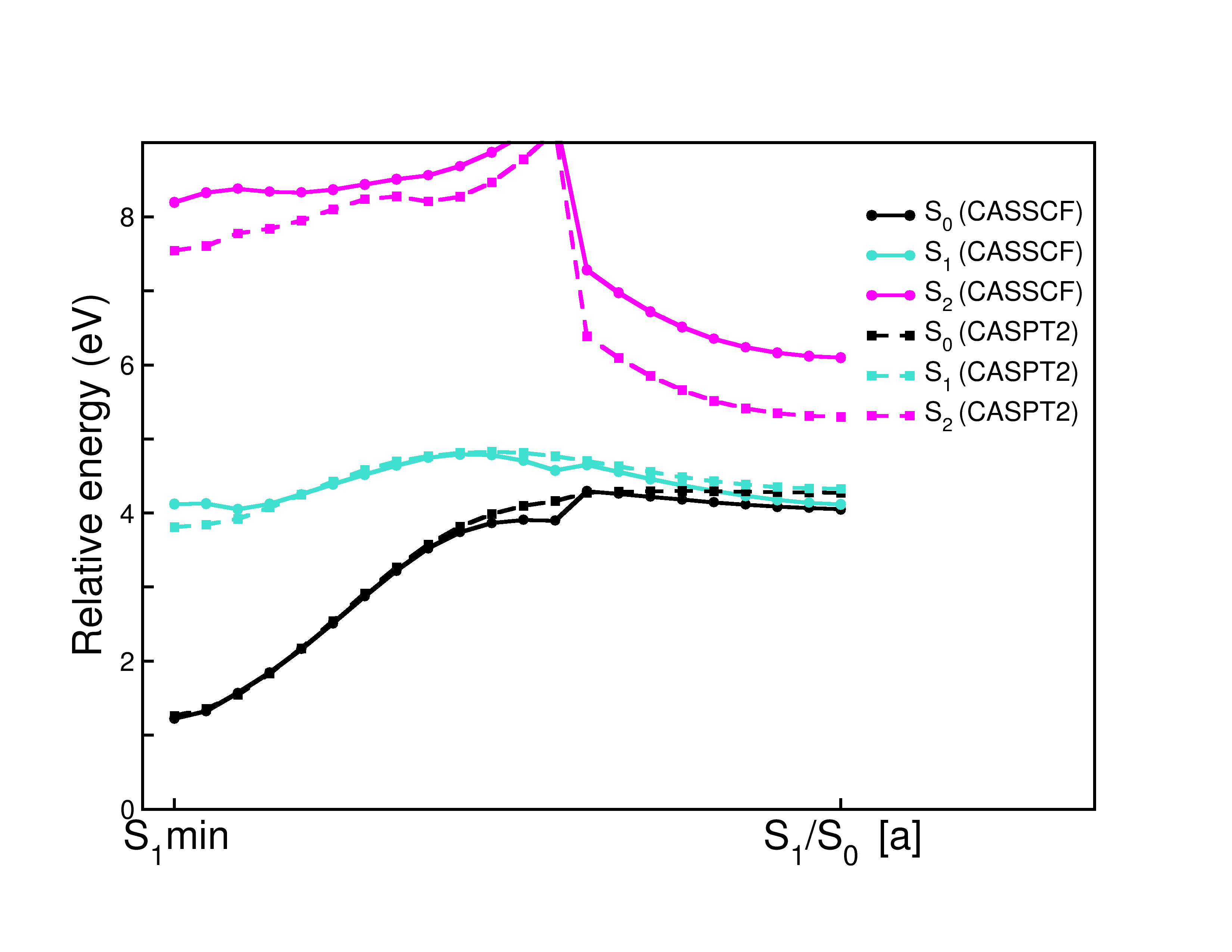}
       \includegraphics[scale=0.3]{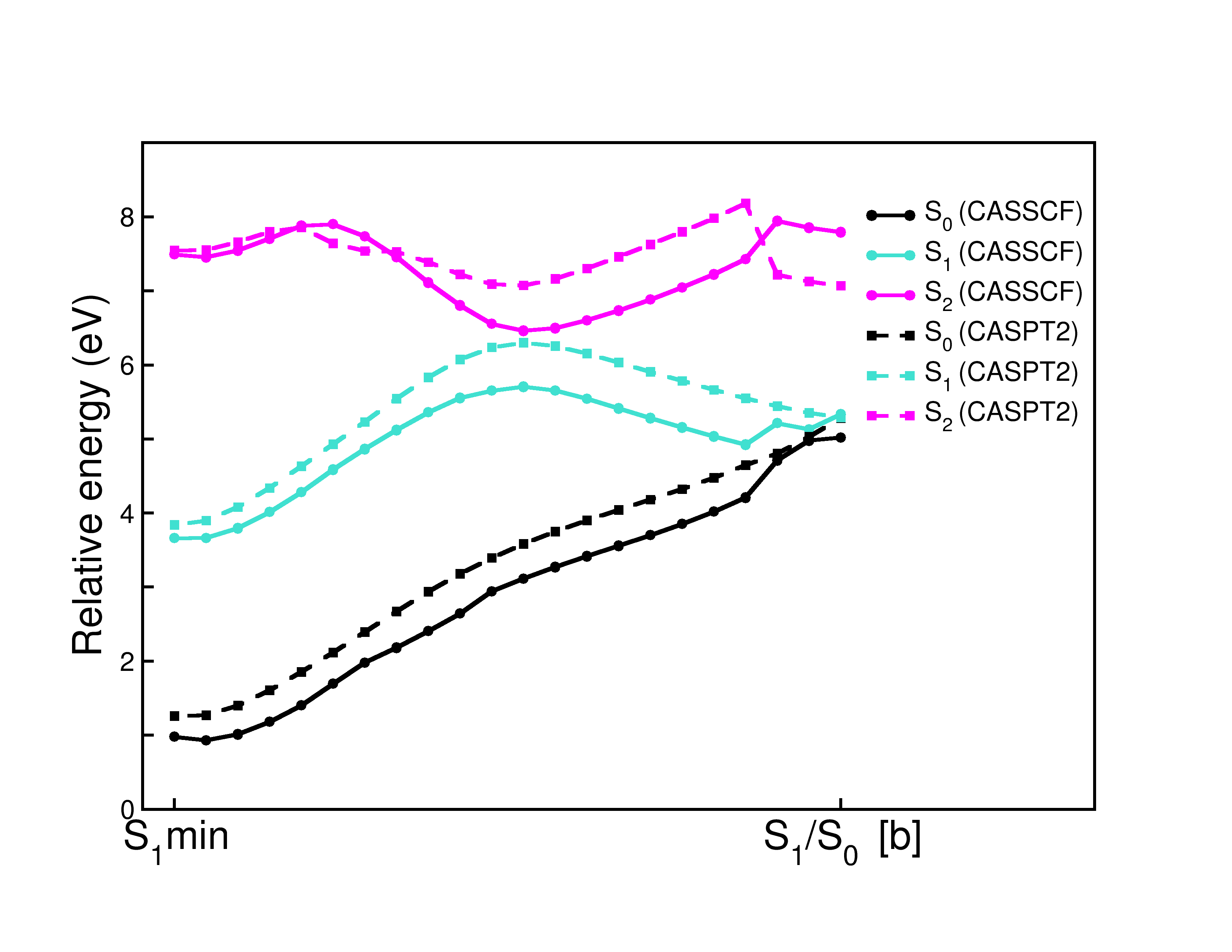}
       \includegraphics[scale=0.3]{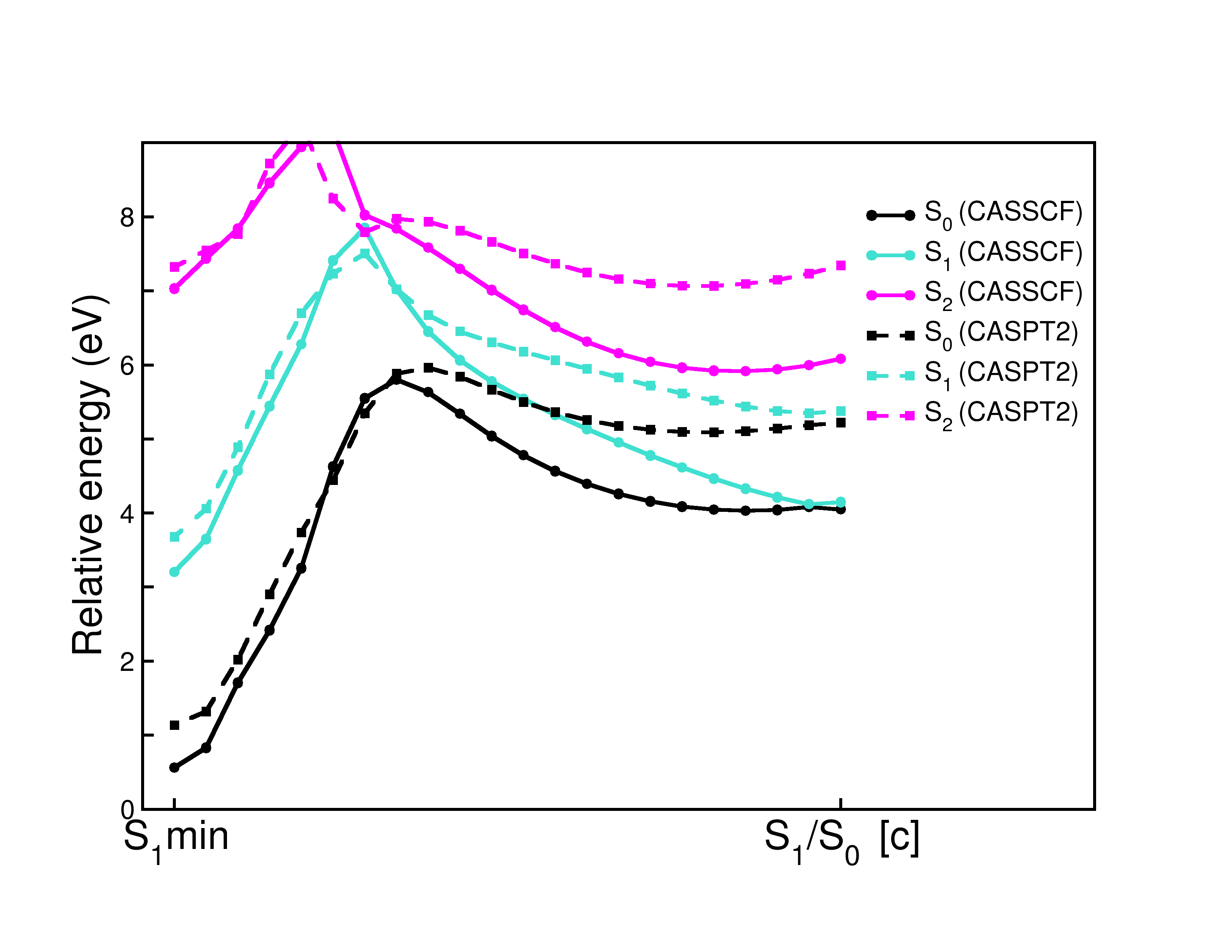}
       \caption{SA-CASSCF and XMS-CASPT2 single-point energies for intermediate structures generated between critical points using a linear interpolation in internal coordinates, calculated at the AS3/6-31+G* level of theory. Energies have been shifted by the \S0 energy at the \C{2v} geometry at the respective level of theory. }
       \label{fig:caspt2_631G}
\end{figure}

Comparisons of the potential energy surfaces with SA-CASSCF and XMS-CASPT2 (Molpro's XMS=1 option, with an IPEA shift of 0.3) between minima and MECIs are provided in Figs.~\ref{fig:caspt2} and \ref{fig:caspt2_631G}, for the aug-cc-pVDZ and 6-31+G* basis sets respectively.

\clearpage

\begin{table}[]
\sisetup{round-mode=places, round-precision=3}
\caption{Vertical excitation energies (VEE) and the magnitude of the transition dipole moment ($|\mu|$) for the three active spaces and two basis sets at the B3LYP/def2-TZVP  \C{2v} geometry, using 3-state SA-CASSCF.}
\label{tab:vert_energies_calc}
\begin{tabular}{@{}l l c c c c @{}}
\toprule
\multirow{2}{*}{Active space} & \multirow{2}{*}{Basis set} & \multicolumn{2}{c}{$\S0 \to \S1$} & \multicolumn{2}{c}{$\S0 \to \S2$}  \\ 
\cmidrule(r){3-4} \cmidrule(l){5-6}
& & VEE (\si{\electronvolt}) & $|\mu|$ (\si{\Debye}) & VEE (\si{\electronvolt}) & $|\mu|$ (\si{\Debye}) \\ \midrule
AS1 (4,4) & 6-31+G* & \num{4.498885898595504} & \num{0.000000040633} & \num{7.270523674420602}  &  \num{0.819273308939}      \\
& aug-cc-pVDZ & \num{4.459869310530898} & \num{0.000000041057} & \num{6.80357193710789}  &  \num{0.7500372557950007}      \\
AS2 (10,9) & 6-31+G* & \num{5.0829701491948} & \num{0.000000067306} & \num{6.919993537254281}    &  \num{0.964863525587}    \\
& aug-cc-pVDZ & \num{5.055819666602838} & \num{0.000000072388} & \num{6.412971213904814}  &  \num{0.9775249339050012}      \\
AS3 (12,11) & 6-31+G* & \num{4.998504290663793} & \num{0.003601357536523357} & \num{6.846156255329743}  &  \num{0.9091657785139129}      \\
& aug-cc-pVDZ & \num{4.744810411021187} & \num{0.00000008178908105609208} & \num{6.231190806961532}  &  \num{0.8369961719730145}      \\
\bottomrule
\end{tabular}
\end{table}

\begin{table}[]
\caption{Vertical excitation energies from literature.}
\label{tab:vert_energies}
\begin{tabular}{@{}l S[table-format=1.2] S[table-format=1.2] @{}}
\toprule
Method                       & {$\S0 \to \S1$ (\si{\electronvolt})} & {$\S0 \to \S2$ (\si{\electronvolt})} \\ \midrule
MS-CASPT2(10,8)/6-31G+*          & 4.1\cite{Xia_cyclobuitanone_mscaspt2}   &    {-}              \\
SA-CASSCF(12,11)/6-31G*          & 4.41\cite{Liu_cyclobutanone_AIMS}           &         {-}            \\
MS-CASPT2(12,11)/6-31G*          & 4.39\cite{Liu_cyclobutanone_AIMS}           &           {-}          \\
CC2/cc-pVTZ+1s1p1d(diffuse)      & 4.48\cite{Kuhlman_cyclobutanone_mctdh_modelPES} & 6.02\cite{Kuhlman_cyclobutanone_mctdh_modelPES}   \\
CCSD/cc-pVTZ+1s1p1d(diffuse)     & 4.45\cite{Kuhlman_cyclobutanone_mctdh_modelPES} & 6.60\cite{Kuhlman_cyclobutanone_mctdh_modelPES}   \\
CCSDR(3)/cc-pVTZ+1s1p1d(diffuse)  & 4.41\cite{Kuhlman_cyclobutanone_mctdh_modelPES} & 6.50\cite{Kuhlman_cyclobutanone_mctdh_modelPES}   \\
Expt.\ (maximum of spectral peak)                        & 4.4\cite{Kao_ringstrain,cyclobutanone_uv-vis}                 &  6.4\cite{cyclobutanone_rydberg_spectrum}            \\ 
\bottomrule
\end{tabular}
\end{table}

\begin{table}[]
    \caption{Vertical excitation energies (VEE) and the magnitude of the transition dipole moment ($|\mu|$) at the B3LYP/def2-TZVP \C{2v} geometry, using 5-state (3S+2T) SA-CASSCF(AS3)/aug-cc-pVDZ.}
    \centering
    \begin{tabular}{c c c}
    \toprule
    Structure & VEE (eV) & $|\mu|$ (Debye) \\
    \midrule
         T$_1$ & 4.419 & \\
         S$_1$ & 4.600 & 0.000 \\ %
         T$_2$ & 5.987 & \\
         S$_2$ & 6.059 & 0.821 \\
    \bottomrule
    \end{tabular}
    \label{tab:vee_and_aee}
\end{table}

\begin{table}[]
    \caption{Relative energies of minimum energy conical intersections (MECI) and crossing points (MECP) at the \C{2v} geometry. The energy of a crossing is defined as the average energy of the two states involved and the spin--orbit coupling (SOC) is calculated using the Breit--Pauli Hamiltonian. All crossings were optimized at the SA-CASSCF/6-31+G* level of theory, with a state-averaging over S$_0$, S$_1$, and S$_2$ (S$_0$, T$_1$, S$_1$, T$_2$ and S$_2$) for the MECIs (MECPs).  Each of the three \S1/\S0 and \S2/\S1 MECIs are labelled (a,b,c) with corresponding geometries shown in Figs.~\ref{fig:meci_s1s0} and \ref{fig:meci_s1s2} respectively. }
    \centering
    \begin{tabular}{c c c}
    \toprule
    Structure & Energy (eV) & SOC (cm$^{-1}$) \\
    \midrule
        S$_2$/S$_1$ [a]     & 4.182 & \\
        S$_2$/S$_1$ [b]     & 5.545 & \\
        S$_2$/S$_1$ [c]     & 4.546 & \\
        S$_2$/T$_2$     & 4.244 & 5.37\\ 
        S$_1$/S$_0$ [a] & 3.771 & \\
        S$_1$/S$_0$ [b] & 4.860 & \\
        S$_1$/S$_0$ [c] & 3.727 & \\
 
        S$_1$/T$_1$     & 4.514 & 0.00 \\ %
        T$_1$/S$_0$     & 1.875 & 0.30\\
    \bottomrule
    \end{tabular}
    \label{tab:crossings}
\end{table}

\begin{comment}
\section{Machine learning} %
\begin{itemize}
    \item show training curves
\end{itemize}
\end{comment}

\clearpage

\section{Initial distribution} %
\begin{comment}
\begin{itemize}
    \item describe how Wigner distribution is obtained for double well
    %
    \item assume that adiabatic gas expansion leaves molecules close to 0 K - no, on the contrary, it's warmed up
    \item temperature effects?
    \item finite-pulse length: does Todd do anything special for this?
\end{itemize}
\end{comment}

The ground-state potential energy surface was explored at the level of B3LYP/def2-TZVP.
These density-function theory calculations were carried out in ORCA.\cite{Neese2012orca}
At this level of theory, cyclobutanone was found to show a double-well structure in the puckering mode.
The simple harmonic approximation therefore becomes unreliable. 
In fact, cyclobutanone was known to have a double well from early infrared and Raman spectroscopy. \cite{Frei1961cyclobutatanone,Durig1966cyclobutanone}

\begin{table}[]
    \centering
    \begin{tabular}{ccc}
        \toprule
        Rotational constant & Calculated & Expt. \\
        \midrule
        $A$ & 10867.5 & 10784.5 \\
        $B$ & 4807.4 & 4806.7 \\
        $C$ & 3555.3 & 3558.5 \\
         \bottomrule
    \end{tabular}
    \caption{Rotational constants (in MHz) calculated from rigid-rotor approximation using the $\C{2v}$ saddle-point structure compared with experimentally determined values for the ground vibrational state from Ref.~\citenum{Alonso1992rotational}.}
    \label{tab:rotational}
\end{table}

The saddle point of the double well is the most useful reference geometry. %
The predicted rotational constants using the rigid-rotor approximation around this geometry are given in Table~\ref{tab:rotational}, which are in good agreement with the experimentally determined values except for $A$.
However, note that $A$ is strongly dependent on the vibrational state (for example, it drops to 10738.0 MHz for the first vibrationally excited state) which explains why the calculated result based on the rigid-rotor approximation is too large.
The harmonic frequencies at this geometry are presented in Table~\ref{tab:frequencies}.
Frequencies greater than 2500\,cm$^{-1}$ were scaled by 0.96 to account for anharmonic effects as well as deficiencies of the approximate DFT treatment.
This is seen to give a systematic improvement in the comparison to experiment.
Note that for the harmonic analysis as well as all the dynamics calculations, we use the atomic mass of the most abundant isotope. %

\begin{table}[]
    \centering
    \begin{tabular}{cccc}
    \hline
    Calculated & Scaled & Expt. & symmetry \\
    \hline
        403 & - & 395 & $B_2$ \\ 
        458 & - & 454 & $B_1$ \\
        624 & - & 902* & $A_2$ \\
        673 & - & 670 & $A_1$ \\
        742 & - & 735 & $B_2$ \\
        834 & - & 850 & $A_1$ \\
        919 & - & - & $B_1$ \\
        949 & - & - & $A_2$ \\
        965 & - & 956 & $A_1$ \\
        1080 & - & 1124 & $B_1$ \\
        1098 & - & 1073 & $B_2$ \\
        1189 & - & 1242 & $B_1$ \\
        1222 & - & 1209 & $B_2$ \\
        1224 & - & 1200* & $A_2$ \\
        1239 & - & - & $A_1$ \\
        1273 & - & 1332 & $B_1$ \\
        1428 & - & 1402 & $B_1$ \\
        1444 & - & 1470 & $A_1$ \\
        1500 & - & 1479 & $A_1$ \\
        1860 & - & 1816 & $A_1$ \\
        3051 & 2929 & 2933 & $B_1$ \\
        3055 & 2933 & 2933 & $A_1$ \\
        3081 & 2955 & 2978 & $A_1$ \\
        3093 & 2970 & 2978 & $B_2$ \\
        3100 & 2977 & 2975* & $A_2$ \\
        3128 & 3002 & 3004 & $B_2$ \\
        \hline
    \end{tabular}
    \caption{Harmonic normal-mode frequencies (in wavenumbers) of ground-state C$_{2v}$ saddle point according to B3LYP/def2-TZVP (excluding the low-frequency puckering mode).
    Experimental results are from gas-phase infrared spectroscopy of Frei and G\"unthard \cite{Frei1961cyclobutatanone}, except where indicated by a *, for which liquid-phase Raman spectrum are given instead.
    }
    \label{tab:frequencies}
\end{table}

The harmonic approximation is clearly not sufficient for the puckering mode.
However, one can easily obtain the anharmonic Wigner function for this single mode directly from the DVR function using Fourier transforms of products of Hermite polynomials, much of which can be worked out analytically.

We first performed a scan along the puckering normal mode and obtained the vibrational wavefunctions for this single degree of freedom using Hermite DVR\@.\cite{Light1985DVR}
Unfortunately, this gave a vibrational spacing of about 142\,cm$^{-1}$, which is significantly larger than the experimentally observed result of 35\,cm$^{-1}$.
This implies that the approximation to treat the normal modes as separable is at fault.

\begin{figure}
    \centering
    \includegraphics[width=\textwidth]{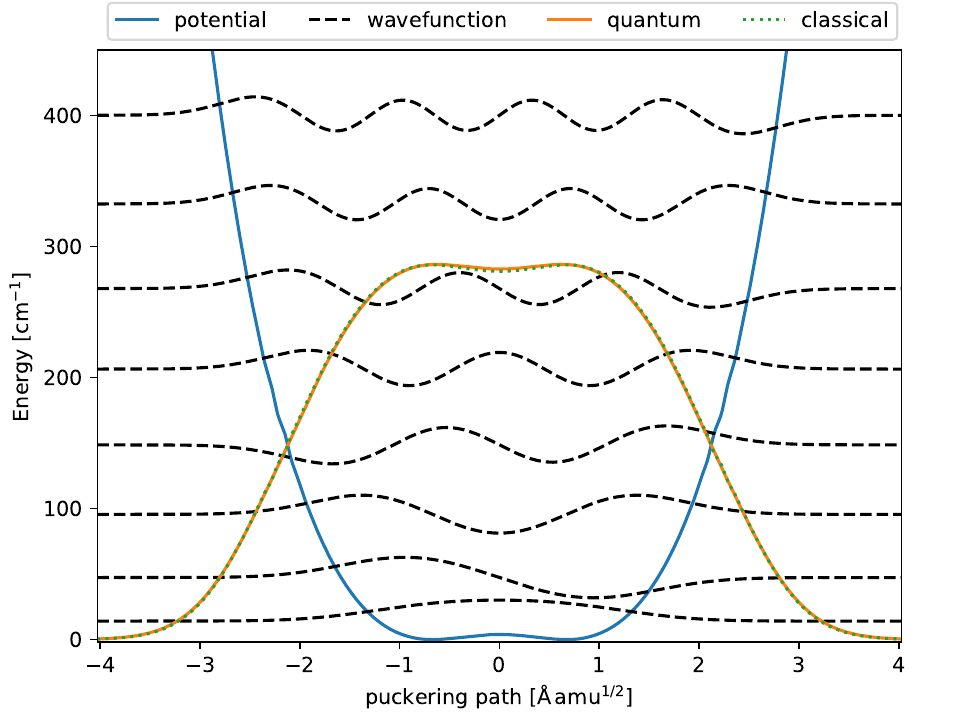}
    \caption{Potential energy (blue) along the puckering path at B3LYP/def2-TZVP level.  A one-dimensional DVR leads to the wavefunctions shown by the black dotted lines.  
    The resulting quantum Boltzmann distribution at $T=\SI{325}{K}$ is shown in orange (with arbitrary units).
    This function is well approximated by a classical Boltzmann distribution, shown with a green dotted line.}
    \label{fig:puckering}
\end{figure}

Next, we computed a ``minimum-energy'' pathway as described in the main text.
Unlike the scan along the normal mode, this path is curved.
The potential along the ``puckering path'' is plotted in Fig.~\ref{fig:puckering}.
Using DVR we computed the vibrational energy levels and wavefunctions (shown in Table~\ref{tab:puckering} and Fig.~\ref{fig:puckering}).
With this approach, the calculated levels match much better with experiment.

\begin{table}
    \centering
    \begin{tabular}{ccc}
    \toprule
    transition & DVR & Expt. \\
    \midrule
    $1\leftarrow0$ & 33.24 & 35.30 \\
    $2\leftarrow1$ & 48.17 & 57.03 \\
    $3\leftarrow2$ & 52.95 & 64.99 \\
    $4\leftarrow3$ & 57.66 & 72.17 \\
    $5\leftarrow4$ & 61.34 & 77.77 \\
    $6\leftarrow5$ & 64.68 & 81.85 \\
    $7\leftarrow6$ & 68.12 & 85.33 \\
    \bottomrule
    \end{tabular}
    \caption{Calculated vibrational transitions (in wavenumbers) for the puckering mode based on the one-dimensional DVR along the minimum-energy pathway compared with experimental data from far-infrared spectroscopy.\cite{Durig1966cyclobutanone}}
    \label{tab:puckering}
\end{table}

In order to compute the Wigner function along the puckering path, we employ a finite-basis representation of harmonic-oscillator eigenstates,
\begin{align}
    \chi_n(x) = \frac{1}{\sqrt{2^n n!}} \left(\frac{a}{\pi}\right)^{1/4} H_n(\sqrt{a}x) \, \eu{-ax^2/2},
\end{align}
for $n=0,1,2,\dots$, where $a$ is an arbitrary scaling factor.
Using DVR, the ground-state wavefunction in the puckering mode is found as a linear combination of harmonic-oscillator states as
\begin{align}
    \psi(x) = \sum_n c_n \chi_n(x).
\end{align}
The Wigner function is then defined as
\begin{subequations}
\begin{align}
    W(x,p) &= \frac{1}{\pi\hbar} \int \psi^*(x+s) \psi(x-s) \,\eu{2\iu ps/\hbar} \, \rmd s
    \\ &= \frac{1}{\pi\hbar} \sum_{n,m} c_n^* c_m w_{nm}(x,p) ,
\end{align}
\end{subequations}
where
\begin{subequations}
\begin{align}
    w_{nm}(x,p) &= \int \chi_n(x+s) \chi_m(x-s) \, \eu{2\iu ps/\hbar} \, \rmd s
    \\ &= \frac{1}{\sqrt{2^n n!}\sqrt{2^{m} m!}} \left(\frac{a}{\pi}\right)^{1/2} \int H_n(\sqrt{a}(x+s)) H_m(\sqrt{a}(x-s)) \, \eu{-ax^2-as^2 + 2\iu ps/\hbar} \, \rmd s
    \\ &= \frac{1}{\sqrt{2^n n!}\sqrt{2^{m} m!}} \left(\frac{a}{\pi}\right)^{1/2} \int H_n\left(\sqrt{a}\big(x+\frac{\iu p}{\hbar a}+s\big)\right) H_m\left(\sqrt{a}(x-\frac{\iu p}{\hbar a}-s)\right) \eu{-ax^2-p^2/\hbar^2a-as^2} \, \rmd s
    \\ &= (-1)^m \sqrt\frac{2^n m!}{2^m n!} \left(\sqrt{a}x+\frac{\iu p}{\hbar\sqrt{a}}\right)^{n-m} L_m^{n-m}\left(2ax^2+\frac{2p^2}{\hbar^2a}\right) \eu{-ax^2-p^2/\hbar^2a} ,
\end{align}
\end{subequations}
for $m<n$ and $L_m^{n-m}$ is the generalized Laguerre polynomial.
For the final step, we were inspired by Eq.~(7.377) from Gradshteyn and Ryzhik,\cite{Gradshteyn} although their formula appears to have an error which had to be corrected by exchanging the arguments of the two Hermite polynomials.

The Wigner function given above is valid only in the low-temperature limit.
However, we wish to initialize the molecule at $T=325\,\mathrm{K}$.
Therefore, not just the ground state, but the lowest 25 vibrational states were computed (with energies up to about 8$k_\mathrm{B}T$)
and the thermal Wigner function defined as a Boltzmann weighted sum over these states:
$W(x,p) = \sum_\nu W^{(\nu)}(x,p) \, \eu{-\beta E_\nu} / Z$, where $Z=\sum_\nu \eu{-\beta E_\nu}$ and $\nu$ labels the vibrational state.

\begin{figure}
    \centering
    \includegraphics[width=\textwidth]{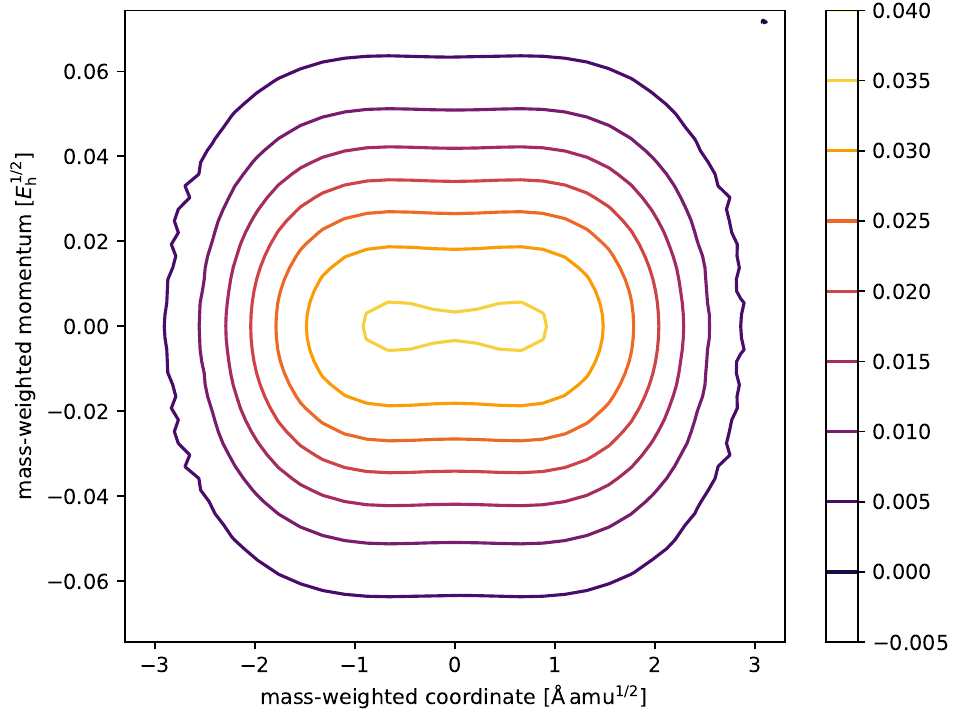}
    \caption{Wigner function along the puckering path. %
    The small wiggles are probably due to the finite truncation and representation on a discrete grid, but are not expected to significantly influence the results.
    }
    \label{fig:wigner_contour}
\end{figure}

Because of the relatively high temperature in this case, it was found to give results quite similar to a simple classical approximation $W_\mathrm{cl}(x,p) = \eu{-\beta V(x)} \eu{-\beta p^2/2} / (2\pi\hbar Z_\mathrm{cl})$, where
$Z_\mathrm{cl} = (2\pi\hbar)^{-1} \iint \eu{-\beta V(x)} \eu{-\beta p^2/2} \, \rmd x \, \rmd p$.
However, the quantum-mechanical approach is more general and was used in our simulations.
The thermal Wigner function is shown in Fig.~\ref{fig:wigner_contour}.

Coordinates and momenta were sampled from the Wigner function in such a way that in the final ensemble all trajectories will contribute with equal weight.
We do this as generating samples is cheap, but running trajectories is computationally expensive and we wish to avoid running an expensive trajectory only to find that it contributes with a small weight.
In principle it is possible for the Wigner function to have negative values, in which case the assigned weights would be negative.
However, in practice, we did not encounter any negative values in our distribution due to the relatively high temperature.

\begin{figure}
    \centering
    \includegraphics[width=\textwidth]{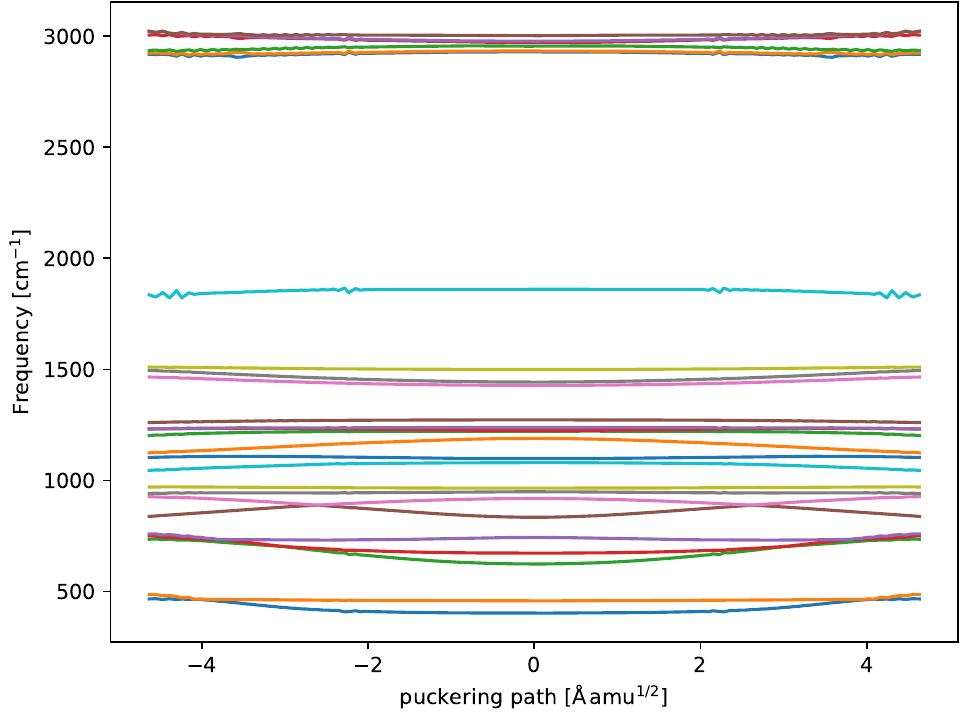}
    \caption{Vibrational frequencies perpendicular to the puckering path as calculated by B3LYP/def2-TZVP within the harmonic approximation.}
    \label{fig:frequencies}
\end{figure}

Finally, the perpendicular modes were sampled.
This was done by first interpolating Hessians along the puckering path.
Then for each sample in the ensemble, we projected out translations and rotations as well as the vector along the puckering path and diagonalized the mass-weighted Hessian to obtain frequencies and perpendicular normal modes.
Frequencies greater than 2500\,$\mathrm{cm}^{-1}$ were scaled by 0.96 to account for anharmonic effects as well as deficiencies of the approximate DFT treatment.
Coordinates and momenta corresponding to each of these modes were sampled from thermal Wigner functions using the harmonic approximation
before being transformed back into Cartesian coordinates.
The perpendicular frequencies obtained along the path are shown in Fig.~\ref{fig:frequencies}.  It is seen that they are much larger than the puckering frequency as well as relatively slowly varying, which justifies our approximation to adiabatically separate them from the puckering path.

Angular momentum was included 
by sampling three mass-weighted angular momenta $p_\theta$, $p_\phi$ and $p_\psi$ from a normal distribution with variance $1/\beta$.
These scalars were used to provide the magnitude of the angular momentum vector (computed for each geometry along the puckering path) that was added to the sampled momentum.

\begin{figure}
    \centering
    \includegraphics[width=\textwidth]{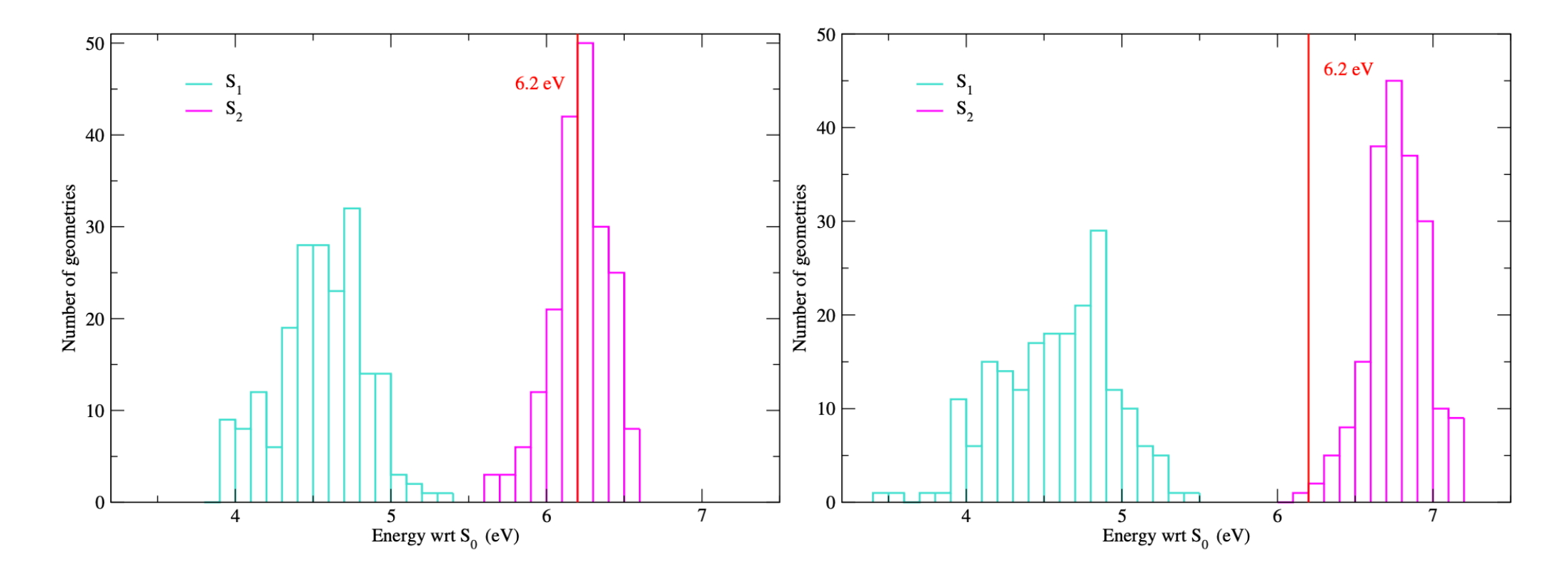}
    \caption{Histogram of the S$_1$ and S$_2$ electronic energies of all structures in the initial distribution with respect to the ground state (aug-cc-pVDZ on the left and 6-31+G* on the right). The red vertical line marks 6.2 eV which corresponds to the experimental excitation energy.}
    \label{fig:ener-distribution}
\end{figure}

\begin{figure}
    \centering
    \includegraphics[width=\textwidth]{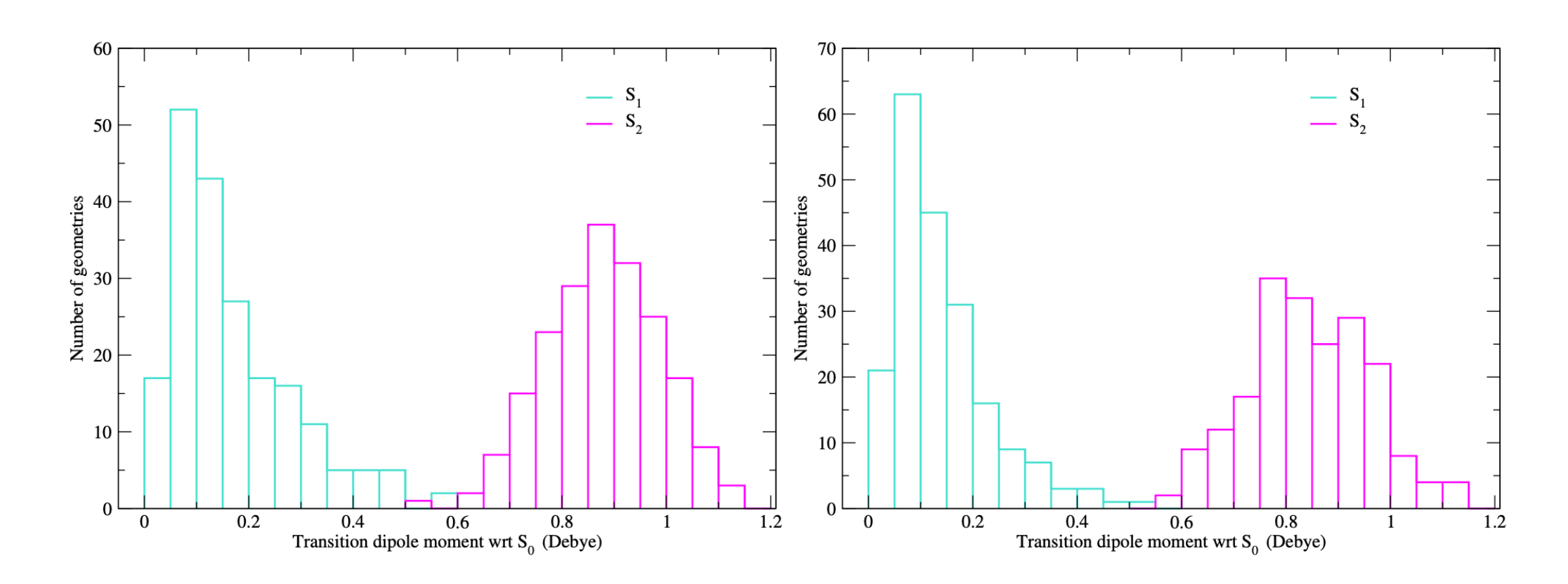}
    \caption{Histogram of the norms of the transition dipole moments of the S$_1$ and S$_2$ electronic states (with respect to the ground state) of all structures in the initial distribution (aug-cc-pVDZ on the left and 6-31+G* on the right).}
    \label{fig:tdm-distribution}
\end{figure}

\begin{figure}
    \centering
    \includegraphics[width=\textwidth]{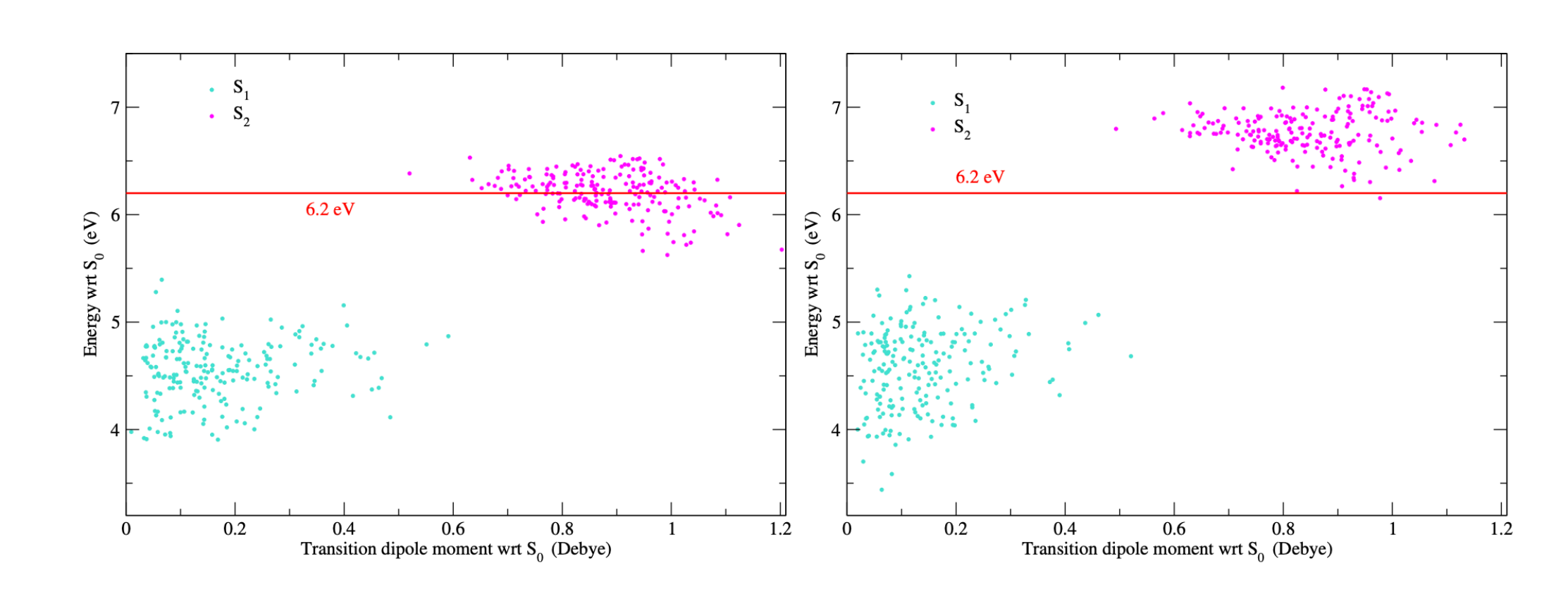}
    \caption{Scatter plot of the energies and norms of the transition dipole moments of the S$_1$ and S$_2$ electronic states (with respect to the ground state) for all structures in the initial distribution (aug-cc-pVDZ on the left and 6-31+G* on the right). The red horizontal line marks 6.2 eV which corresponds to the experimental excitation energy.}
    \label{fig:ener-tdm_scatter}
\end{figure}

For each geometry in our ensemble, we ran a single-point CASSCF calculation to determine the excited-state energies and transition dipole moments.
These results are presented in Figs.~\ref{fig:ener-distribution}, \ref{fig:tdm-distribution} and \ref{fig:ener-tdm_scatter}.
It is seen that only the \S2 state is in resonance with the excitation energy of 200\,nm, which is equivalent to 6.2\,eV\@.
It is also seen that the transition dipole moment between \S0 and \S2 does not vary dramatically across the ensemble.

We found no correlation between the norm of the transition dipole moment and the puckering mode and instead found that it depends on the high-frequency modes.
As these are treated quantum mechanically via the Wigner function, it would not be valid to simply weight each geometry in the ensemble by the square of the transition dipole moment.
In fact, the correct way to account for the variation of the transition dipole moment is to include it in the calculation of the Wigner function.
However, for simplicity, in our treatment, we effectively treat it as a constant.

We also neglect the effects of the pulse shape and allow all trajectories to start in \S2 regardless of the \S2--\S0 energy gap.
We can see no rigorous justification for simply rejecting samples whose \S2--\S0 energy gap is outside the bandwidth of the laser.
The only rigorous approach would require us to simulate the pulse in real time.
We believe it will be possible to extend MASH to treat such problems in future work.

\clearpage
\section{Numerical implementation of MASH}
We give here the technical details of how the multi-state MASH calculations were performed. Note that the method described here is the uncoupled-sphere MASH method (unSMASH). We consider here only the case of a system starting in a pure adiabatic state (i.e.\ no initial adiabatic coherences).
\subsection{Initial Sampling}
For a system with $N$ adiabatic states with an initial active state, $\activestate$, there are a total of $N-1$ effective Bloch spheres, $\{\bm{S}^{(n,j)}|\,j=1,\dots,N \text{ and } j\neq \activestate\}$. Each of the spheres is normalised such that it has a radius of one,
\begin{subequations}
\begin{equation}
     S^{(n,j)}_x = \sin(\theta^{(n,j)})\cos(\phi^{(n,j)})   
\end{equation}
\begin{equation}
     S^{(n,j)}_y = \sin(\theta^{(n,j)})\sin(\phi^{(n,j)})   
\end{equation}
\begin{equation}
     S^{(n,j)}_z = \cos(\theta^{(n,j)}),   
\end{equation}
\end{subequations}
and are defined here such that $S^{(n,j)}_z>0$ corresponds to being on state $\activestate$. Note that, as is true for the two-state Bloch sphere
\begin{equation}
    \begin{pmatrix}
        S_x^{(b,a)} \\
        S_y^{(b,a)} \\
        S_z^{(b,a)} 
    \end{pmatrix} =
  \begin{pmatrix}
        S_x^{(a,b)} \\
        -S_y^{(a,b)} \\
        -S_z^{(a,b)} \label{eq:si-spin-label-swap}
    \end{pmatrix}.
\end{equation}

The initial $\theta^{(n,j)}$ and $\phi^{(n,j)}$ are sampled from the distribution
\begin{equation}
  \rho(\theta^{(n,j)},\phi^{(n,j)}) = \frac{\sin(\theta^{(n,j)}) \, |\cos(\theta^{(n,j)})|  \, h(\cos(\theta^{(n,j)})) }{\int_0^\pi \mathrm{d} \theta \int_0^{2\pi} \mathrm{d} \phi \, \sin(\theta) \, |\cos(\theta)|  \, h(\cos(\theta)) }, \label{eq:si-initial-sphere-dist}
\end{equation}
which is implemented practically by uniformly sampling $u^{(n,j)}\in[0,1)$ and $v^{(n,j)}\in(0,1]$ and setting
\begin{subequations}
    \begin{equation}
        \phi^{(n,j)} = 2\pi u^{(n,j)}
    \end{equation}
    \begin{equation}
        \theta^{(n,j)} = \acos( \sqrt{v^{(n,j)}} ).
    \end{equation}
\end{subequations}

\subsection{Equations of motion}
Here we give a detailed description of the algorithm used to evolve the MASH equations of motion. We note that alternative (but formally equivalent) versions of the algorithm are possible. In particular, as discussed in the main text, the present algorithm predominantly makes use of the non-adiabatic coupling vectors:
\begin{equation}
    d^{(i,j)}_\nu(\bm{q}) = \braket{\psi_i}{\frac{\partial \psi_j}{\partial q_\nu}}
\end{equation}
rather than the overlaps
\begin{equation}
    O_{i,j}(t,t+\delta t) = \braket{\psi_i(t)}{\psi_j(t+\delta t)}.
\end{equation}
This is because the overlaps were found to be significantly more expensive to calculate than the non-adiabatic coupling vectors. However, it would be straightforward to modify the algorithm to predominantly make use of the overlaps if this were not the case. Of course, as is well documented,\cite{Meek2014FSSHIntegrator,Jain2016AFSSH,Mai2020FSSHChapter} it is necessary to use the overlaps rather than the non-adiabatic coupling vectors in the vicinity of a conical intersection, and for this reason the overlaps are used whenever one of the energy gaps drops bellow a predetermined threshold, set here as $V_{\rm cut}=2000\,\mathrm{cm}^{-1}$. 

The algorithm is as follows:
\begin{enumerate}
    \item Propagate the positions and momenta using a velocity Verlet step from $t$ to $t+\delta t$ on the current active surface:
    \begin{subequations}
    \begin{equation}
        p_\nu' \leftarrow p_\nu(t)-  \frac{\delta t}{2}\, \partial_\nu V_{\activestate}\big(\bm{q}(t)\big)
    \end{equation}
    \begin{equation}
        q_\nu(t+\delta t) \leftarrow q_\nu(t) + \frac{p_\nu'}{m_\nu} \, \delta t
    \end{equation}
    At this point the electronic structure code is called, and the new position is used to calculate the adiabatic potentials $\bm{V}$, the force on the current active state, $\bm{F}=- \frac{\partial V_{\activestate}}{\partial \bm{q}}$, and the $N-1$ derivative couplings between the active state and all other states, $\bm{d}_{\activestate,j}$. Additionally, if any of the $|\Delta V_{i,j}(t)|<V_{\rm cut}$  or   $|\Delta V_{i,j}(t+\delta t)|<V_{\rm cut}$ then the overlaps $O_{i,j}(t,t+\delta t)$ are calculated. Note the sign of the adiabatic wavefunctions are determined using the scheme described in Sec.~\ref{si-sign_fix}.

The momenta are then updated under the new force (note these are not yet the final momenta)
    \begin{equation}
        p_\nu'' \leftarrow p_\nu'- \frac{\delta t}{2}\, \partial_\nu V_{\activestate}\big(\bm{q}(t+\delta t)\big)
    \end{equation}
    \end{subequations}
    \item Propagate the electronic variables from from $t$ to $t+\delta t$ according to 
    \begin{subequations}
    \begin{equation}
        \bm{S}^{(\activestate,j)} \leftarrow \exp({\mathbf{\Omega}}^{(\activestate,j)} \delta t ) \bm{S}^{(\activestate,j)}(t)
    \end{equation}
    where 
    \begin{equation}
  \mathbf{\Omega}^{(\activestate,j)} = \begin{pmatrix} 0 & -\overline{\Delta V}_{n,j}/\hbar & \overline{T}_{n,j} \\
\overline{\Delta V}_{n,j}/\hbar & 0 & 0 \\
-\overline{T}_{n,j} & 0 & 0
 \end{pmatrix}
\end{equation}
in which the averaged adiabatic energy gap, $\overline{\Delta V}_{\activestate,j}$, is given by
\begin{equation}
  \overline{\Delta V}_{\activestate,j} = \frac{\Delta V_{\activestate,j}(\bm{q}(t)) + \Delta V_{\activestate,j}(\bm{q}(t+\delta t)) }{2}
\end{equation}
($\Delta V_{\activestate,j}=V_{\activestate}-V_{j}$) and the averaged non-adiabatic coupling, $\overline{T}_{n,j}$, is calculated either directly from the non-adiabatic coupling vectors as
\begin{equation}
    \overline{T}_{n,j} = \sum_\nu\frac{ p_\nu(t)\, d^{(\activestate,j)}_\nu\!\big(\bm{q}(t)\big)  + p_\nu''\,d^{(\activestate,j)}_\nu\!\big(\bm{q}(t+\delta t)\big)  }{m_\nu}
\end{equation}
or, if either  $|\Delta V_{\activestate,j}(t)|<V_{\rm cut}$  or   $|\Delta V_{\activestate,j}(t+\delta t)|<V_{\rm cut}$ then it is calculated from the matrix logarithm of the orthogonalised overlap matrix\cite{Meek2014FSSHIntegrator,Jain2016AFSSH} 
\begin{equation}
    \overline{T}_{n,j} = \frac{2}{\delta t} \asin( \tilde{O}^{({n,j})}_{1,2}(t,t+\delta t) ). 
\end{equation}
Here $\tilde{O}^{({n,j})}_{1,2}(t,t+\delta t)$ is the upper right element of the L\"owdin orthogonalisation of the $2\times2$ overlap matrix
\begin{equation}
  \mathbf{O}^{({n,j})} = \begin{pmatrix} \langle \psi_\activestate(t) | \psi_\activestate(t+\delta t) \rangle & \langle\psi_\activestate(t) | \psi_j(t+\delta t) \rangle \\ \langle\psi_j(t) | \psi_\activestate(t+\delta t) \rangle &\langle \psi_j(t) | \psi_j(t+\delta t) \rangle
 \end{pmatrix}  
\end{equation}
i.e.~given the SVD decomposition of $\mathbf{O}^{({n,j})}$ 
\begin{equation}
  \mathbf{O}^{({n,j})} =   \mathbf{U}^{({n,j})} \mathbf{\Sigma}^{({n,j})} (\mathbf{V}^{({n,j})})^T
\end{equation}
 then
\begin{equation}
  \tilde{\mathbf{O}}^{({n,j})} =   \mathbf{U}^{({n,j})} (\mathbf{V}^{({n,j})})^T.
\end{equation}
    \end{subequations}
    \item Check for hops: 
    If all $S_z^{(\activestate,j)}>0$ then no hop has occurred, and one simply sets $p_\nu(t+\delta t)\leftarrow p_{\nu}''$ and $\bm{S}^{(\activestate,j)}(t+\delta t)\leftarrow \bm{S}^{(\activestate,j)}$ before continuing to the next time step. However, if  $S_z^{(\activestate,j)}<0$ then there is an attempted hop, from state $\activestate$ to state $j$. In the case that more than one $S_z^{(\activestate,j)}<0$, then we take the the attempted hop to be to the one for which $|S_z^{(\activestate,j)}|$ is largest.  
    
    Now there are two possibilities
    \begin{enumerate}[i.]
        \item If $E_{\rm kin}^{(d)}=\frac{1}{2} \frac{(\tilde{\bm{p}} \cdot \tilde{\bm{d}})^2}{\tilde{\bm{d}}\cdot\tilde{\bm{d}}}   >\Delta V_{j,\activestate}(t+\delta t)$,
where $\tilde{p}_\nu=p_\nu''/\sqrt{m_\nu}$ and $\tilde{d}_\nu=d^{(n,b)}_\nu(t+\delta t)/\sqrt{m_\nu}$ are the mass-weighted momentum and derivative coupling vectors respectively, then the system has enough energy to hop, and the hop is successful. The mass weighted momentum is then rescaled along the non-adiabatic coupling vector according to 
\begin{subequations}
    \begin{equation}
\tilde{\bm{p}}(t+\delta t) \leftarrow \tilde{\bm{p}} + \left(\sqrt{\frac{E_{\rm kin}^{(d)}+\Delta V_{\activestate,j}(t+\delta t)}{E_{\rm kin}^{(d)}}}-1\right)\tilde{\bm{d}}\,\frac{\tilde{\bm{p}}\cdot\tilde{\bm{d}}}{\tilde{\bm{d}}\cdot \tilde{\bm{d}}},
\end{equation}
to give the new momenta $p_\nu(t+\delta t)=\tilde{{p}}_\nu(t+\delta t)\sqrt{m_\nu}$,
and the active surface is updated 
\begin{equation}
    \activestate(t+\delta t)\leftarrow j.
\end{equation}
As a hop has occurred we must now define a new set of $N-1$ effective Bloch spheres between the new active surface and the other states.
For the basic unSMASH method this is done as follows, we first define 
\begin{equation}
    n_{\rm i} = \activestate(t)
\end{equation}
and 
\begin{equation}
    n_{\rm f} = \activestate(t+\delta t)
\end{equation}
then we define the new set of spheres according to the rule 
\begin{equation}
    \bm{S}^{(\activestate_{\rm f},k)}(t+\delta t) \leftarrow\begin{cases} \bm{S}^{(\activestate_{\rm i},k)} & \text{ if } k\neq \activestate_{\rm i} \\
    \bm{S}^{(\activestate_{\rm f},k)} & \text{ if } k = \activestate_{\rm i} 
    \end{cases}.
\end{equation}
where it is helpful to make use of Eq.~\ref{eq:si-spin-label-swap} which relates spheres whose labels are interchanged. 

        \item Or, if $E_{\rm kin}^{(d)}<\Delta V_{j,\activestate}(t+\delta t)$, then the trajectory does not have enough energy to hop, and the hop is rejected (a frustrated hop). In this case we reverse the mass weighted momentum along the derivative coupling vector to give
        \begin{equation}
    \tilde{\bm{p}}(t+\delta t) \leftarrow  \tilde{\bm{p}} - 2 \tilde{\bm{d}}\,\frac{\tilde{\bm{p}}\cdot\tilde{\bm{d}}}{\tilde{\bm{d}}\cdot \tilde{\bm{d}}}.
\end{equation}
to give the new momenta $p_\nu(t+\delta t)=\tilde{{p}}_\nu(t+\delta t)\sqrt{m_\nu}$. As the hop is rejected the active surface remains unchanged $n(t+\delta t)\leftarrow n$, however, the sphere associated with the hop is still in the wrong hemisphere. As was shown in the original MASH paper, analytically solving the equations of motion in the vicinity of an attempted hop shows that at a frustrated hop $S_z$ never changes sign. To mimic this here we simply set   
\begin{equation}
    S_z^{(n,j)}(t+\delta t) \leftarrow -S_z^{(n,j)}
\end{equation}
leaving $S_x^{(n,j)}(t+\delta t) \leftarrow S_x^{(n,j)}$ and $S_y^{(n,j)}(t+\delta t) \leftarrow S_y^{(n,j)}$, with all other spheres left unchanged $\bm{S}^{(\activestate,j)}(t+\delta t)\leftarrow \bm{S}^{(\activestate,j)}$. The only exception to this is the case that at the start of stage 3 more than one $S_z^{(\activestate,j)}<0$, in this case the $S_z$ for these spheres are also adjusted according to $S_z^{(n,j)}(t+\delta t) \leftarrow -S_z^{(n,j)}$.  
\end{subequations}
    \end{enumerate}
\end{enumerate}

\subsection{Sign-fixing} \label{si-sign_fix}
We note that as with all non-adiabatic dynamics methods it is essential that a consistent phase is used for each of the adiabatic wavefunctions, such that the wavefunctions vary continuously along the trajectory. As all wavefunctions are real, this means that the signs of each wavefunction should be chosen consistently for successive time steps.
Given a set of adiabatic wavefunctions with arbitrary signs at a series of times, {$\{   \phi_j(\alpha \delta t) \, | \, \alpha \in \mathbb{Z} \}$}, then we we define the set of wavefunctions with consistent signs as  $\{ \psi_j(\alpha \delta t)= \sigma_j(\alpha \delta t) \phi_j(\alpha \delta t) \, | \, \alpha \in \mathbb{Z} \}$, where $\sigma_j(0)=1$.  
Away from conical intersections, in regions where the overlaps are not calculated, we impose the continuity of the adiabatic wavefunctions by requiring that
\begin{equation}
     \mathrm{sign}\!\left[ \bm{d}^{(n,j)}\big(\bm{q}((\alpha+1) \delta t)\big) \cdot  \bm{d}^{(n,j)}\big(\bm{q}(\alpha \delta t)\big) \right] = 1
\end{equation}
in terms of the derivative coupling calculated using the uncorrected wavefunctions, $\bm{d}_{\phi}^{(n,j)}$, this means
\begin{equation}
      \sigma_j((\alpha+1) \delta t)  \sigma_n((\alpha+1) \delta t) \times  \mathrm{sign}\!\left[ \bm{d}_{\phi}^{(n,j)}\big(\bm{q}((\alpha+1) \delta t)\big) \cdot  \bm{d}^{(n,j)}\big(\bm{q}(\alpha \delta t)\big) \right] = 1.
\end{equation}
Now since it is only the relative signs that are important we can define $\sigma_n((\alpha+1) \delta t)=\sigma_n(\alpha \delta t)$, and hence we have
\begin{equation}
    \sigma_j((\alpha+1) \delta t) \leftarrow \sigma_n(\alpha \delta t) \times \mathrm{sign}\!\left[ \bm{d}_{\phi}^{(n,j)}\big(\bm{q}((\alpha+1) \delta t)\big) \cdot  \bm{d}^{(n,j)}\big(\bm{q}(\alpha \delta t)\big) \right],
\end{equation}
from which we can obtain the sign corrected derivative coupling
\begin{equation}
    \bm{d}^{(n,j)}\big(\bm{q}((\alpha+1) \delta t)\big) \leftarrow \bm{d}_{\phi}^{(n,j)}\big(\bm{q}((\alpha+1) \delta t)\big) \times \sigma_j((\alpha+1) \delta t) \times \sigma_n((\alpha+1) \delta t).
\end{equation}
When the overlaps are available we simply define the signs to maintain a positive diagonal in the overlap matrix 
\begin{equation}
    \sigma_j((\alpha+1) \delta t) \leftarrow \sigma_j(\alpha \delta t) \times \mathrm{sign}\Big[ \braket{\phi_j(t)}{\phi_j(t+\delta t)}\Big].
\end{equation}
With this scheme we obtain a unique set of $\sigma_j(\alpha \delta t)$, that give a continuous set of wavefunctions as the time step $\delta t\to 0$. The overlap matrices are then simply calculated from the output of the electronic structure code as
\begin{equation}
  O_{i,j}(t,t+\delta t)=\braket{\psi_i(t)}{\psi_j(t+\delta t)} = \sigma_i(t) \sigma_j(t+\delta t) \braket{\phi_i(t)}{\phi_j(t+\delta t)}.
\end{equation}
\subsection{Interface to electronic structure package}
The unSMASH integrator was interfaced with Molpro 2023\cite{MOLPRO2023} for the calculation of all necessary electronic properties. This interface was designed to minimise the cost of the CASSCF calculations by requesting Molpro calculate only the necessary quantities at each step (e.g.\ gradients, NACVs or overlaps). Additionally to aid in the convergence of the SCF calculation, each step used the previous wavefunction as an initial guess.

\clearpage

\section{Analysis of hopping points}\label{sec:hop_bin}
\begin{figure}
       \centering
       \includegraphics[scale=0.4]{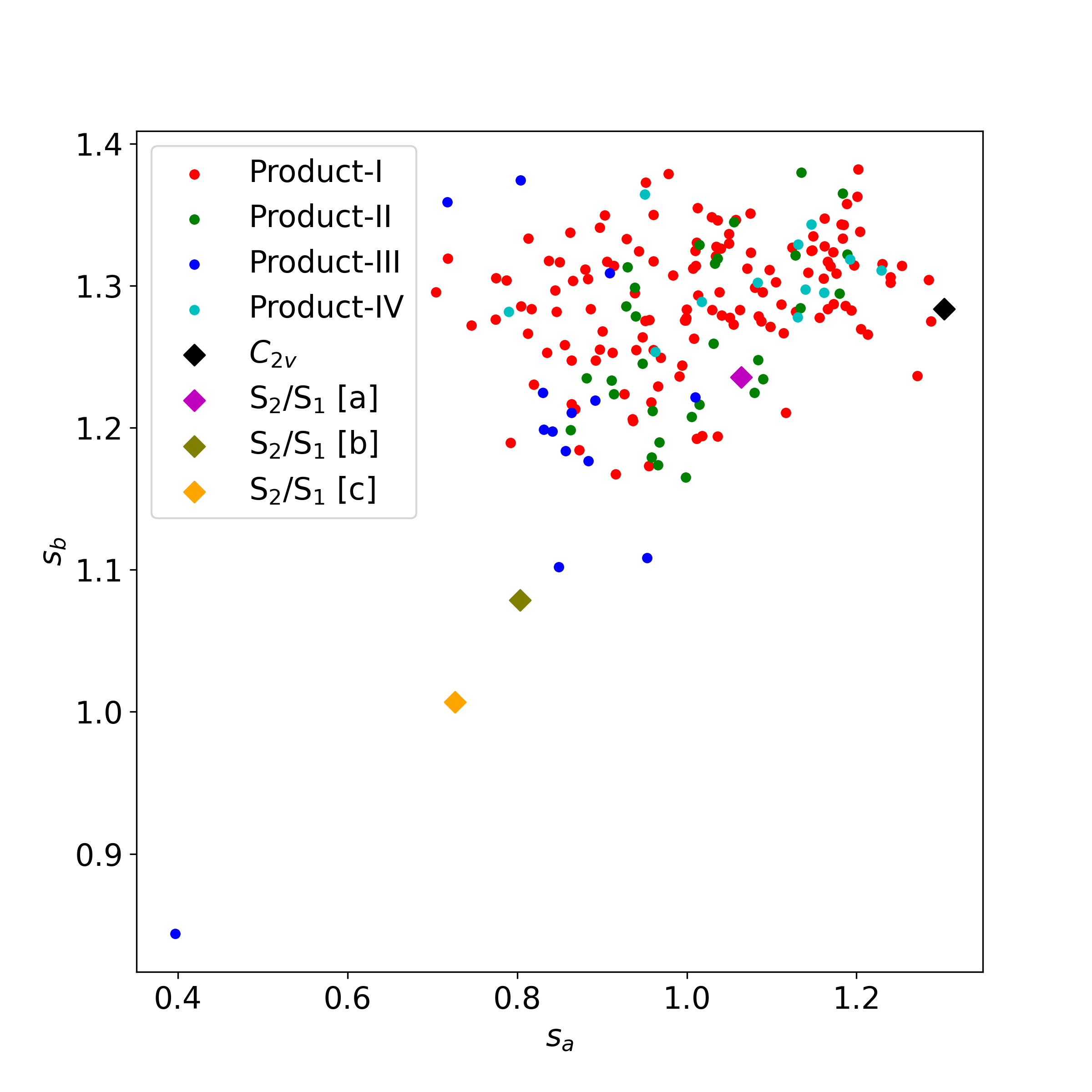}
       \includegraphics[scale=0.4]{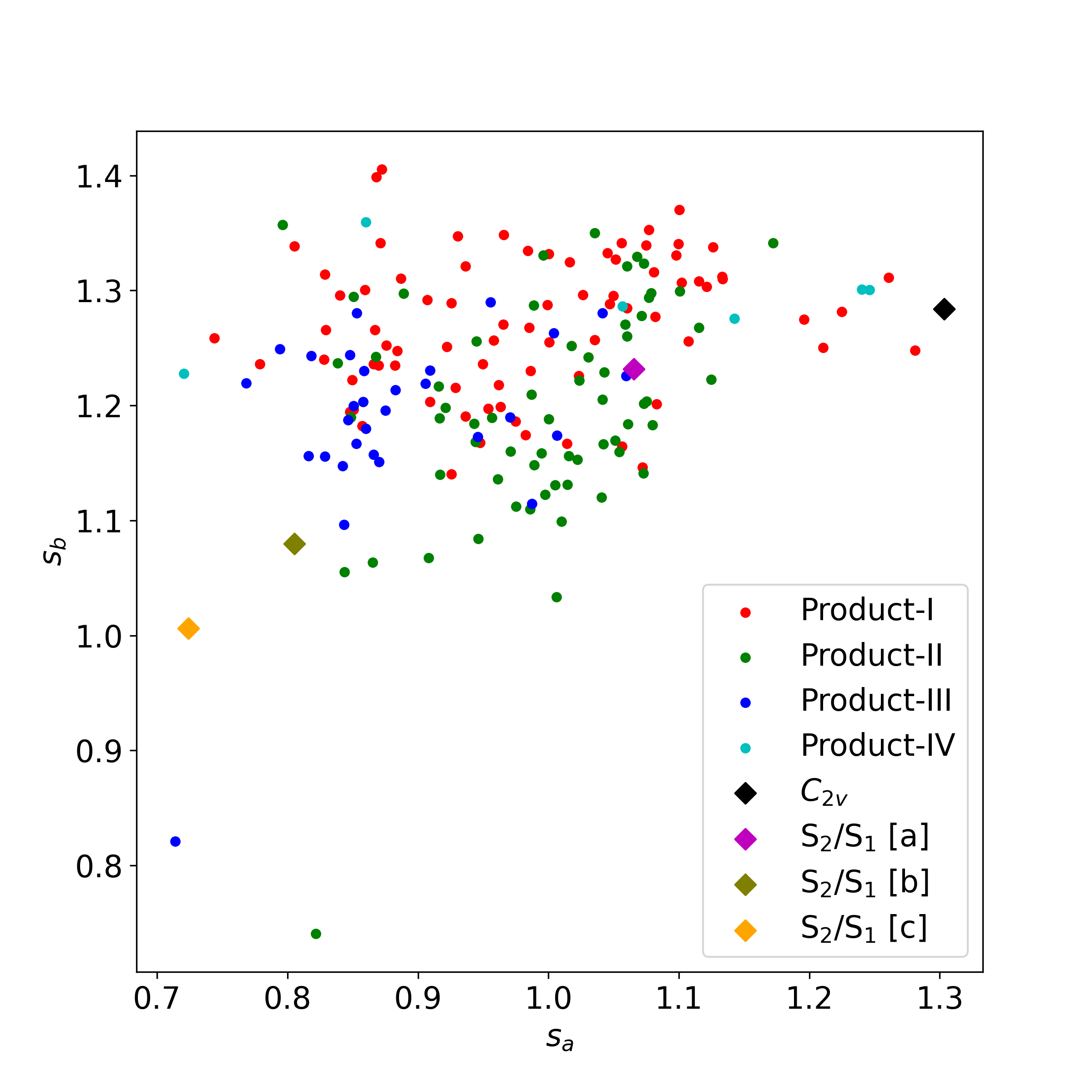}
       \includegraphics[scale=0.4]{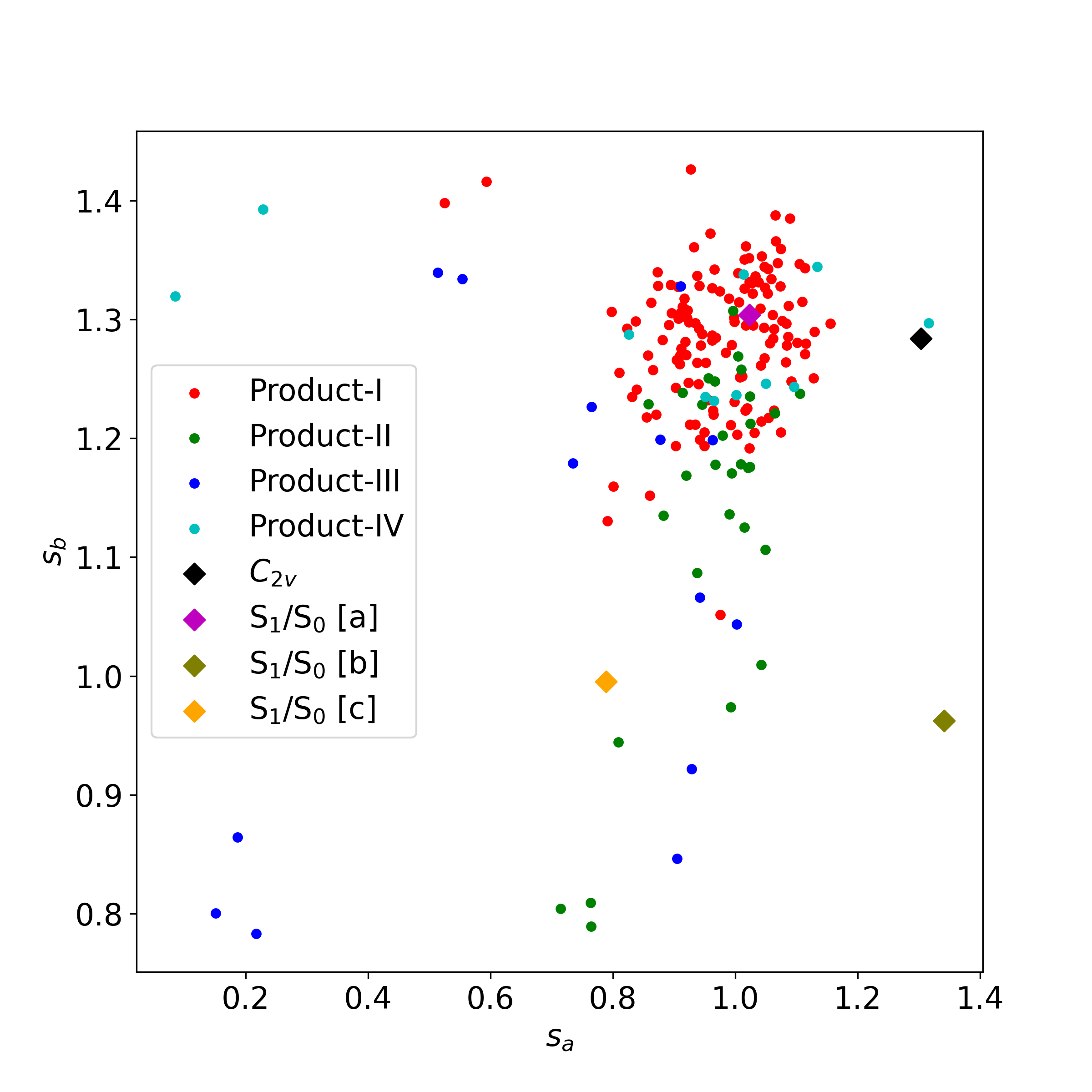}
       \includegraphics[scale=0.4]{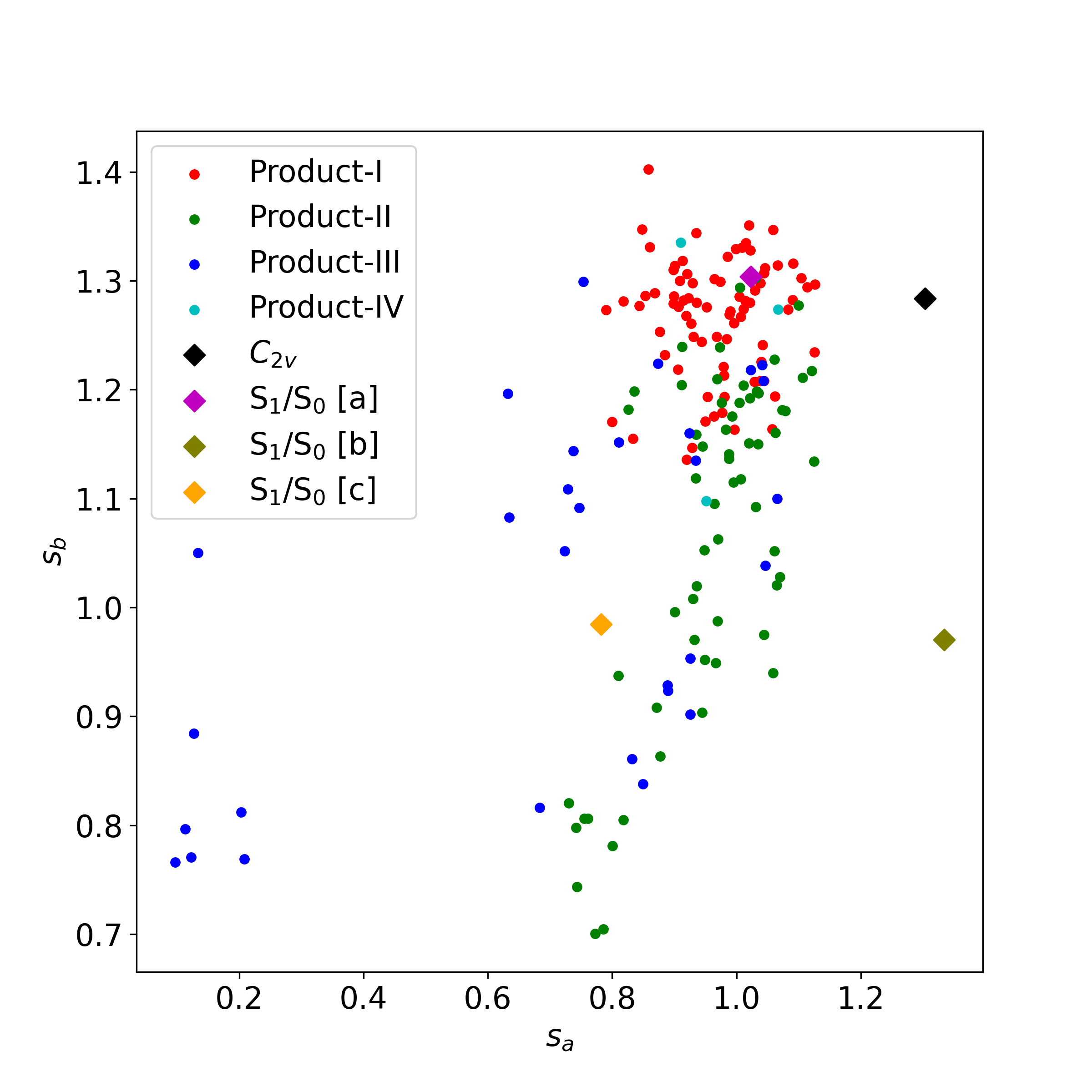}
       \caption{Final hops between $\S2$ and $\S1$ (top) and between $\S1$ and $\S0$ (bottom) have been projected along the $r_a$ and $r_b$ coordinates (aug-cc-pVDZ on the left and 6-31+G* on the right). Note Product-IV includes all types of reaction products not in groups I-III.}
       \label{fig:hops_bin}
\end{figure}

To determine if there is a correlation between the products and the MECIs, we have projected the geometries of final hops in all trajectories along coordinates $s_a$ and $s_b$, which are defined as sums of inverse C-C bond lengths:
\begin{equation*}
    s_a = 1/r_{\alpha\beta_1} + 1/r_{\alpha\beta_2} 
\end{equation*}
\begin{equation*}
    s_b = 1/r_{\beta_1\gamma} + 1/r_{\beta_2\gamma} \\
\end{equation*}
where, $r_{\alpha\beta_1}$ is the bond length between the \textalpha-C and one of the \textbeta-C atoms, and so on. We have verified that $r_a$ and $r_b$ can differentiate between the various critical points (solid diamonds in Fig.~\ref{fig:hops_bin}). The hopping points are coloured according to the product they result in. Note that product IV refers to trajectories where the fragments correspond to none of the products defined in the main text (this only happens in 12 (6) trajectories at the aug-cc-pVDZ (6-31+G*) level of theory).

For the \S2 to \S1 hops, the \S2/\S1 [a] MECI is mostly preferred, which is consistent with it being the lowest energy MECI (Table~\ref{tab:crossings}). The vicinity of \S2/\S1 [b] is visited more with the 6-31+G* basis set than with the aug-cc-pVDZ basis set, particularly when the trajectories lead to product II and III. The \S2/\S1 [c] is never visited by the hopping points, even though it is not that high in energy, but likely because it involves the breaking of too many bonds (Fig.~\ref{fig:meci_s1s2}).

For the \S1 to \S0 hops, \S1/\S0 [a] is mostly preferred, and \S1/\S0 [b] is never visited by any of the hopping points, which is consistent with it being the highest energy MECI (Table~\ref{tab:crossings}). We also see that product I nearly always hops near \S1/\S0 [a] but products II and III can be formed by hopping over a much wider region, which may be near \S1/\S0 [a] or \S1/\S0 [c]. This correlation between the formation of product II and III and visiting the region near \S1/\S0 [c] is stronger with the 6-31+G* basis set than with aug-cc-pVDZ.

\section{Triplet populations}\label{sec:triplets}
\begin{figure}
    \centering
    \includegraphics[width=0.8\columnwidth]{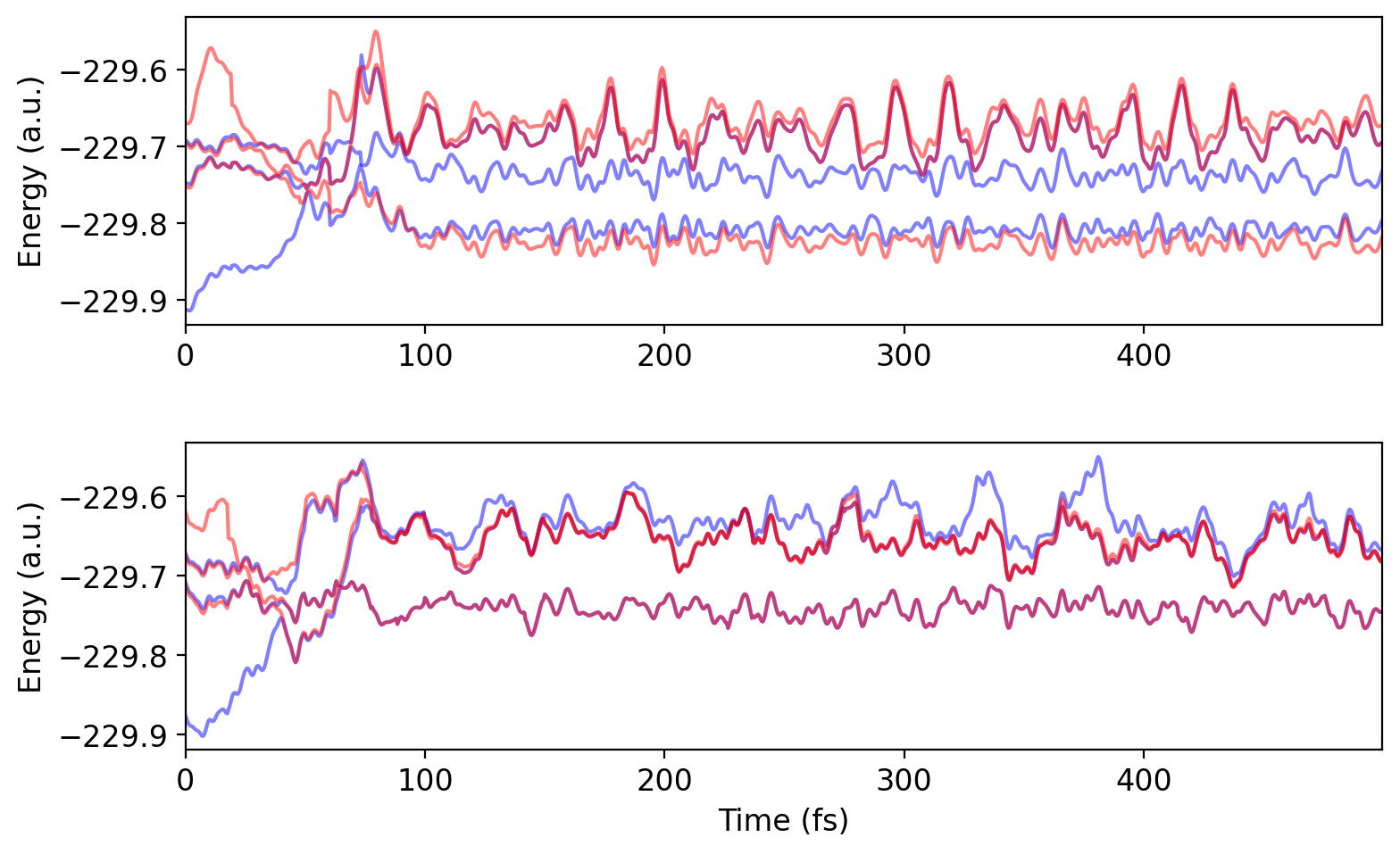}
    \caption{Energies calculated for the 3 lowest singlet (blue) and triplet states (red) calculated along 2 trajectories from the final set with a 6-state SA-CASSCF using (12,11) active space and aug-cc-pVDZ basis set.}
    \label{fig:triplet-es}
\end{figure}

As described in the main text, we do not include the triplet states in the simulation. In order to check whether this approximation holds, using 6-state (3 singlet and 3 triplet states) SA-CASSCF with a (12,11) active space and aug-cc-pVDZ basis set, we calculated the energies, overlaps and SOCs between the singlet and triplet states over 10 randomly selected MASH trajectories. Figure \ref{fig:triplet-es} shows the energies of the singlet and triplet states along 2 of these trajectories.
Note that the 6-state SA-CASSCF calculations yield slightly different energies for the singlet states compared to the 3-state calculations used to generate the trajectories, although the behaviour is qualitatively similar.

\begin{figure}
    \centering
    \includegraphics[width=\columnwidth]{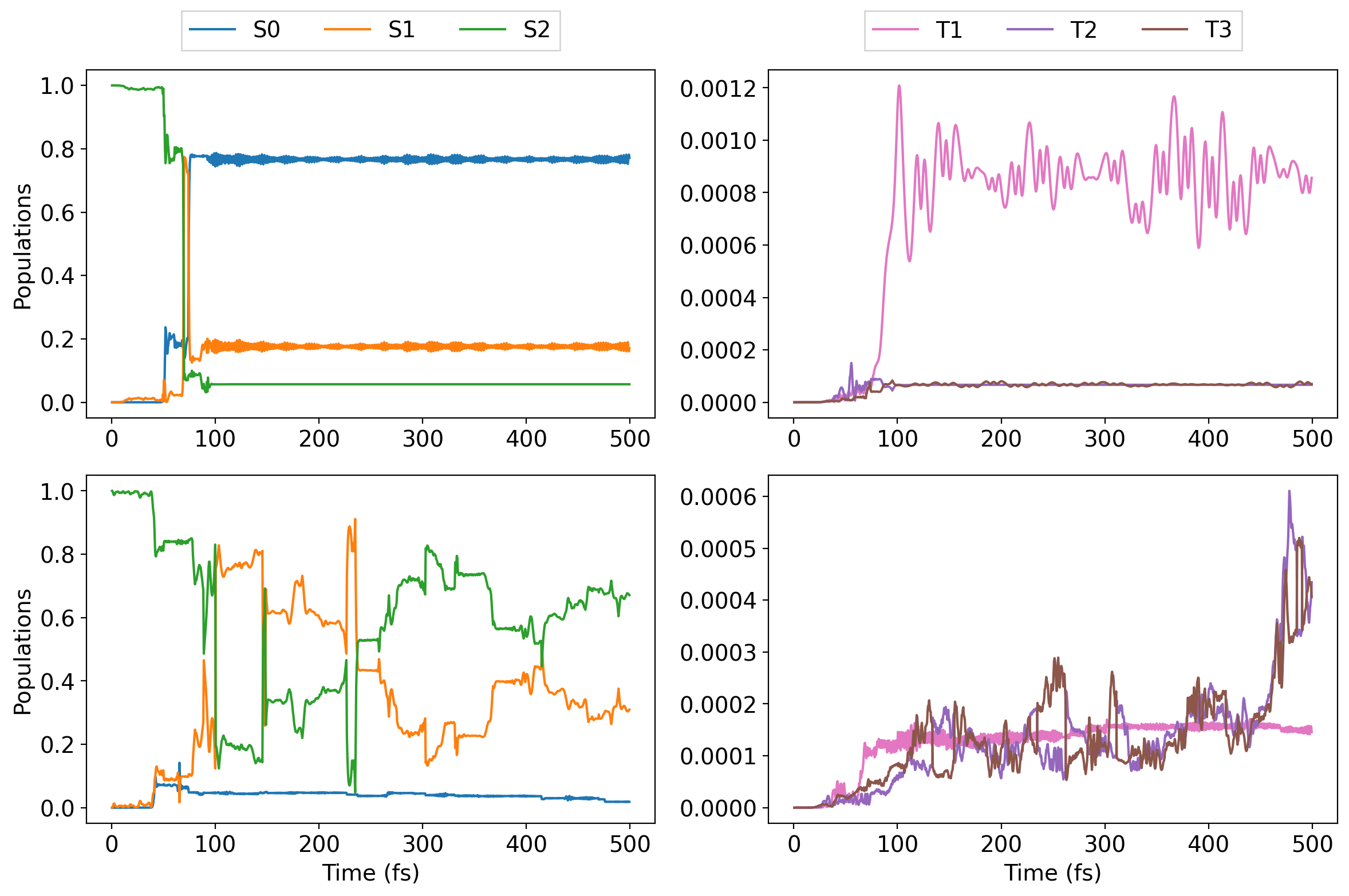}
    \caption{Singlet (left) and triplet (right) state populations calculated as described in Section \ref{sec:triplets} along two representative trajectories.}
    \label{fig:triplet-pops}
\end{figure}

We then simulated the electronic dynamics along each of the selected trajectories (i.e., neglecting the back reaction), using the overlaps and the SOCs to determine the time-evolution of an electronic wavefunction initialized in a pure \S2 state. 
Figure \ref{fig:triplet-pops} shows the singlet and total triplet populations for two of the selected trajectories. The total triplet population was found to be less than $0.6$\% along all trajectories considered.

\clearpage
\section{Electron Diffraction} %

The electron-diffraction signal was simulated 
within the single-scattering independent-atom approximation \cite{Centurion2022review}
using the following formulae for the differential cross sections $I(s)$:
\begin{subequations}
\begin{align}
    I_\mathrm{atom}(s) &= \sum_{i=1}^{\Natom} |f_i(s)|^2 + |g_i(s)|^2
    \\
    I_\mathrm{mol}(s) &= \sum_{i=1}^{\Natom} \sum_{j=i+1}^{\Natom} \Re\left[f_i^*(s)f_j(s) + f_j^*(s)f_i(s) + g_i^*(s)g_j(s) + g_j^*(s)g_i(s)\right] \Braket{ \frac{\sin(s r_{ij})}{s r_{ij}} }
    \\
    sM(s) &= s \frac{I_\mathrm{mol}(s)}{I_\mathrm{atom}(s)}
    \\
    \mathrm{PDF}(r) &= \int_{s_\mathrm{min}}^{s_\mathrm{max}} sM(s) \sin(s r) \, \eu{-\alpha s^2} \, \rmd s ,
\end{align}
\end{subequations}
where $f_i(s)$ and $g_i(s)$ are the atomic form factors for direct and spin-flip scattering events on atom $i$.
Note that the spin-flip $g_i(s)$ factors were included in our calculations even though they are significantly smaller than the corresponding $f_i(s)$ and could thus probably be safely neglected.
The average is taken over the nuclear distribution obtained from the ensemble of trajectories at given time, where $r_{ij}$ is distance between atoms $i$ and $j$.
No extra broadening is necessary as the Wigner distribution used in our study accounts for zero-point energy effects.
The Fourier transform used to generate PDF is damped by a Gaussian with parameters given in Table~\ref{tab:ED}.

\begin{table}[h]
    \centering
    \begin{tabular}{cc}
        \toprule
        quantity  & value \\
        \midrule
        $s_\mathrm{min}$ & $0.01\,\AA^{-1}$ \\
        $s_\mathrm{max}$ & $10\,\AA^{-1}$ \\
        $\alpha$ & $0.05\,\AA^2$ \\
        $E_\mathrm{kin}$ & 3.7\,MeV \\
        \bottomrule
    \end{tabular}
    \caption{Parameters for the electron-diffraction calculations}
    \label{tab:ED}
\end{table}

Note that various definitions of the pair distribution function (PDF) are used in the literature.
Our definition is equivalent to that used in Ref.~\citenum{Wolf2019CHD} and is related to that used in Ref.~\citenum{Centurion2022review} by $P(r)=r\mathrm{PDF}(r)$.

The atomic form factors were generated using
the ELSEPA package.\cite{Salvat2021elsepa}
The default settings were used, that is a Fermi nuclear charge distribution, Dirac--Fock electron density and Furness--McCarthy exchange potential.
The de Broglie wavelength of the relativistic scattering electron is
$\lambda = h/p$, where $E^2 = (pc)^2 + (mc^2)^2$ and $E = E_\mathrm{kin} + mc^2$.
We assume the electron has kinetic energy of $E_\mathrm{kin}=3.7\,\mathrm{MeV}$ as in Ref.~\citenum{Wolf2019CHD}, which gives $\lambda=0.00297\AA$.
The momentum transfer for an elastic collision is $\hbar s$, where
\begin{align}
    s = \frac{4\pi}{\lambda} \sin\frac{\theta}{2} ,
\end{align}
and $\theta$ is the scattering angle (from 0 to $\pi$).

\begin{figure}
    \centering
    \includegraphics{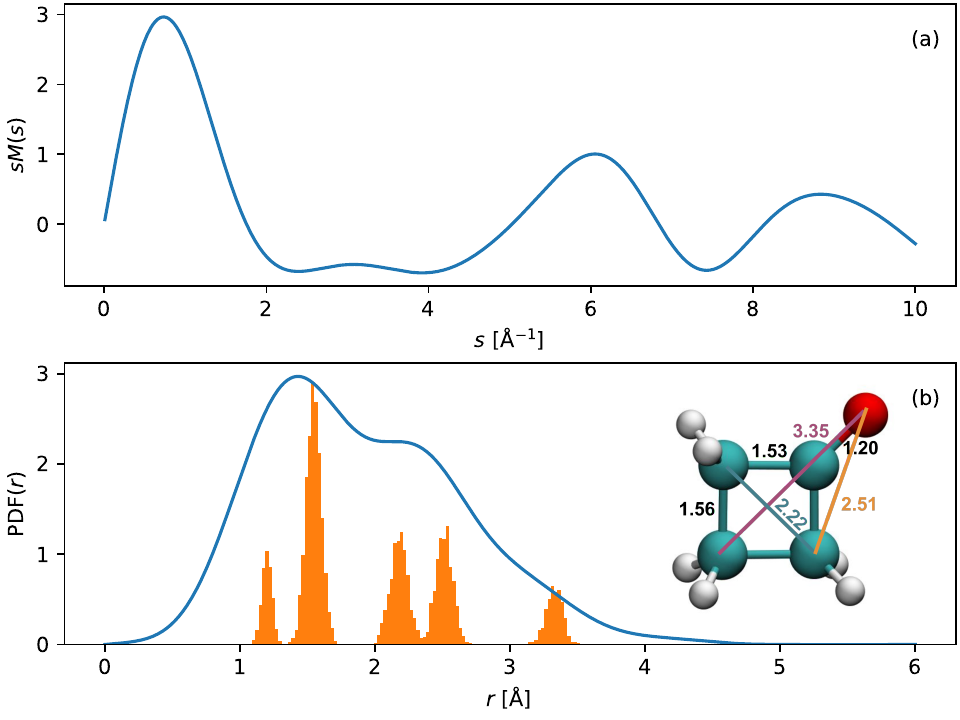}
    \caption{(a) Calculated electron-diffraction signal for the initial distribution.  (b) Calculated PDF based on transforming the signal from panel (a).  Also shown in orange is a histogram of the atom pair distribution (carbons and oxygens only) without the smearing caused by the limited range of $s$ accessible to experiment.
    The inset shows the atom pair distances in the $\C{2v}$ geometry.
    }
    \label{fig:EDmin}
\end{figure}

The steady-state signal generated from the initial distribution is given in Fig.~\ref{fig:EDmin}.
There are only very minor differences in the simulated PDF between the ensemble average and the result from a single geometry.

\clearpage

\section{Additional Numerical Results}
We give here a series of additional numerical results to support the findings of the main paper. This includes additional results for the calculations performed using the 6-31+G* basis. For these calculations we used the same set of 200 initial geometries, momenta and effective Bloch-spheres as for the aug-cc-pVDZ calculations. In total there were 199 trajectories that ran for 500\,fs, trajectories for which spin-contamination resulted in a failure of the SCF convergence were again finished using SS-CASSCF. 

\begin{figure}
    \centering
    \includegraphics[width=0.495\textwidth]{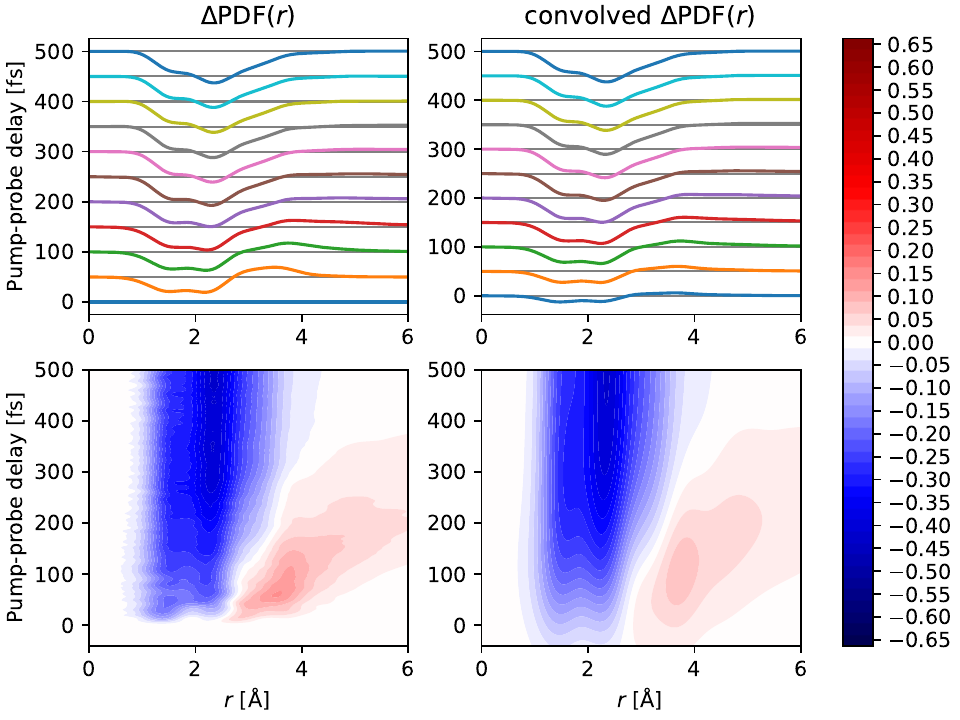}
    \includegraphics[width=0.495\textwidth]{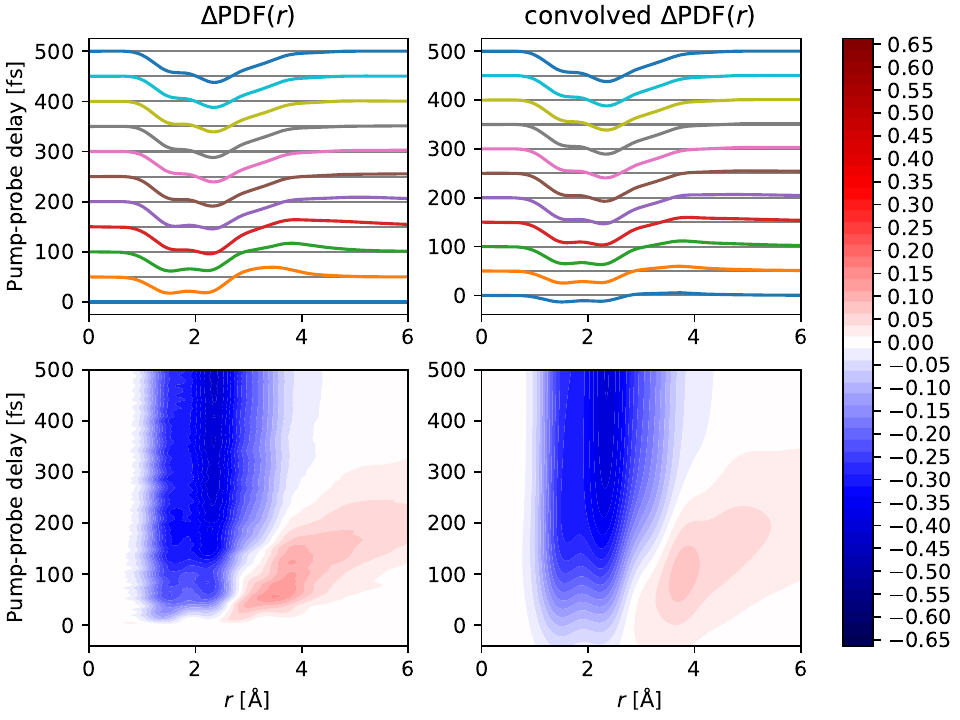}
    \caption{Simulated ultrafast electron-diffraction results for the trajectories that correspond to rxn.~products I at $t=500$\,fs, left calculated using the aug-cc-pVDZ basis set and the right with the 6-31+G* basis set.
    The panels on the left of each subfigure show the change in the probability density function relative to the initial configuration.
    The panels on the right show the same data convolved with a 160\,fs (FWHM) Gaussian to simulate the instrument response function.
    Blue is loss, red is gain, with equally-spaced contour levels showing the height of the $\Delta \mathrm{PDF}(r)$ signal relative to the maximum peak height in the steady state PDF. }
    \label{fig:gued-rxn-1}
\end{figure}

\begin{figure}
    \centering
    \includegraphics[width=0.495\textwidth]{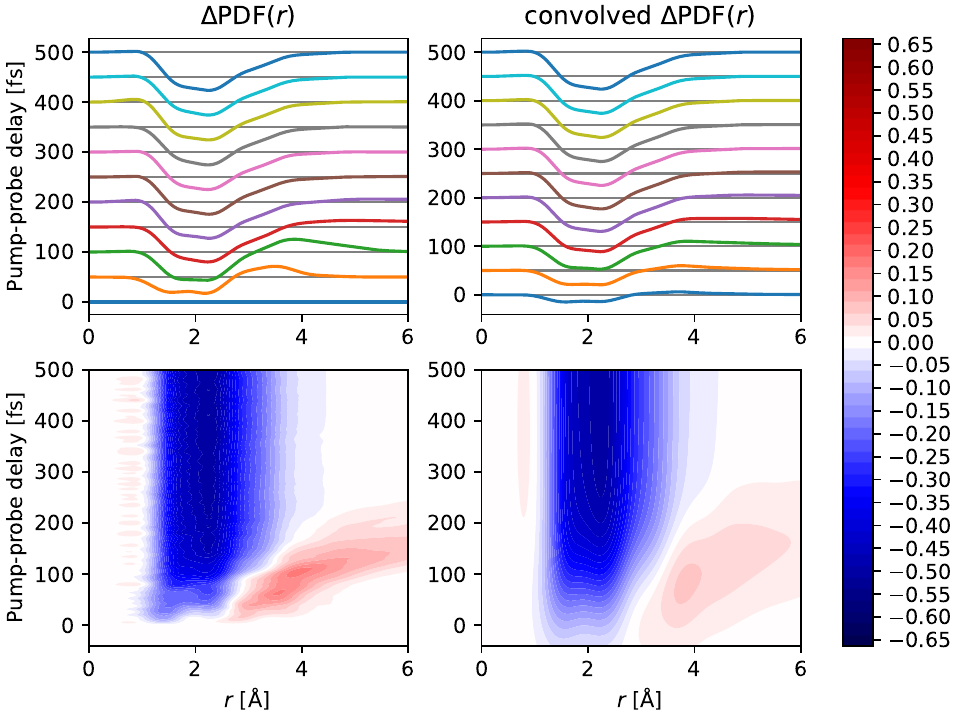}
    \includegraphics[width=0.495\textwidth]{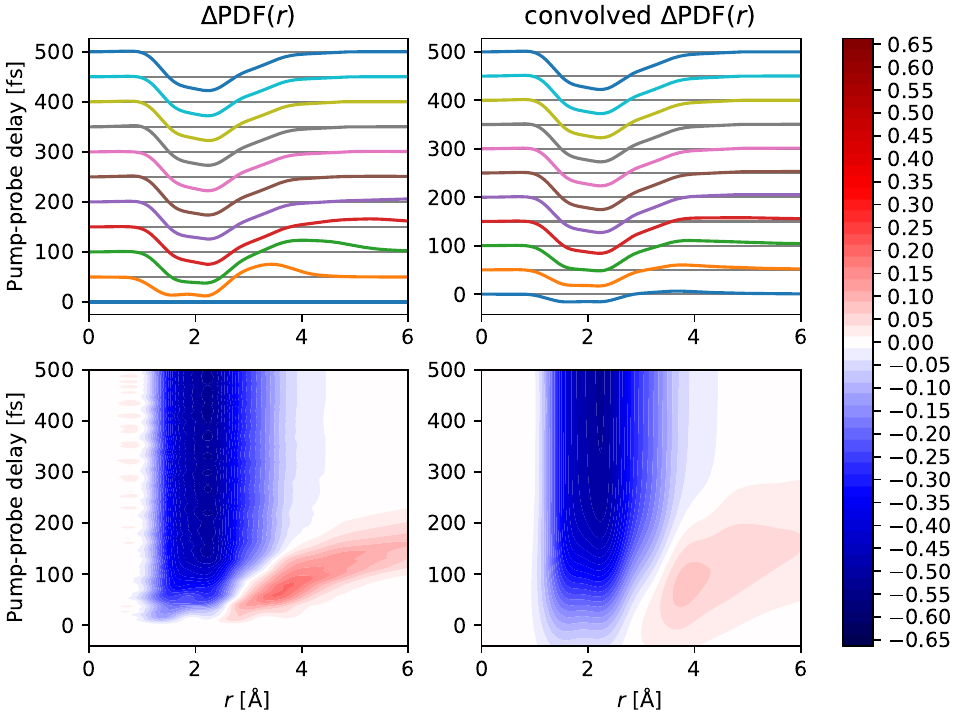}
    \caption{Simulated ultrafast electron-diffraction results for the trajectories that correspond to rxn.~products II at $t=500$\,fs, left calculated using the aug-cc-pVDZ basis set and the right with the 6-31+G* basis set.
    The panels on the left of each subfigure show the change in the probability density function relative to the initial configuration.
    The panels on the right show the same data convolved with a 160\,fs (FWHM) Gaussian to simulate the instrument response function.
    Blue is loss, red is gain, with equally-spaced contour levels showing the height of the $\Delta \mathrm{PDF}(r)$ signal relative to the maximum peak height in the steady state PDF. }
    \label{fig:gued-rxn-2}
\end{figure}

\begin{figure}
    \centering
    \includegraphics[width=0.495\textwidth]{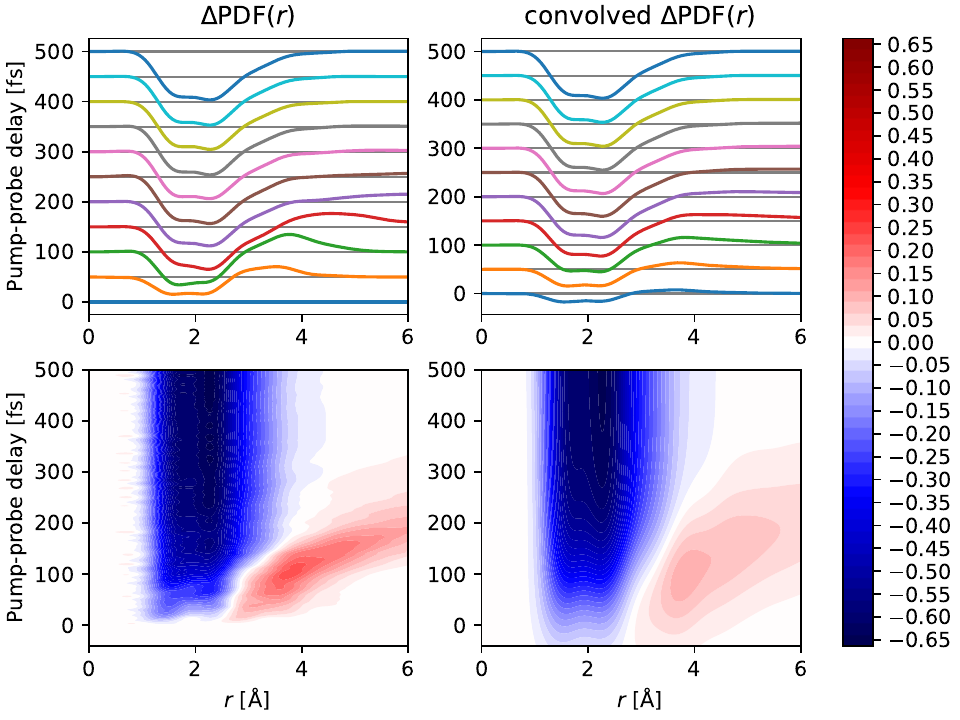}
    \includegraphics[width=0.495\textwidth]{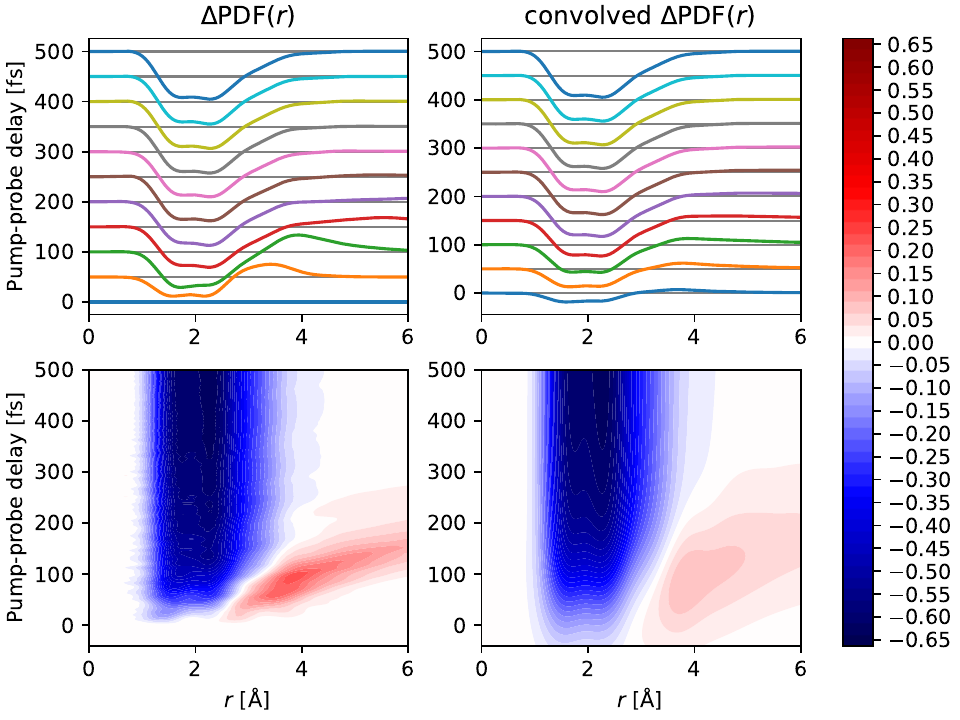}
    \caption{Simulated ultrafast electron-diffraction results for the trajectories that correspond to rxn.~products III at $t=500$\,fs, left calculated using the aug-cc-pVDZ basis set and the right with the 6-31+G* basis set.
    The panels on the left of each subfigure show the change in the probability density function relative to the initial configuration.
    The panels on the right show the same data convolved with a 160\,fs (FWHM) Gaussian to simulate the instrument response function.
    Blue is loss, red is gain, with equally-spaced contour levels showing the height of the $\Delta \mathrm{PDF}(r)$ signal relative to the maximum peak height in the steady state PDF. }
    \label{fig:gued-rxn-3}
\end{figure}

Figures~\ref{fig:gued-rxn-1}--\ref{fig:gued-rxn-3} show the ultrafast electron diffraction signal resolved by the final reaction products for both the aug-cc-pVDZ basis and also the 6-31+G*. 

The total yields of each molecular fragment are given in Table~\ref{tab:product-yields-6-31+G*}, from which we see that the products produced are broadly consistent between different choices of basis set. Table~\ref{sitab:reaction-products-6-31+G*} shows the yield of the main reaction pathways at 500\,fs for the calculations with the 6-31+G* basis set. From this (including also those undissociated molecules in the C3 pathway) we can calculate the fraction of all C3 and C2 trajectories that dissociate via C3 as 0.5 with a 95\% Wilson score confidence interval of $(43\%,57\%)$ corresponding to a C3/C2 ratio of 1 with the a statistical confidence interval at 95\% of $(0.75,1.35)$.

\begin{figure}[t]
    \centering
    \includegraphics{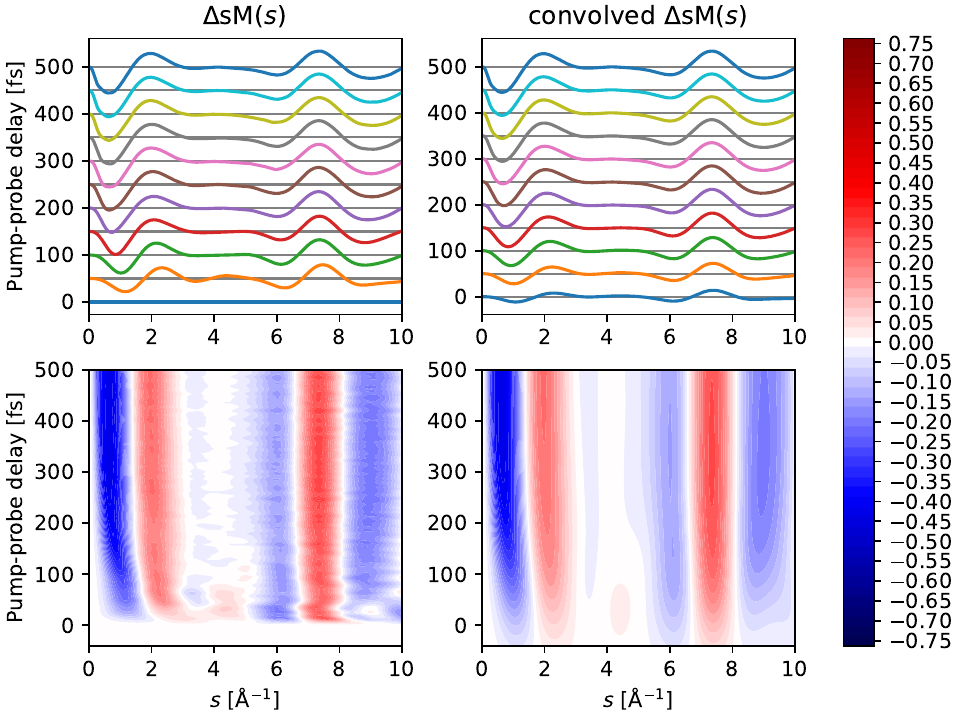}
    \caption{Predicted undamped GUED difference signal in momentum ($s$) space, for the aug-cc-pVDZ trajectories.}
    \label{fig:aug-cc-pVDZ-sMs}
\end{figure}

\begin{figure}[t]
    \centering
    \includegraphics{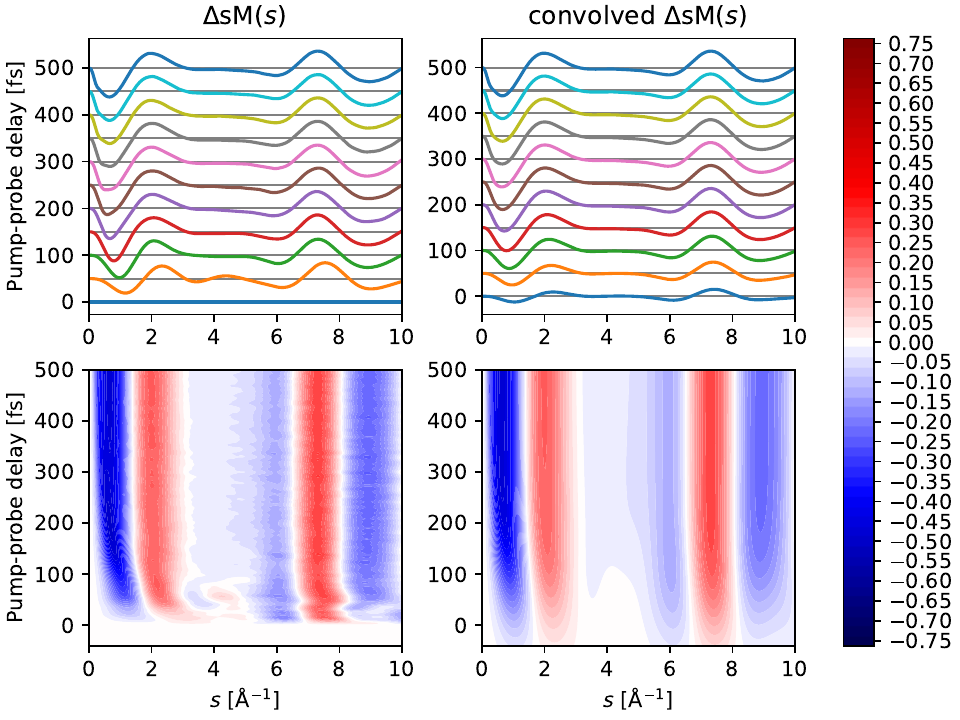}
    \caption{Predicted undamped GUED difference signal in momentum ($s$) space, for the 6-31+G* trajectories.}
    \label{fig:6-31+Gd-sMs}
\end{figure}

{Figures \ref{fig:aug-cc-pVDZ-sMs} and \ref{fig:6-31+Gd-sMs} show the predicted GUED signal $\Delta sM(s,t) = \frac{sM(s,t)-sM_{\rm ss}(s)}{\min(\mathrm{PDF}_{\rm ss}(r))}$ in momentum space. Integration using Eq.~(20d) recovers the difference signal $\Delta \mathrm{PDF}(r)$. We give this for completeness as it is independent of the choices of $\alpha$, $s_{\rm min}$ and $s_{\rm max}$ needed to convert to position space.}

\begin{figure}
\centering
  \centering
  \includegraphics{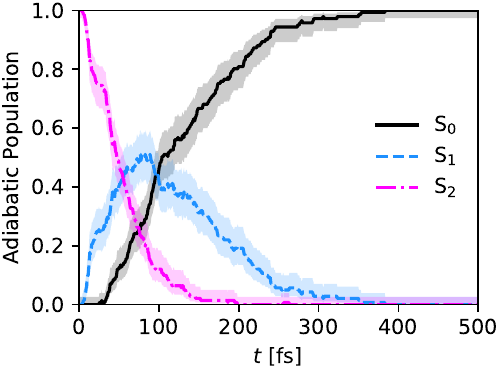}
  \includegraphics{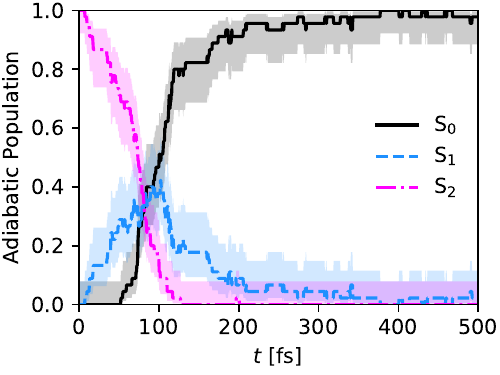}
\caption{Average (unconvolved) adiabatic populations as a function of time for the 141 C3 trajectories (those that produce \ch{C3H6}) left and the 45 C2 trajectories (those that produce \ch{C2H4}) right (calculated with aug-cc-pVDZ basis). Shaded region shows an approximate 95\% confidence interval (the Wilson score interval).\cite{Wilson1927BinomialErrors}  
    }
\label{supfig:popsC3vsC2}
\end{figure}

\begin{figure*}
    \centering
    \includegraphics[width=\textwidth]{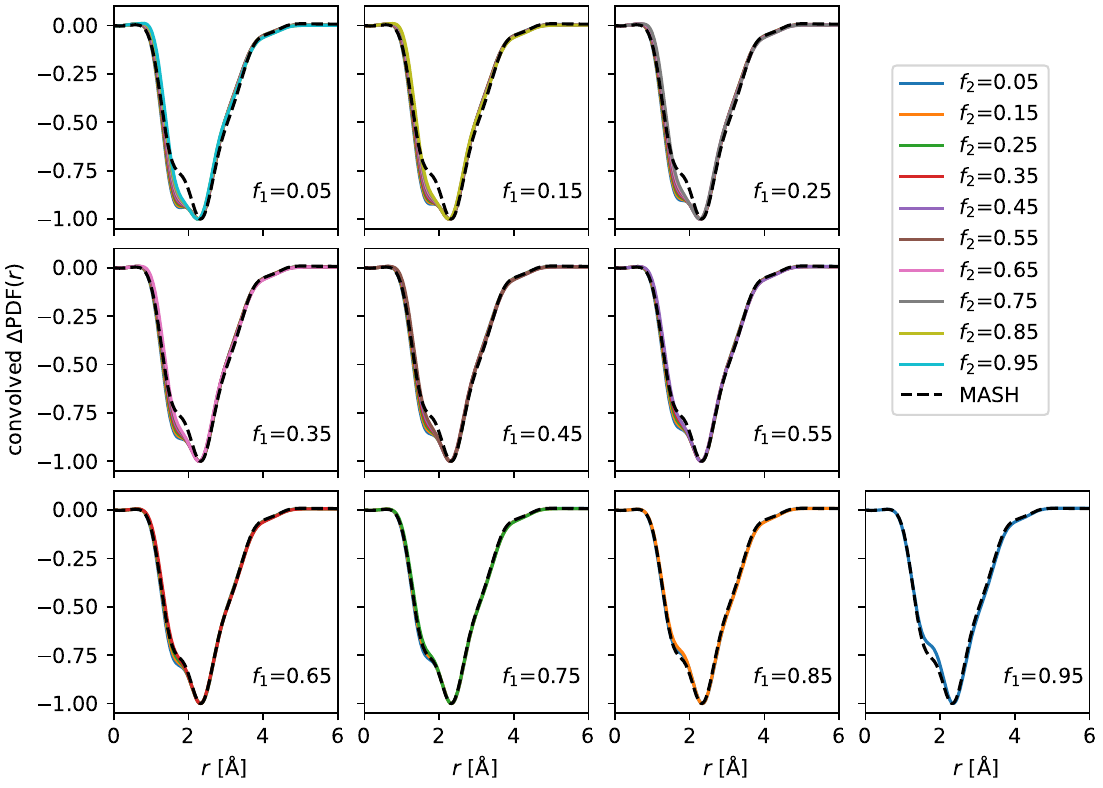}
    \caption{As Fig.~6 from the main text except normalised so that $\min(\Delta\text{PDF}(r))=-1$.}
    \label{fig:gued-weighted-averages-normalised}
\end{figure*}

\begin{figure*}[t]
    \centering
    \includegraphics[width=\textwidth]{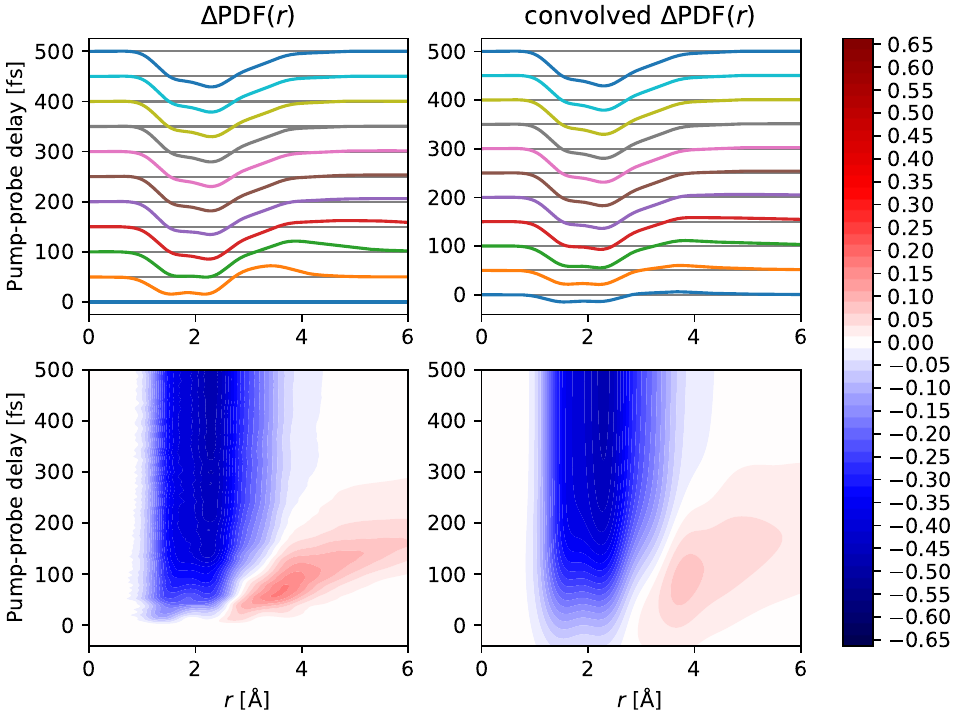}
    \caption{Simulated ultrafast electron-diffraction results calculated using the 6-31+G* basis set.
    The panels on the left show the change in the probability density function relative to the initial configuration.
    The panels on the right show the same data convolved with a 160\,fs (FWHM) Gaussian to simulate the instrument response function.
    Blue is loss, red is gain, with equally-spaced contour levels showing the height of the $\Delta \mathrm{PDF}(r)$ signal relative to the maximum peak height in the steady state PDF.    
    }
    \label{fig:gued-6-31+G*}
\end{figure*}

\begin{figure*}[t]
    \centering
    \includegraphics[width=\textwidth]{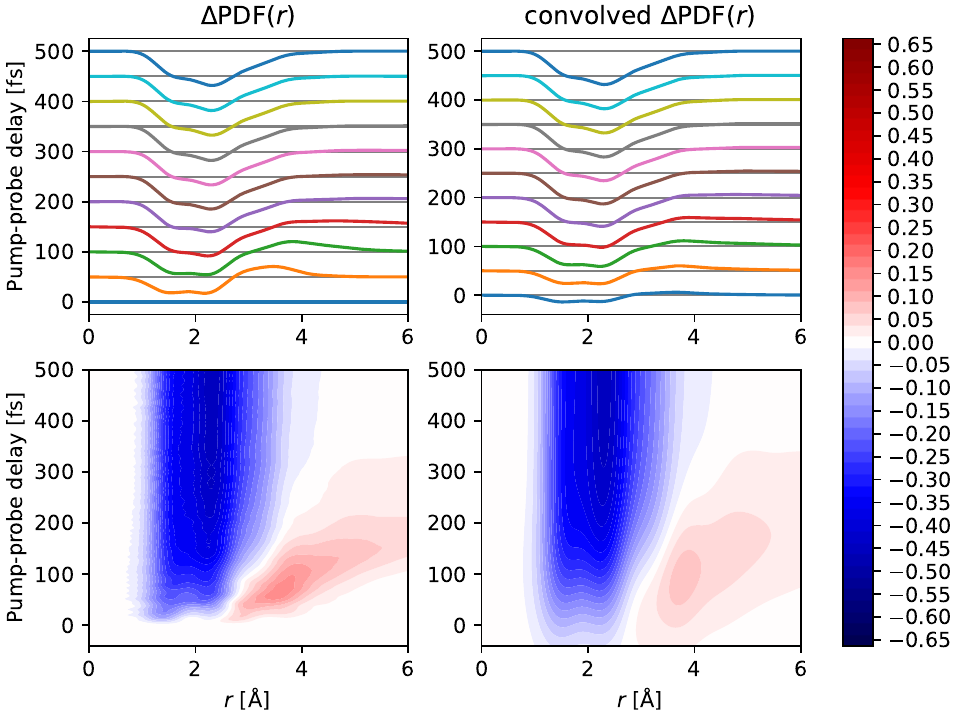}
    \caption{Simulated ultrafast electron-diffraction results calculated as an equally weighted average over results from the aug-cc-pVDZ and 6-31+G* basis sets.
    The panels on the left show the change in the probability density function relative to the initial configuration.
    The panels on the right show the same data convolved with a 160\,fs (FWHM) Gaussian to simulate the instrument response function.
    Blue is loss, red is gain, with equally-spaced contour levels showing the height of the $\Delta \mathrm{PDF}(r)$ signal relative to the maximum peak height in the steady state PDF.    
    }
    \label{fig:gued-6-31+G*}
\end{figure*}

\begin{table}[t]
\centering
\begin{tabular}{lrr}
\toprule
\textbf{Fragment} & \textbf{Count} & \textbf{Yield} \\
 \midrule
 \ch{CO} & 125 & 63\% \\
 \ch{C3H6} & 94 & 47\% \\
 \ch{C2H4} & 99 & 50\% \\
 \ch{C2H2O} & 69 & 35\% \\
 \ch{CH2} & 30 & 15\% \\
 \ch{C4H6O} & 5 & 2.5\% \\
 \ch{H} & 1 & 0.5\% \\
\ch{C3H5} & 1 & 0.5\% \\
\bottomrule
\end{tabular}
\caption{Total product yields  500\,fs after initial excitation for the calculations with the 6-31+G* basis set. Note the total number of initial \ch{C4H6O} molecules is 199. Reaction products are identified by using a cutoff radius of $2\AA$.}
\label{tab:product-yields-6-31+G*}
\end{table}

\begin{table}[t]
    \centering
    \begin{tabular}{lc|c|c}
    \toprule
         & I & II & III  \\
         \textbf{Products} & \ch{C3H6} + \ch{CO} & \ch{C2H4} + \ch{C2H2O} &  \ch{C2H4} + \ch{CH2} + \ch{CO} \\
         \textbf{Count} & 94 (47\%)  & 69 (35\%) & 30 (15\%) \\
    \bottomrule
    \end{tabular}
    \caption{Main reaction products at 500\,fs for the calculations with a 6-31+G* basis set. Reaction products are identified by using a cutoff radius of $2\AA$. %
}
    \label{sitab:reaction-products-6-31+G*}
\end{table}

\begin{figure}[t]
    \centering
    \includegraphics{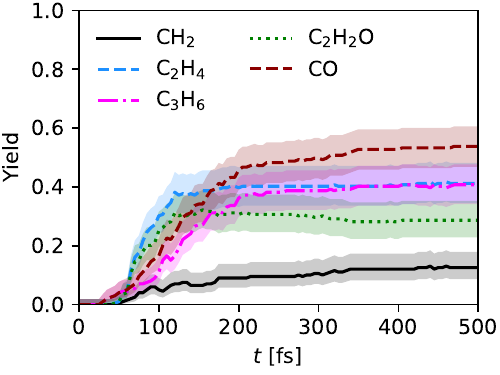}
    \caption{Average (unconvolved) fragment yields, for the 5 most common fragments, as a function of time for the 199 6-31+G* trajectories. Shaded region shows an approximate 95\% confidence interval (the Wilson score interval).\cite{Wilson1927BinomialErrors}  
    }
    \label{fig:fragment_yields_t_dep-6-31+G*}
\end{figure}

\begin{figure}[t]
    \centering
    \includegraphics{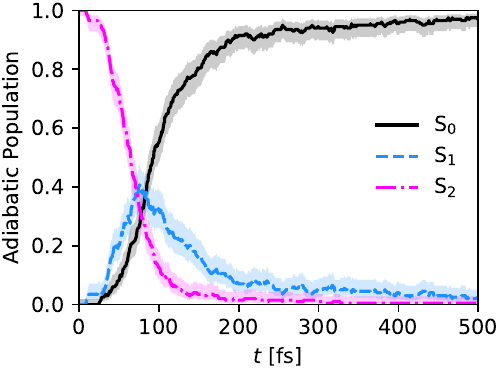}
    \caption{Average (unconvolved) adiabatic populations as a function of time for the 199 trajectories calculated with the 6-31+G* basis set. Shaded region shows an approximate 95\% confidence interval (the Wilson score interval).\cite{Wilson1927BinomialErrors}  
    }
    \label{fig:pops-si-6-31+G*}
\end{figure}

\begin{figure}[t]
    \centering
    \includegraphics{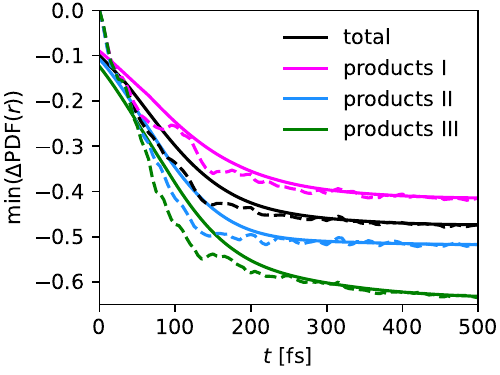}
    \caption{Minimum value of the simulated electron diffraction signal given as a change in probability density function relative to the initial configuration calculated using the 6-31+G* basis. 
    Dashed lines show the unconvolved results from the 199 trajectories, and solid lines show the convolved results using a 160\,fs (FWHM) Gaussian to simulate the instrument response function.
    Note the height of the $\Delta \mathrm{PDF}(r)$ signal in all cases is given relative to the maximum peak height in the steady state PDF\@. Note that here the curves are normalised such that a weighted average over all the reactions would recover the total signal.  
    }
    \label{fig:min_vals_elec_diff-si}
\end{figure}

\begin{figure*}
    \centering
    \includegraphics[width=\textwidth]{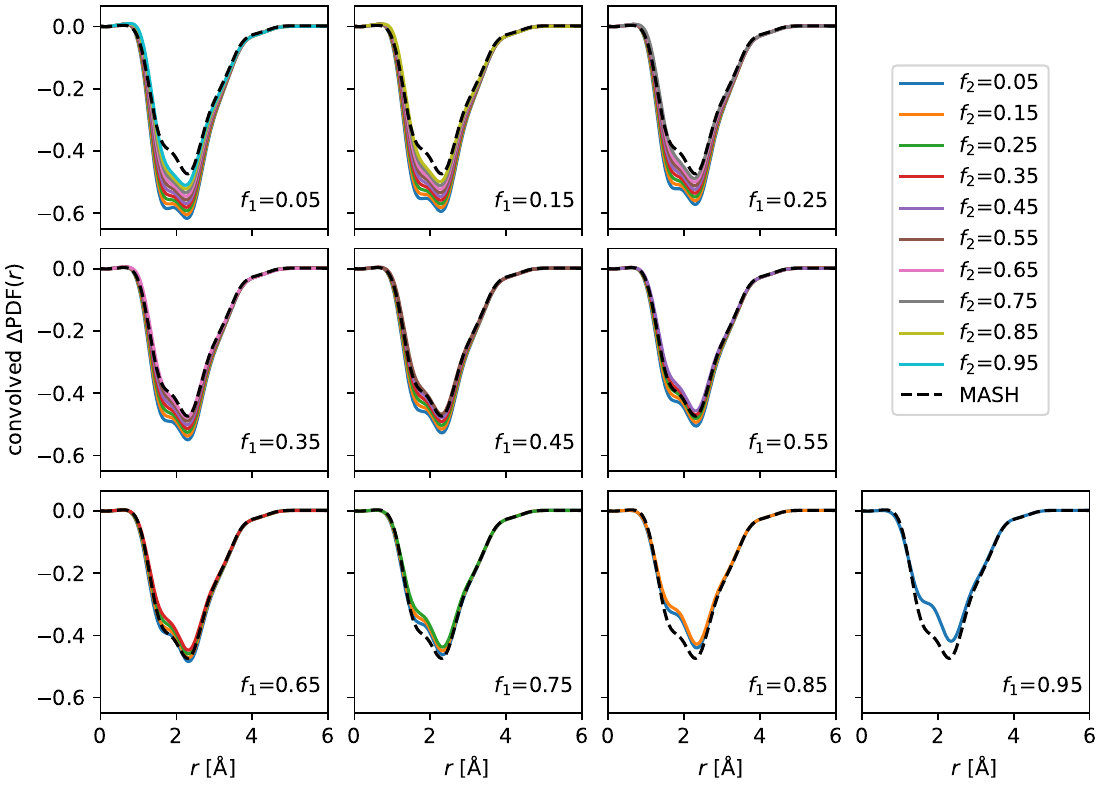}
    \caption{Weighted average over the three dominant reaction products of the simulated ultrafast electron-diffraction results at 500\,fs. $f_1$ corresponds to the fraction of products I and $f_2$ to the fraction of products II used in the weighted average, with the remaining fraction corresponding to products III\@.  
    Results are shown here  for the signal convoluted with with a 160\,fs (FWHM) Gaussian to simulate the instrument response function. Dashed line shows the prediction from the 6-31+G* trajectories. %
    }
    \label{fig:gued-weighted-averages}
\end{figure*}

\begin{figure}[t]
    \centering
    \includegraphics{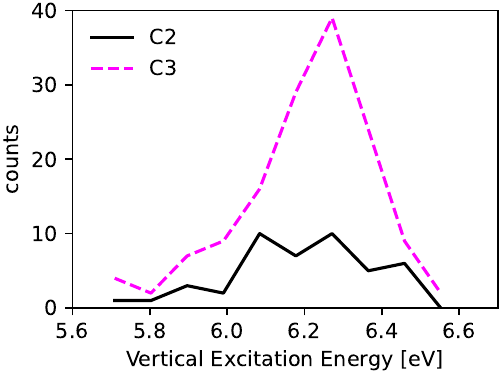}
    \caption{Histogram showing the vertical excitation energy for trajectories that produce products I (C3) and those that produce products II+III (C2).}
    \label{fig:VEE_histogram_C2vsC3}
\end{figure}

\begin{comment}
\begin{figure}
    \centering
    \includegraphics{fig/ElectronDiffraction/gued.pdf}
    \caption{Example time-dependent signals calculated from a single FSSH trajectory.
    The column on the right is convoluted/convolved with a 150 fs (FWHM) Gaussian to simulate the instrument response function.
    Red is loss, blue is gain, with equally-spaced but arbitrary contour levels.
    (could also show sMs at separate times - SI only)}
    \label{fig:gued}
\end{figure}
\end{comment}

\begin{comment}
\begin{itemize}
    \item compare with Xray tabulated data
    \item compare with including gs too
    \item compare with phase shifts
    \item compare with absorption
    \item comparison with nuclear shape
    \item ZPE
    \item exchange effects
    \item valence bonding / electronic-state dependence (compare with Iatom using same method)
    \item not just at minimum, but at products too
    \item show PDF with infinite smax too or maybe without fs at all
\end{itemize}
\end{comment}

\clearpage

\section*{References}
\bibliography{references,other,extra_refs,molpro} %